\documentclass[11pt,a4paper]{article}
\pdfoutput=1
\usepackage{jheppub}
\usepackage{graphicx}
\usepackage{amsthm}
\usepackage{amsmath}
\usepackage{tcolorbox}
\usepackage{makecell}
\usepackage{hyperref}
\usepackage{xcolor}
\usepackage{multirow}
\definecolor{lightblue}{rgb}{0.68,0.85,0.9}

\newcommand{\mz}{\mathtt{z}}
\newcommand{\mzb}{\bar{\mathtt{z}}}

\theoremstyle{definition}

\newtheorem{prop}{Proposition}
\newtheorem{conj}{Conjecture}

\title{On Solving Dual Conformal Integrals in Coulomb-branch Amplitudes and Their Periods}

\author[a,b]{Song He,}\emailAdd{songhe@itp.ac.cn}
\author[a]{Xuhang Jiang}\emailAdd{xhjiang@itp.ac.cn}
\affiliation[a]{Institute of Theoretical Physics, Chinese Academy of Sciences, Beijing 100190, China}
\affiliation[b]{School of Fundamental Physics and Mathematical Sciences, Hangzhou Institute for Advanced Study and ICTP-AP, UCAS, Hangzhou 310024, China}

\abstract
{We define and study infinite families of all-loop planar, dual conformal invariant (DCI) integrals, which contribute to four-point Coulomb-branch amplitudes and correlators in ${\cal N}=4$ supersymmetric Yang-Mills theory, by solving ``boxing" differential equations via \texttt{HyperlogProcedures}~\cite{hyperlogprocedures}; The resulting single-valued harmonic polylogarithmic functions (SVHPL) are nicely labeled by ``binary" strings of $0$ and $1$ without consecutive $1$'s. These functions are special cases of the so-called generalized ladders studied in~\cite{Drummond:2012bg}, where extended Steinmann relations (no consecutive $1$'s) are imposed due to planarity. Our results can be viewed as ``two-dimensional" extensions of the well-known ladder integrals to many more infinite families of DCI integrals: the ladders have strings with a single $1$ followed by all $0$'s, and the other extreme, which nicely evaluate to the ``zigzag" SVHPL functions with alternating $1$'s and $0$'s, are nothing but the four-point DCI integrals from the very special family of anti-prism $f$-graphs (while all other binary DCI integrals lie in between these two extreme cases). We also study periods of these integrals: while their periods are in general complicated single-valued multiple zeta values (SVMZV), the ``zigzag" DCI integrals from anti-prism gives exactly the famous ``zigzag" periods proportional to $\zeta_{2L{+}1}$, and empirically it provides a numerical lower-bound for $L$-loop periods of any binary string, with the upper-bound given by that of the ladder (also proportional to $\zeta_{2L{+}1}$). Based on $f$-graphs as a tool for studying these periods, we discuss several interesting facts and observations about these (motivic) SVMZV and relations among them to all loops, and enumerate a basis for them up to $L=10$.}

\begin{document}
\maketitle

\section{Introduction}
Recent years have witnessed enormous progress in the study of scattering amplitudes and other physical quantities in quantum field theory, especially in the most symmetric of all four-dimensional gauge theories, the ${\cal N}=4$ supersymmetric Yang-Mills (SYM) theory in the planar limit. Much of the progress comes from the discovery of remarkable new structures within the `theoretical data' from hard computations, which leads to insights and powerful new tools for computation, and more data and more discoveries. 

The four-point amplitude in planar ($\mathcal{N}\!=\!4$) SYM has long served as an important benchmark in our perturbative computations. It was first determined at the integrand level via generalized unitarity through six loops \mbox{\cite{Bern:1997nh,Anastasiou:2003kj,Bern:2005iz,Bern:2006ew,Bern:2007ct,Bern:2012di}}, to eight loops using the `soft-collinear bootstrap' in \mbox{\cite{Bourjaily:2011hi,Bourjaily:2015bpz}}; note that the four-point planar integrand (as well as those for higher-point squared amplitudes) can all be unified in and extracted from the integrand of four-point stress-tensor correlation function (exploiting the correspondence between amplitude squared/Wilson loops and correlation functions~\cite{Alday:2010zy,Adamo:2011dq, Alday:2007hr,Drummond:2007aua,Brandhuber:2007yx,Bern:2008ap,Drummond:2008aq,Arkani-Hamed:2010zjl,Mason:2010yk,Eden:2010zz,Eden:2010ce,Caron-Huot:2010ryg,Eden:2011yp,Eden:2011ku}), thus in~\cite{Bourjaily:2016evz} the planar integrands of four-point correlator and amplitude were determined up to ten loops by a set of graphical rules based on the so-called $f$-graphs. More recently, a new universal graphical rule based on ``cusp limit" was described in~\cite{He:2024cej} which brought this benchmark to eleven~\cite{He:2024cej} and twelve loops~\cite{Bourjaily:2025iad}. These results thus provide invaluable and rich theoretical data for the perturbative study of correlator, amplitudes and related Feynman integrals.

Among other things, the planar integrands of four-point amplitudes in terms of dual conformal invariant integrals (DCI integrals) after taking lightlike limits of $f$-graphs, which exhibit extremely rich combinatorial~\cite{He:2024cej, Bourjaily:2025iad} and even geometrical structures~\cite{He:2024xed, He:2025rza}, provide an exciting playground for studying Feynman integrals with nice properties which evaluate to {\it transcendental} functions and numbers that are of great interests for many reasons. By evaluating these DCI integrals contributing to four-point amplitudes or even the more general conformal integrals for four-point correlator or the integrated correlators viewed as periods, one learns valuable lessons not only about mathematical structures of the full amplitudes and (integrated) correlators, but also lead to new, powerful tools for studying more general Feynman integrals and periods with implications in perturbative QFT and mathematics such as number theory. One such example is the intimate relations between Feynman integrals and multiple zeta values~\cite{Broadhurst:1995km,Kreimer:1996js,Broadhurst:1996az,Broadhurst:1996ur,Broadhurst:1996kc}. The \textit{multiple zeta values} (MZVs), which are also called Euler-Zagier sums, Euler sums, or multiple harmonic series in the literature, appear ubiquitously in the results of single-scale Feynman integrals or their Laurent expansion with respect to dimensional parameter $\epsilon$ in dimensional regularization \cite{Broadhurst:1996kc,Kreimer:2000zh,Brown:2008um,Schnetz:2008mp,Baikov:2010hf}\footnote{Of course not all single-scale Feynman integrals evaluate to MZVs~\cite{Brown:2009ta,Brown:2010bw}.}. They are defined as
\begin{equation}
    \zeta_{n_1,n_2,\ldots,n_{r}}=\sum_{1\le k_1<k_2<\ldots<k_{r}}\frac{1}{k_{1}^{n_1}k_{2}^{n_2}\ldots k_{r}^{n_{r}}}, n_{i}\in \mathbb{N}, n_{r}\ge 2.
\end{equation}
$w=n_1+n_2+\ldots+n_{r}$ is called the weight and $r$ is called the depth of MZV. Even for multi-scale cases, many families of non-trivial Feynman integrals can be evaluated as multiple polylogarithms~\cite{Goncharov:1998kja,Goncharov:2001iea} and MZVs further emerge as the special values of these multiple polylogarithms~\cite{Borwein:1999js,Goncharov:2001zfh,zhao2016}. MZVs are not all independent. There are many relations, some still conjectural, among the MZVs of the same weight~\cite{hoffman1992,HOFFMAN1997477,Borwein:1996yq,Borwein:1999js,Zudilin2003,Charlton:2024bmk}. The independent basis of MZVs and reduction of general MZVs have been studied both by computer algebra methods~\cite{Broadhurst:1996az,Bailey:1999nv,Blumlein:2009cf} and by the motivic version of MZVs~\cite{Goncharov:2001iea,Goncharov:2005sla,Brown:2011ik,brown2012mixed,Brown:2013gia,Brown:2021hzr}. The study of MZVs emerging from Feynman integrals also inspires the study of Feynman integrals by treating them as periods of motives associated to graph polynomials~\cite{belkale2003matroids,Bloch:2005bh,Brown:2008um,Brown:2009ta}, which has led to the development of powerful method  for direct integration, such as the automated package \texttt{HyperInt}~\cite{Panzer:2013cha,Panzer:2014caa}.

Moreover, for various classes of physical quantities or Feynman integrals with simple kinematics, only the single-valued version of multiple (harmonic) polylogarithms (SVHPL)
~\cite{Brown:2004ugm} and single-valued MZVs appear~\cite{Dixon:2012yy,Schnetz:2013hqa,Drummond:2012bg,Schlotterer:2012ny,Stieberger:2013wea,Leurent:2013mr,Stieberger:2014hba,Gurdogan:2020ppd,Alday:2024ksp}. They form a special subclass of the multiple polylogarithms and MZVs respectively. The quantities we will study in this paper belong to this subclass. As we have discussed earlier, we are studying a special class of dual conformal integrals (DCI integrals) that will appear in the four-point correlator $\langle \mathcal{O}_{2}(x_{1})\mathcal{O}_{2}(x_{2})\mathcal{O}_{2}(x_{3})\mathcal{O}_{2}(x_4)\rangle$ where $\mathcal{O}_{2}=Y_{I}Y_{J}\mathrm{tr}(\Phi^{I}\Phi^{J}),Y^2=0$ in $\mathcal{N}=4$ SYM. These DCI integrals can be generated from $f$-graphs which serve as an ansatz for the integrands of the correlator\footnote{Not all $f$-graphs will contribute to the correlator, only a small subset of them will when it goes to very high loops and these $f$-graphs have nonzero coefficients in the solution of initial ansatz. And due to the duality between correlators and amplitudes~\cite{Alday:2010zy,Adamo:2011dq,Eden:2010zz,Eden:2010ce,Caron-Huot:2010ryg,Eden:2011yp,Eden:2011ku}, some of the DCI integrals generated from these $f$-graphs with nonzero coefficients are Feynman integrals in four-point Coulomb branch amplitudes of $\mathcal{N}=4$ SYM.}. We will also study their periods, defined as integration of all the external points $x_i$ with some given measure. This is equivalent to integrating all the points in $f$-graphs out. Physically, these periods are part of the integrated four-point correlators~\cite{Dorigoni:2021guq,Wen:2022oky,Zhang:2024ypu} which are studied perturbatively in the weak coupling limit. And mathematically, they are single-valued MZVs and bridge these $f$-graphs to the study of MZVs. These SVHPLs and their periods will be studied by using \texttt{HyperlogProcedures}~\cite{hyperlogprocedures}, which implements the method of graphical functions~\cite{Schnetz:2013hqa,Borinsky:2021gkd,Schnetz:2021ebf,Borinsky:2022lds,Schnetz:2024qqt}. The $f$-graphs correspond to the completion of Feynman graphs in the terminology of \cite{Schnetz:2013hqa} and they are more naturally related to periods than Feynman graphs. They will serve as another tool for us to study the relations among periods of different DCI integrals. 

Recently in~\cite{He:2025vqt}, we have bootstrapped all contributing DCI integrals at four loops and most of them at five loops can also be evaluated using different methods. We also studied various interesting structures to even higher loops, such as leading singularities as prefactors of pure functions, differential equations they satisfy, as well as the so-called magic identities by connecting these integrals to the Coloumb-branch amplitudes known from integrability~\cite{Coronado:2018cxj, Caron-Huot:2021usw}. It is clear from such a preliminary study that we have only touched the tip of an iceberg regarding these transcendental functions from evaluating DCI/conformal integrals contributing to four-point amplitudes/correlator, and a natural question is if we could study some all-loop DCI integrals similar to the famous ladder integrals~\cite{Broadhurst:1985vq, Usyukina:1992jd, Usyukina:1993ch, Broadhurst:1993ib}. In this paper, we point out that indeed one can find and solve infinite families of DCI integrals, which originate from certain $f$-graphs and contribute to four-point Coloumb-branch amplitudes, to very high loop orders. The way we compute such all-loop DCI integrals is by solving the second-order, ``boxing" differential equations (see also~\cite{Drummond:2006rz,Drummond:2010cz,Arkani-Hamed:2021iya}) using graphical functions~\cite{Schnetz:2013hqa,Borinsky:2021gkd,Schnetz:2021ebf,Borinsky:2022lds,Schnetz:2024qqt}, and the resulting SVHPL functions are labeled by length-$L$ strings of just $1$ and $0$'s depending on the corresponding $f$-graphs (which we call binary DCI integrals). Very nicely, the $L$-loop ladder integrals correspond to the string with a single $1$ followed by $L{-}1$ $0$'s, while the other extreme corresponds to what we call ``zigzag" DCI integrals, which are labeled by alternating $1$'s and $0$'s ({\it e.g.} $1010$, $101010$, $10101010$ for $L=4,6,8$). Generically such SVHPL functions have been classified in~\cite{Drummond:2012bg} which are known as ``generalized ladder integrals", but we emphasize that most of those conformal integrals do not correspond to planar DCI integrals: it turns out that the most general DCI integrals we will encounter correspond to those without consecutive $1$'s, which follows from the important physical constraints known as (extended) Steinmann relations~\cite{Steinmann:1960,Steinmann2:1960,Cahill:1973qp,Caron-Huot:2018dsv,Caron-Huot:2019bsq} (see~\cite{Caron-Huot:2020bkp} for a review). We will refer to these ES-satisfying SVHPL functions, which are the results of solving boxing differential equations (known as ``inverse boxing") of our binary DCI integrals, as binary Steinmann SVHPL functions; these functions are enumerated by the Fibonacci sequence. Based on these binary Steinmann functions, we can then study periods of the underlying $f$-graphs, which we computed completely up to $L=9$ and also prove or conjecture some interesting patterns to all loops.


This paper is organized as follows. In Sec.~\ref{sec:zigzagdef}, we study the so-called antiprism $f$-graphs and solve the zigzag dual conformal integrals generated from them. We will introduce and solve the boxing differential equations for these integrals, and show that the antiprism $f$-graphs are indeed directly related to the zigzag periods. While these functions and periods are well known in the context of $\phi^4$ theory~\cite{Brown:2012ia}, we will review them now in the context of antiprism $f$-graphs, which will serve as a starting point for the discussions in later sections. In Sec.~\ref{sec:inverseboxing}, we study the wider class of infinite families of binary dual conformal integrals: they can be solved by boxing differential equations or equivalently the graphical rules in~\cite{Schnetz:2013hqa,Borinsky:2021gkd,Borinsky:2022lds}, which then evaluate to this nice subset of Steinmann-satisfying SVHPL functions. Note that all binary DCI integrals are generated from $f$-graphs, thus our results can be viewed as a step in understanding more integrated results of $f$-graphs/DCI integrals contributing to the four-point correlator/amplitude in $\mathcal{N}=4$ SYM. In Sec.~\ref{sec:periods}, we study the periods of these binary DCI integrals (or rather the corresponding $f$-graphs) by \texttt{HyperlogProcedures}~\cite{hyperlogprocedures} and find some interesting phenomena in their periods. Especially, $f$-graphs can serve as a useful tool to study relations between periods of the integrals generated from them. In Sec.~\ref{sec:conclusion}, we summarize and point out some further directions. We leave some more detailed discussions to the appendices. In App.~\ref{app:fourier}, we prove how the antiprisms are related to zigzag periods in a diagrammatic way. In App.~\ref{app:notbinary}, we discuss (examples of) $f$-graphs that cannot be related to binary DCI integrals introduced earlier. In App.~\ref{app:weight}, the transcendental weight properties of those $f$-graphs that can be related to binary DCI integrals are discussed. They are classified into weight-preserving and weight-dropping types and understood from the differential equations. In App.~\ref{app:proof}, we give a proof to an important proposition of this paper. In App.~\ref{app:Bxxx1}, we show how to relate binary Steinmann SVHPLs whose words end with '1' to the $f$-graphs we are discussing. It finishes the construction for the ``canonical series'' of $f$-graphs for binary Steinmann SVHPLs.


\section{An invitation: from ladders to zigzag DCI integrals}\label{sec:zigzagdef}

In this section, we present a new family of all-loop dual conformal integrals called ``zigzag DCI": it is the unique four-point DCI from taking lightlike limits of the special class of even-loop $f$-graphs called ``antiprism"~\cite{Bourjaily:2016evz}, which at even $L$ loops has been conjectured (and checked up to $L=12$ or $16$ points) to be the unique $f$-graph with the largest coefficient in all the $L$-loop $f$-graphs. 
\subsection{The special family of ``zigzag" DCI integrals from the antiprism \texorpdfstring{$f$-graphs}{f-graphs}}\label{sec:antiprism}
Among all the $f$-graphs which contribute to the integrand of four-point correlator $\langle\mathcal{O}_{2}\mathcal{O}_{2}\mathcal{O}_{2}\mathcal{O}_{2}\rangle$ in $\mathcal{N}=4$ SYM, the so-called antiprism $f$-graphs (with $2m$ points) form a very special infinite family. It has been conjectured and checked up to 12 loops~\cite{Bourjaily:2016evz,Bourjaily:2025iad} that their coefficients are given by signed Catalan numbers, $(-)^m C_{m{-}3}$, {\it e.g.} $1, -1, 2, -5, 14, -42$ for $m=3,4,5,6,7,8$ or $L=2m-4=2,4,6,8,10,12$. Moreover, they are highly symmetric $f$-graphs without any numerators, and the first few examples for $m=3,4,5,6$ are given in \eqref{eq:antiprism}. 
\begin{equation}\label{eq:antiprism}
    \vcenter{\hbox{\includegraphics[scale=0.3]{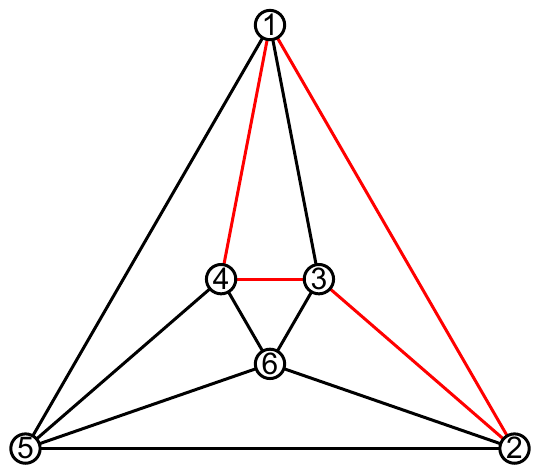}}},\, \vcenter{\hbox{\includegraphics[scale=0.3]{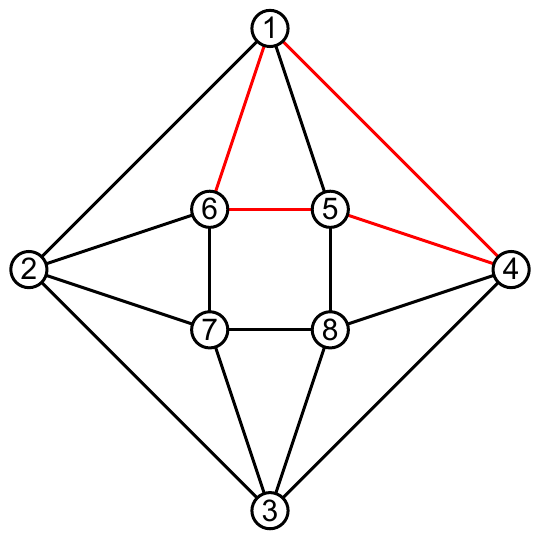}}}, \, \vcenter{\hbox{\includegraphics[scale=0.3]{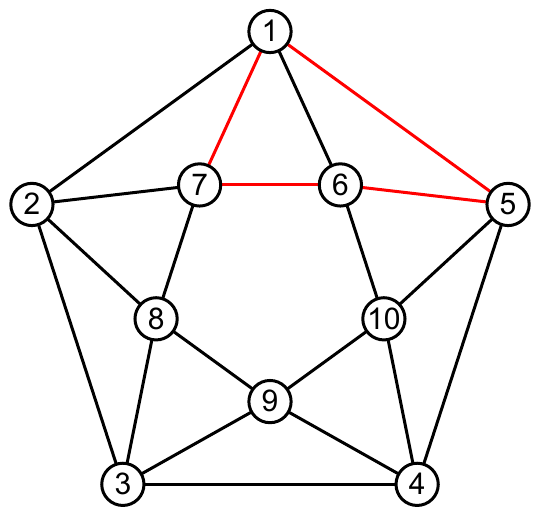}}}, \, \vcenter{\hbox{\includegraphics[scale=0.3]{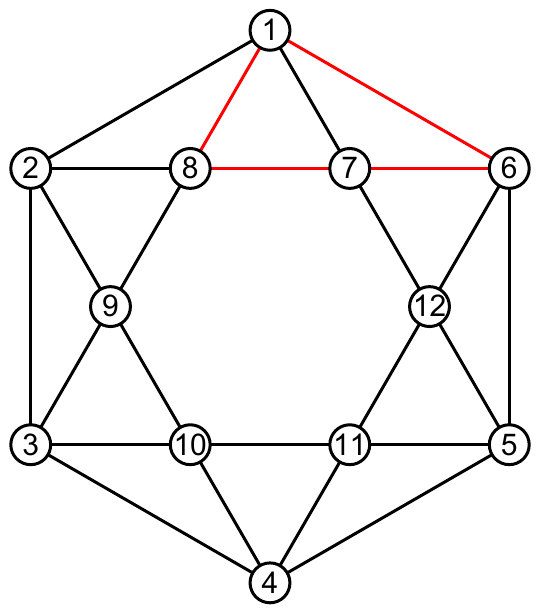}}}, \, \ldots
\end{equation}
Remarkably, as pointed out in~\cite{Bourjaily:2016evz}, the contribution from these special anti-prism $f$-graphs to the four-point amplitudes are also rather special: starting from $m=5$ (ten points or six loops for four-point amplitudes), there is a unique inequivalent four-cycle, which gives a unique DCI integral~\footnote{For $m=4$ or four loops, there is another possibility which gives the $2\times 2$ fishnet integral~\cite{He:2025vqt}.}. This is obtained by taking the four points $a,b,c,d$ on the cycle in order as external points and multiply the following factor:
\begin{equation}
    \xi_{4}=x_{ab}^{2}x_{bc}^{2}x_{cd}^{2}x_{da}^{2}x_{ac}^{4}x_{bd}^{4}.
\end{equation}
This is the way we extract the four-point amplitudes from $f$-graphs after taking the 4d light-like limit $x_{ab}^{2},x_{bc}^{2},x_{cd}^{2},x_{da}^{2}\to 0$. However, we can also study its 10d light-like limit and get the so-called Coulomb branch amplitudes \cite{Caron-Huot:2021usw}. That is, $x_{ab}^{2},x_{bc}^{2},x_{cd}^{2},x_{da}^{2}$ will acquire masses. They are usually taken as mass regulators for infrared divergence of the original massless amplitudes in $\mathcal{N}=4$ SYM. So we do \textit{not} assume $x_{a},x_{b},x_{c},x_{d}$ are null-separated in what follows\footnote{Besides, we usually suppress a factor $x_{ac}^{2}x_{bd}^{2}$ when drawing the diagrams of conformal integrals in dual space. That is, all the external points have degree one in the diagrams. Sometimes, we will supplement this factor to emphasize that the conformal weights of external points are actually 0.}.
We will study this series of DCI integrals which are obtained by taking the fully triangulated four cycle without inner points as indicated by the red cycles in \eqref{eq:antiprism}. The first few examples corresponding to the above $f$-graphs are listed in \eqref{eq:antiprismDCI integrals}
\begin{equation}\label{eq:antiprismDCI integrals}
    \vcenter{\hbox{\includegraphics[scale=0.35]{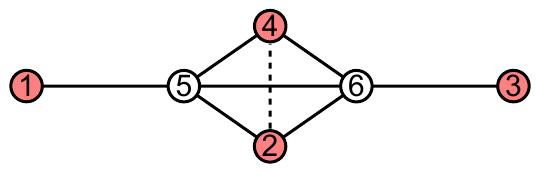}}},\, \vcenter{\hbox{\includegraphics[scale=0.35]{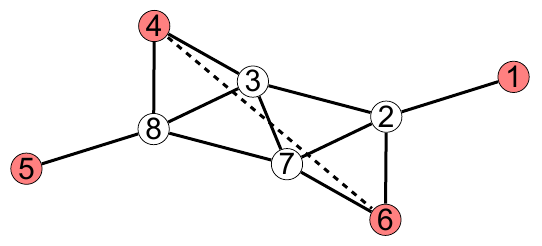}}}, \, \vcenter{\hbox{\includegraphics[scale=0.35]{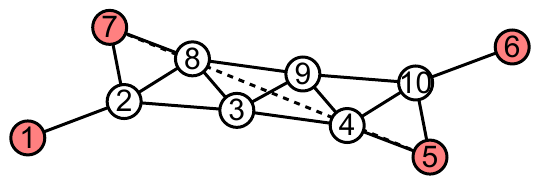}}}, \, \vcenter{\hbox{\includegraphics[scale=0.4]{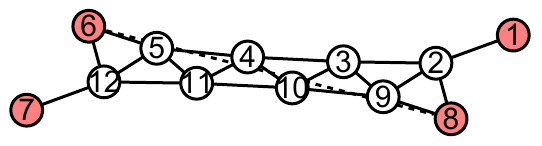}}}, \, \ldots
\end{equation}
where the \textit{dashed lines} are always used to represent numerators. Their diagrams in momentum space take the following forms
\begin{equation}
    \begin{aligned}
        \vcenter{\hbox{\vbox{\hbox to 3cm{\hfill \includegraphics[scale=0.3]{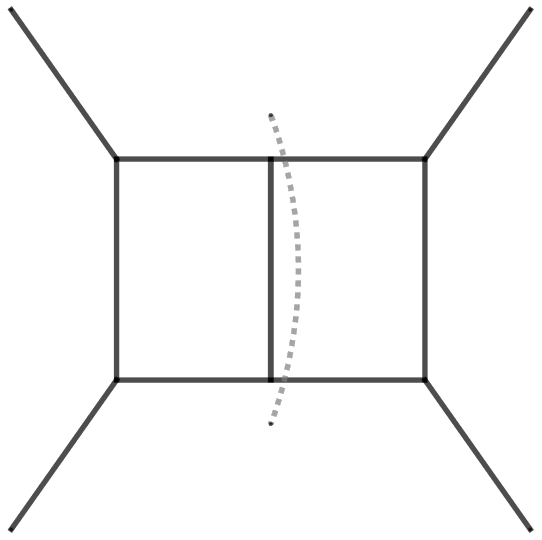} \hfill}}}}, \vcenter{\hbox{\vbox{\hbox to 3cm{\hfill \includegraphics[scale=0.3]{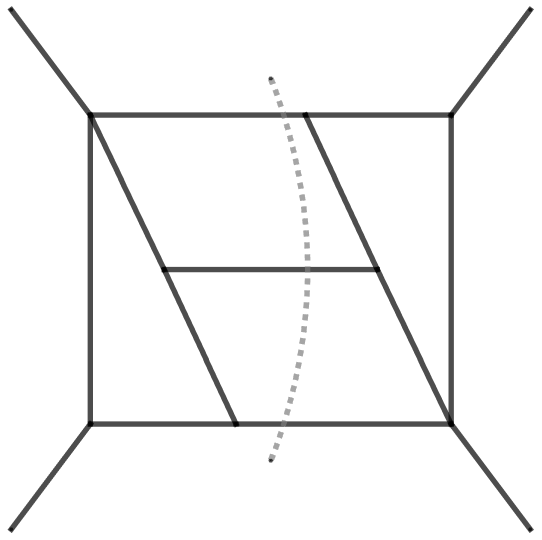} \hfill}}}},
        \vcenter{\hbox{\vbox{\hbox to 3cm{\hfill \includegraphics[scale=0.3]{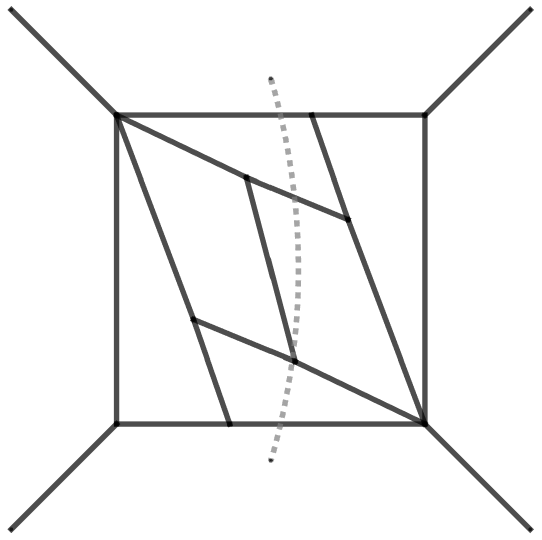} \hfill}}}}, \, \vcenter{\hbox{\vbox{\hbox to 3cm{\hfill \includegraphics[scale=0.3]{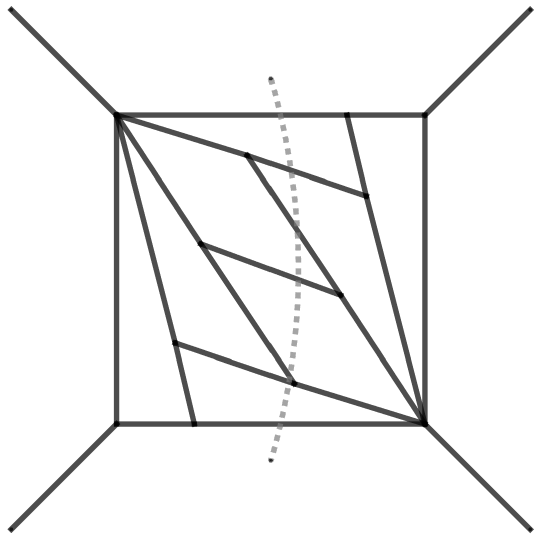} \hfill}}}},\ldots
    \end{aligned}
\end{equation}
We will show that they are exactly the zigzag dual conformal integrals constructed in \cite{Brown:2012ia} and their integrated version give the famous zigzag periods~\cite{Broadhurst:1995km}. This is done by solving the boxing differential equations of above DCI integrals.
\subsection{Solving boxing differential equations}
Let us first review the boxing differential equations satisfied by ladder type integrals. Higher-loop ladder integrals can be related to lower-loop ones by boxing differential equations. Thus we can solve them recursively. 

\begin{figure}
        \centering
        \includegraphics[width=0.3\textwidth]{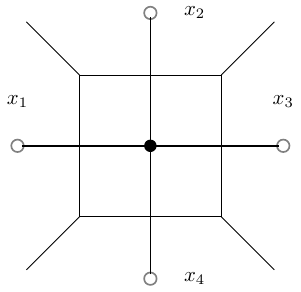}
        \caption{box integral which evaluates to Bloch-Wigner function with leading singularity $1/(z-\bar{z})$.}\label{fig:box}
    \end{figure}

\begin{figure}
        \centering
        \includegraphics[width=0.3\linewidth]{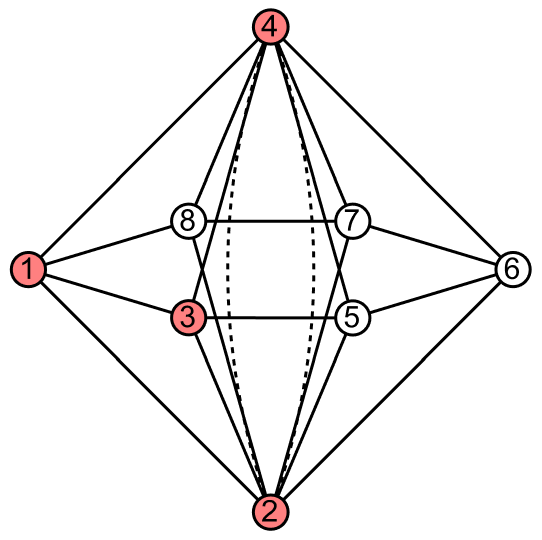}
        \caption{Above is one typical type (which corresponds to the canonical type of ladder integrals in Sec.~\ref{sec:canonical}) of $f$-graph that corresponds to four-loop ladder. The hexagon in the middle plane can be generalized to $n$-gon, with the two dashed line also being generalized to $n-4$, which corresponds to $(n-2)$-loop ladder integral. The four points colored red are taken as external points here.}
        \label{fig:ladderfgraph}
    \end{figure}

We start from the four-point box depicted in Fig.~\ref{fig:box}.
    It is calculated to be
    \begin{equation}
        \Phi_{1}(z,\bar{z})=\frac{4D_{2}(z,\bar{z})}{z-\bar{z}}
    \end{equation}
    where $D_{2}$ is the Bloch-Wigner function, it can be expressed as SVHPL: $\frac{1}{4}(\texttt{I}_{z,1,0,0}-\texttt{I}_{z,0,1,0})$ where \texttt{I} denotes the SVHPL functions\footnote{This is the notation from package \texttt{HyperlogProcedures}~\cite{Shlog}. We will come back to a different notation $\mathcal{L}_{10}(z)=\texttt{I}_{z,1,0,0}$ in Sec.~\ref{sec:binaryDCI integrals}. }. Its explicit expression is
    \begin{equation}
        D_2(z)=\log|z|\mathrm{Im}\log(1-z) + \mathrm{Im}\mathrm{Li}_{2}(z).
    \end{equation}
    And it has the property
    \begin{equation}
        \begin{aligned}
            &D_{2}(z)=-D_{2}(\bar{z}) \\
            &D_{2}(z)=D_{2}(1-\frac{1}{z})=D_{2}(\frac{1}{1-z})=-D_{2}(\frac{1}{z})=-D_{2}(1-z)=-D_{2}(\frac{-z}{1-z}).
        \end{aligned}
    \end{equation}
    In above expression, $z$ and $\bar{z}$ is defined to be
    \begin{equation}
        z\bar{z}= u\equiv \frac{x_{12}^2x_{34}^2}{x_{13}^2x_{24}^2}, \, (1-z)(1-\bar{z})=v\equiv\frac{x_{14}^2x_{23}^2}{x_{13}^2x_{24}^2}.
    \end{equation}
    We first discuss the mapping between boxing operator of $x_i$ and differential operation on $z,\bar{z}$. For example, if we extend the box from the direction of $x_4$, then the boxing operator will be
    \begin{equation}\label{eq:boxingladder}
        \frac{x_{41}^2x_{43}^{2}}{x_{13}^{2}}\square_{4}\,\frac{\Phi_{2}(u,v)}{x_{13}^{2}x_{24}^2}=-4\frac{\Phi_{1}(u,v)}{x_{13}^{2}x_{24}^2}
    \end{equation}
    where $\square_{4}\equiv \partial_{x_{4\mu}}\partial_{x_{4}^{\mu}}$ and $\Phi_{2}(u,v)$ is the double box along the direction of $x_{4}$. Transforming above identity to differential operator of $u,v$ results in
    \begin{equation}\label{eq:boxingladder1}
        \triangle^{(2)}_{u,v}\Phi_{2}(u,v)=-\frac{1}{uv}\Phi_{1}(u,v).
    \end{equation}
    where the differential operator is defined as
    \begin{equation}
        \triangle^{(c)}_{u,v}\equiv c\partial_{u}+c\partial_{v}+(u+v-1)\partial_{u}\partial_{v}+u\partial^{2}_{u}+v\partial^{2}_{v} \, .
    \end{equation}
    It can be rewritten in $z,\bar{z}$ as
    \begin{equation}
        \triangle^{(c)}_{u,v}=\partial_{z}\partial_{\bar{z}}-\frac{c-1}{z-\bar{z}}\left(\partial_{z}-\partial_{\bar{z}}\right) .
    \end{equation}
    If we further define
    \begin{equation}
       f^{(L)}(z,\bar{z})\equiv (z-\bar{z})\Phi_{L}(u,v) \, ,
    \end{equation}
    and generate \eqref{eq:boxingladder1} to arbitrary loop order, then finally, we have the corresponding differential operator of $z,\bar{z}$ to be
    \begin{equation}
        z(1-z)\partial_{z}\bar{z}(1-\bar{z})\partial_{\bar{z}}f^{(L)}(z,\bar{z})=-f^{(L-1)}(z,\bar{z}).
    \end{equation}
    This is the recursion for the ladder family. The ladder integrals $f^{(L)}$ actually correspond to many $f$-graphs, here we only give a typical one as an example in Fig.~\ref{fig:ladderfgraph}.
    Now let us recall why we introduce a factor $\frac{1}{x_{13}^{2}x_{24}^{2}}$ in \eqref{eq:boxingladder}: There is actually a numerator $x_{13}^{4}x_{24}^{2}$ in the definition of 2-loop ladder $\Phi_{2}$. We add a prefactor $1/(x_{13}^{2}x_{24}^{2})$ to cancel the $x_{24}^{2}$ in the numerator. However, nothing could stop us from normalizing the integrand with $x_{12}^{2}x_{34}^{2}$ instead of $x_{13}^{2}x_{24}^{2}$. We only need to normalize this integral with an additional $z\bar{z}$ to get the same SVHPL function.
    This is what actually happens when we deriving the DCI integrals of antiprism from boxing. After performing the boxing twice, we should normalize the integral with $1/u$ to ensure that we can act the boxing operator again. In other words, there is a mismatch of a factor $u$ between the $2(m-1)$-point antiprism DCI integrals with the results of twice boxing of $2m$-point antiprism DCI integrals. Let us denote the $2m$-point antiprism DCI integrals integral as
    \begin{equation}
        \Phi^{ap}_{2m}(u,v)=\frac{f^{ap}_{2m}(z,\bar{z})}{z-\bar{z}}
    \end{equation}  
    Then differential equation will be
    \begin{equation}\label{eq:boxdiff}
        \begin{aligned}
            \left[z(1-z)\partial_{z}\bar{z}(1-\bar{z})\partial_{\bar{z}}\right]\left[z(1-z)\partial_{z}\bar{z}(1-\bar{z})\partial_{\bar{z}}\right]f^{ap}_{2m}(z,\bar{z})&=z\bar{z}f^{ap}_{2(m-1)}(z,\bar{z}). \\
            &\Downarrow \\
            \left[(1-z)\partial_{z}(1-\bar{z})\partial_{\bar{z}}\right]\left[z(1-z)\partial_{z}\bar{z}(1-\bar{z})\partial_{\bar{z}}\right]f^{ap}_{2m}(z,\bar{z})&=f^{ap}_{2(m-1)}(z,\bar{z}).
        \end{aligned}
    \end{equation}
    We show an example in the following figure:
    \begin{equation}\label{eq:boxingdiag}
        \begin{aligned}
        &\vcenter{\hbox{\includegraphics[scale=0.35]{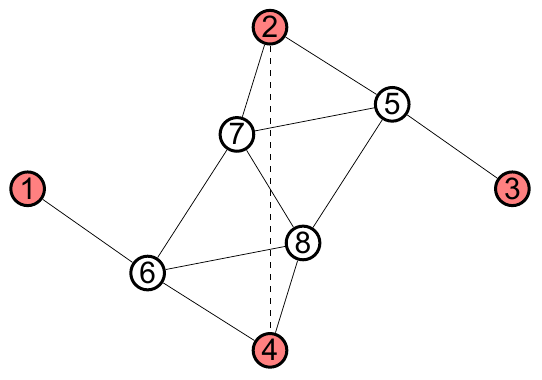}}}\xrightarrow{\frac{x_{14}^2x_{12}^{2}}{x_{24}^2}\square_{1}}\vcenter{\hbox{\includegraphics[scale=0.3]{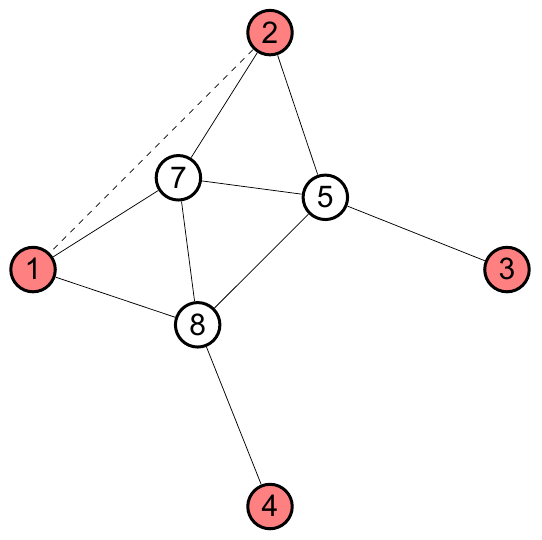}}}\xrightarrow{\frac{x_{14}^2x_{43}^{2}}{x_{13}^2}\square_{4}}\vcenter{\hbox{\includegraphics[scale=0.3]{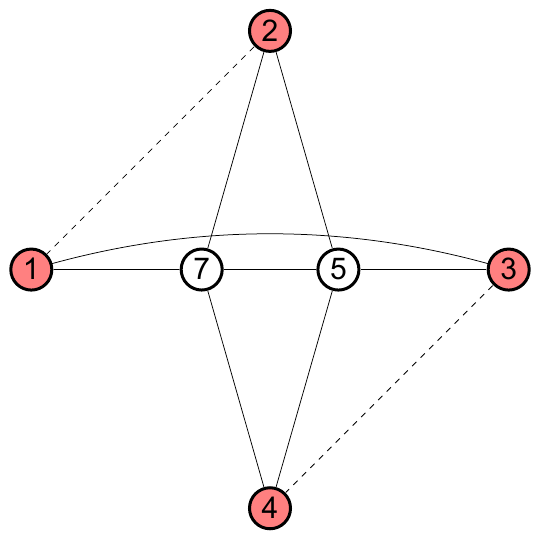}}} \\
        &=\frac{x_{12}^2x_{34}^2}{x_{13}^2x_{24}^2}\times\vcenter{\hbox{\includegraphics[scale=0.3]{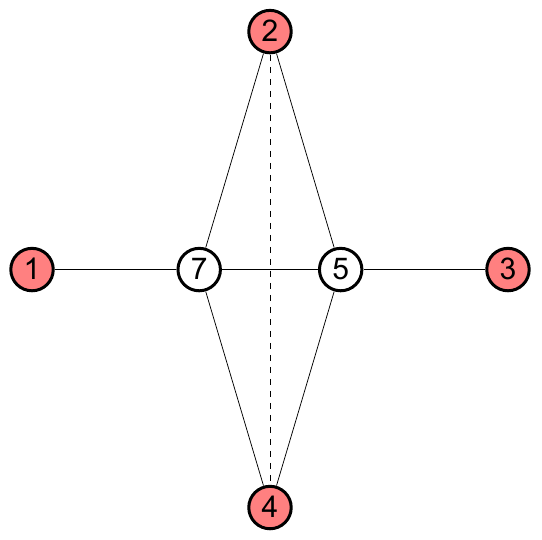}}}
        \end{aligned}
    \end{equation}
    For the DCI integrals, we can always permute its external legs to get an equivalent expression. This is equivalent to performing a modular transformation to $z$. We choose such a transformation to simplify the differential equation:
    \begin{equation}\label{eq:ztrans}
        z\to \frac{\mathtt{z}}{\mathtt{z}-1} \,\, (\bar{z}\to \frac{\bar{\mathtt{z}}}{\bar{\mathtt{z}}-1}).
    \end{equation}
    Here we use a new variable $\mathtt{z}$ to distinguish from the original $z$ and for an arbitrary function $f$
    \begin{equation}
        f(z,\bar{z})=f(\frac{\mathtt{z}}{\mathtt{z}-1},\frac{\bar{\mathtt{z}}}{\bar{\mathtt{z}}-1})=\tilde{f}(\mathtt{z},\bar{\mathtt{z}}).
    \end{equation}
    Then we arrive at
    \begin{equation}
        [(1-\mathtt{z})\partial_{\mathtt{z}}(1-\bar{\mathtt{z}})\partial_{\bar{\mathtt{z}}}][\mathtt{z}\partial_{\mathtt{z}}\bar{\mathtt{z}}\partial_{\bar{\mathtt{z}}}]\tilde{f}^{ap}_{2m}(\mathtt{z},\bar{\mathtt{z}})=\tilde{f}^{ap}_{2(m-1)}(\mathtt{z},\bar{\mathtt{z}}).
    \end{equation}
    Since $D_{2}(z)=D_{2}(\frac{\mathtt{z}}{\mathtt{z}-1})$=$D_{2}(\mathtt{z})$, the initial condition of this iteration remain unchanged. Next, we want to show that the parity odd condition totally fixes the solution of this infinite series of SVHPL. First, according to the initial condition and differential equations, the solutions must belong to the parity odd SVHPL function space. Supposing there are two different parity odd SVHPL functions $g_{1}(\mathtt{z},\bar{\mathtt{z}})$ and $g_{2}(\mathtt{z},\bar{\mathtt{z}})$ which both satisfy
    \begin{equation}
    \begin{aligned}
        &[(1-\mathtt{z})\partial_{\mathtt{z}}(1-\bar{\mathtt{z}})\partial_{\bar{\mathtt{z}}}][\mathtt{z}\partial_{\mathtt{z}}\bar{\mathtt{z}}\partial_{\bar{\mathtt{z}}}]g_{1}(z,\bar{z})=\tilde{f}^{ap}_{2(m-1)}(\mathtt{z},\bar{\mathtt{z}}). \\
        &[(1-\mathtt{z})\partial_{\mathtt{z}}(1-\bar{\mathtt{z}})\partial_{\bar{\mathtt{z}}}][\mathtt{z}\partial_{\mathtt{z}}\bar{\mathtt{z}}\partial_{\bar{\mathtt{z}}}]g_{2}(z,\bar{z})=\tilde{f}^{ap}_{2(m-1)}(\mathtt{z},\bar{\mathtt{z}}).
    \end{aligned}    
    \end{equation}
    Then we will know
    \begin{equation}
        [(1-\mathtt{z})\partial_{\mathtt{z}}(1-\bar{\mathtt{z}})\partial_{\bar{\mathtt{z}}}][\mathtt{z}\partial_{\mathtt{z}}\bar{\mathtt{z}}\partial_{\bar{\mathtt{z}}}](g_{2}(z,\bar{z})-g_{1}(z,\bar{z}))=0.
    \end{equation}
    And $(g_{2}(\mathtt{z},\bar{\mathtt{z}})-g_{1}(\mathtt{z},\bar{\mathtt{z}}))$ is again a parity odd SVHPL. Possible solution to above equation can only be SVHPL less than weight 4. The number of such SVHPLs is finite. We can run over this set and find that no such parity odd solution exist. Thus this series is uniquely determined. We can extend the definition of antiprism DCI integrals to odd points by
    \begin{equation}\label{eq:recurdiff}
        \begin{aligned}
            \mathtt{z}\partial_{\mathtt{z}}\bar{\mathtt{z}}\partial_{\bar{\mathtt{z}}}\tilde{f}^{ap}_{2m}(\mathtt{z},\bar{\mathtt{z}})&=-\tilde{f}^{ap}_{2m-1}(\mathtt{z},\bar{\mathtt{z}}) \\
            (1-\mathtt{z})\partial_{\mathtt{z}}(1-\bar{\mathtt{z}})\partial_{\bar{\mathtt{z}}}\tilde{f}^{ap}_{2m-1}(\mathtt{z},\bar{\mathtt{z}})&=-\tilde{f}^{ap}_{2(m-1)}(\mathtt{z},\bar{\mathtt{z}})
        \end{aligned}
    \end{equation}
    This infinite series is the SVHPL constructed in \cite{Brown:2012ia}\footnote{For odd-loop series, it is exactly the same. For even-loop series, it differs by a transformation $\mz\to 1-\mz$. Original construction in \cite{Brown:2012ia} divides the even and odd loops into two different function series and $\mz$ will always be taken at $0$ to get the zigzag periods. Here we use a single function series but $\mz$ takes $0$ in odd-loop cases and $1$ in even-loop cases. It is an equivalent approach.}. The even-loop antiprism DCI integrals $\tilde{f}^{ap}_{2m}$ correspond to more than one $f$-graph starting from six loops. One such typical type of $f$-graphs are listed in \eqref{eq:antiprism}. The relation between odd-loop antiprism DCI integrals $\tilde{f}^{ap}_{2m-1}$ and $f$-graphs will be discussed in more detail in App.~\ref{app:Bxxx1}.

\subsection{The periods for antiprism \texorpdfstring{$f$-graphs}{f-graphs}: zigzag periods}
    The zigzag periods are related to above function by (see \cite{Brown:2012ia})
    \begin{equation}
        Z_{n}=\left\{\begin{array}{c}
            \Phi^{ap}_{n+3}(\mathtt{z}=1)=\left.\frac{1}{2}\left(\partial_{\mathtt{z}}\tilde{f}^{ap}_{n+3}(\mathtt{z},\bar{\mathtt{z}})-\partial_{\bar{\mathtt{z}}}\tilde{f}^{ap}_{n+3}(\mathtt{z},\bar{\mathtt{z}})\right)\right|_{\mathtt{z}=1}, \,\, n \text{ is odd} \\
            \Phi^{ap}_{n+3}(\mathtt{z}=0)=\left.\frac{1}{2}\left(\partial_{\mathtt{z}}\tilde{f}^{ap}_{n+3}(\mathtt{z},\bar{\mathtt{z}})-\partial_{\bar{\mathtt{z}}}\tilde{f}^{ap}_{n+3}(\mathtt{z},\bar{\mathtt{z}})\right)\right|_{\mathtt{z}=0}, \, \, n \text{ is even}
        \end{array}\right.
    \end{equation}
    We list the first several $Z_{n}$'s in Tab.~\ref{tab:zn}.
    \begin{table}[htbp]
        \centering
        \begin{tabular}{c|c|c|c|c|c|c|c|c}
        \hline\hline
           n  & 3 & 4 & 5 & 6 & 7 & 8 & 9 & 10 \\
             \hline
          $Z_{n}$ & $6\zeta_{3}$  & $20\zeta_{5}$ & $\frac{441}{8}\zeta_{7}$ & $168\zeta_{9}$ & $\frac{33759}{64}\zeta_{11}$ & $1716\zeta_{13}$ & $\frac{11713845}{2048}\zeta_{15} $ & $19448\zeta_{17}$  \\
        \hline\hline
        \end{tabular}
        \caption{Some examples of $Z_{n}$.}
        \label{tab:zn}
    \end{table}
    Note that $f^{ap}_{n+3}(\mathtt{z},\bar{\mathtt{z}})$ approaches 0 when $z$ approaches the real axis, $\Phi^{ap}_{n+3}(0)$ or $\Phi^{ap}_{n+3}(1)$  is well defined by L'Hospital's rule.
    An interesting thing is that the integrated correlator of antiprism is also the zigzag periods $Z_n$. The integrated correlator can be calculated by \cite{Wen:2022oky}
    \begin{equation}\label{eq:pn}
        P_{n}=\int_{\mathcal{C}}\frac{\mathrm{d}\mathtt{z}\mathrm{d}\bar{\mathtt{z}}}{2\pi i}\frac{(\mathtt{z}-\bar{\mathtt{z}})}{\mathtt{z}\bar{\mathtt{z}}(1-\mathtt{z})(1-\bar{\mathtt{z}})}\tilde{f}_{n}(\mathtt{z},\bar{\mathtt{z}}).
    \end{equation}
   Note that the integration region $\mathcal{C}$ is that $\mathtt{z}\in \mathbb{H}_{-}$ and $\bar{\mathtt{z}}\in \mathbb{H}_{+}$. $\mathbb{H}_{\pm}$ denotes the upper or lower half plane. We can find that $P_{n+2}=Z_{n}$. This is not a coincidence, we can rewrite above integration as
    \begin{equation}
        \begin{aligned}
            P_{n}=&\int_{\mathcal{C}}\frac{\mathrm{d}\mathtt{z}\mathrm{d}\bar{\mathtt{z}}}{2\pi}\left(\frac{1}{\mathtt{z}\bar{\mathtt{z}}(1-\mathtt{z})}\tilde{f}_{n}(\mathtt{z},\bar{\mathtt{z}})-\frac{1}{\mathtt{z}\bar{\mathtt{z}}(1-\bar{\mathtt{z}})}\tilde{f}_{n}(\mathtt{z},\bar{\mathtt{z}})\right). \\
                 =&\int_{\mathcal{C}}\frac{\mathrm{d}\mathtt{z}\mathrm{d}\bar{\mathtt{z}}}{2\pi}\left(\frac{1}{\bar{\mathtt{z}}(1-\mathtt{z})(1-\bar{\mathtt{z}})}\tilde{f}_{n}(\mathtt{z},\bar{\mathtt{z}})-\frac{1}{\mathtt{z}(1-\mathtt{z})(1-\bar{\mathtt{z}})}\tilde{f}_{n}(\mathtt{z},\bar{\mathtt{z}})\right).
        \end{aligned}
    \end{equation}
    Applying \eqref{eq:recurdiff} to the first line when $n$ is even and the second line when $n$ is odd, we can arrive at
    \begin{equation}
        \begin{aligned}
            P_{n}=\left\{\begin{array}{c}
                \int_{\mathbb{H}_{-}}\frac{\mathrm{d}\mathtt{z}}{2\pi i}\frac{\partial_{\mathtt{z}}\tilde{f}_{n+1}(\mathtt{z},\bar{\mathtt{z}})}{(\mathtt{z}-1)}-\int_{\mathbb{H}_{+}}\frac{\mathrm{d}\bar{\mathtt{z}}}{2\pi i}\frac{\partial_{\bar{\mathtt{z}}}\tilde{f}_{n+1}(\mathtt{z},\bar{\mathtt{z}})}{(\bar{\mathtt{z}}-1)}, \,\, n \text{ is odd} \\
                -\int_{\mathbb{H}_{+}}\frac{\mathrm{d}\bar{\mathtt{z}}}{2\pi i}\frac{\partial_{\bar{\mathtt{z}}}\tilde{f}_{n+1}(\mathtt{z},\bar{\mathtt{z}})}{\bar{\mathtt{z}}}+\int_{\mathbb{H}_{-}}\frac{\mathrm{d}{\mathtt{z}}}{2\pi i}\frac{\partial_{{\mathtt{z}}}\tilde{f}_{n+1}(\mathtt{z},\bar{\mathtt{z}})}{{\mathtt{z}}}, \,\, n \text{ is even} 
            \end{array}\right.            
        \end{aligned}
    \end{equation}
    Since $\tilde{f}_{n+1}(\mathtt{z},\bar{\mathtt{z}})$ is a parity-odd SVHPL, we can expect that its Laurent expansion around $\mathtt{z}=0,1,\infty$ is regular, because the singular points are all on the real axis but parity odd SVHPL vanishes on real axis. Thus, $\partial_{\mathtt{z}}\tilde{f}_{n+1}(\mathtt{z},\bar{\mathtt{z}})$ and $\partial_{\bar{\mathtt{z}}}\tilde{f}_{n+1}(\mathtt{z},\bar{\mathtt{z}})$ will not contribute simple pole. Then we arrive at
    \begin{equation}
        \begin{aligned}
            P_{n}=\left\{\begin{array}{c}
                \frac{1}{2}\left.\left(\partial_{\mathtt{z}}\tilde{f}_{n+1}(\mathtt{z},\bar{\mathtt{z}})-\partial_{\bar{\mathtt{z}}}\tilde{f}_{n+1}(\mathtt{z},\bar{\mathtt{z}})\right)\right|_{\mathtt{z}=1}, \,\, n \text{ is odd} \\
                \frac{1}{2}\left.\left(\partial_{\mathtt{z}}\tilde{f}_{n+1}(\mathtt{z},\bar{\mathtt{z}})-\partial_{\bar{\mathtt{z}}}\tilde{f}_{n+1}(\mathtt{z},\bar{\mathtt{z}})\right)\right|_{\mathtt{z}=0}, \,\, n \text{ is even} 
            \end{array}\right.            
        \end{aligned}
    \end{equation}
    Note that for $\mathtt{z}=\infty$, we should perform a transformation $\mathtt{z}=1/t$. However, $\tilde{f}_{n+1}(1/t,1/\bar{t})$ will have the same good behavior as $\tilde{f}_{n+1}(\mathtt{z},\bar{\mathtt{z}})$, it also vanishes on the real axis without any divergence. $\partial_{\mathtt{z}}\tilde{f}_{n+1}(\mathtt{z},\bar{\mathtt{z}})=t^2\partial_{t}\tilde{f}_{n+1}(1/t,1/\bar{t})$. So $1/(\mathtt{z}-1)$ will not contribute a residue at infinity. The same is true for $1/\mathtt{z}$. The $1/2$ is due to the integration region being half plane for $\mz$ or $\mzb$ instead of the whole. Then we have shown that $P_{n+2}=Z_{n}$. At last, we show a diagrammatic way in App.~\ref{app:fourier} to understand this relation between integrated antiprism DCI integrals and zigzag periods studied in the anomalous dimension of $\phi^{4}$ theory which is slightly different from the planar duality used in \cite{Schnetz:2008mp}.
    
    Finally, we give some lower-point examples of $\tilde{f}^{ap}_{n}(\mathtt{z},\bar{\mathtt{z}})$ in Tab.~\ref{tab:zigzagDCI integrals}.
    \begin{table}[htbp]
        \centering
        \begin{tabular}{c|c|c}
            \hline\hline
            L (loop number) & $\tilde{f}_{n}(\mathtt{z},\bar{\mathtt{z}})$& SVHPL results (leading singularity $1/(\mathtt{z}-\bar{\mathtt{z}})$ suppressed) \\
            \hline
            1 & $\tilde{f}_{5}$ & $-\mathrm{I}_{\mz ,0,1,0}+\mathrm{I}_{\mz ,1,0,0}$ \\
            \hline
            2 & $\tilde{f}_{6}$& $\mathrm{I}_{\mz,0,0,1,0,0}-\mathrm{I}_{\mz,0,1,0,0,0}$ \\
            \hline
            3 & $\tilde{f}_{7}$ & \thead{$-20 \mathrm{I}_{\mz,1,0} f_{5}-4 \mathrm{I}_{\mz ,1,0,1,0} f_{3}-\mathrm{I}_{\mz ,1,0,0,1,0,1,0}+\mathrm{I}_{\mz ,1,0,1,0,0,1,0}$} \\
                        \hline
            4 & $\tilde{f}_{8}$ &\thead{$20 f_{5} \mathrm{I}_{\mz ,0,1,0,0}-4 f_{3} \mathrm{I}_{\mz ,0,1,0,0,1,0}+4 f_{3}\mathrm{I}_{\mz ,0,1,0,1,0,0}$\\ $+\mathrm{I}_{\mz ,0,1,0,0,1,0,1,0,0}-\mathrm{I}_{\mz ,0,1,0,1,0,0,1,0,0}$} \\
            \hline
            5 & $\tilde{f}_{9}$ &\thead{$-336 \mathrm{I}_{\mz ,1,0} f_{9}-56 f_{7} \mathrm{I}_{\mz ,1,0,1,0}-4 \mathrm{I}_{\mz ,1,0,1,0,1,0} f_{5}$\\$-\mathrm{I}_{\mz ,1,0,1,0,0,1,0,1,0,1,0}+\mathrm{I}_{z ,1,0,1,0,1,0,0,1,0,1,0}$} \\
            \hline\hline
        \end{tabular}
        \caption{Some examples of lower loop anti-prism where $f_{i}\equiv \zeta_{i}$. $\mathrm{I}_{\mz,\ldots,0}$ is the SVHPL function defined in the package \texttt{HyperlogProcedures}~\cite{Shlog} and it is the same as function $\mathrm{cG}[\ldots,\mz]$ defined in \texttt{PolyLogTools}~\cite{Duhr:2019tlz}.}\label{tab:zigzagDCI integrals}
    \end{table}

\section{Inverse boxing: binary DCI integrals as Steinmann-satisfying SVHPL}\label{sec:inverseboxing}
Now we can readily extend the zigzag dual conformal integrals to a more general integral class which we will name binary DCI integrals. First, we expect that they can be solved recursively by boxing differential equations and related to ladder integrals. This means that they belong to the generalized ladder integral family studied in \cite{Drummond:2012bg}. Second, the extended Steinmann relations must be satisfied since we want to relate them to planar $f$-graphs\footnote{Non-planar $f$-graphs can also generate planar DCI integrals. One example is that the odd-loop zigzag DCI integrals can be generated by taking four cycles of non-planar $f$-graphs. We will also show that how odd-loop zigzag DCI integrals can be obtained from antiprism $f$-graphs by not taking a four cycle but four points on a line segment in App.~\ref{app:Bxxx1}. This means that they are actually not directly contained in four-point amplitudes.}. Third, at most two kinds of differential operators are involved in the recursive solving of boxing differential equations. Any two of $z\bar{z}\partial_{z}\partial_{\bar{z}},(1-z)(1-\bar{z})\partial_{z}\partial_{\bar{z}}, z\bar{z}(1-z)(1-\bar{z})\partial_{z}\partial_{\bar{z}}$ are permitted since we can always perform a conformal transformation to $z$ (which is equivalent to relabeling external points) so that the two differential operators take the form $\mz\mzb\partial_{\mz}\partial_{\mzb},(1-\mz)(1-\mzb)\partial_{\mz}\partial_{\mzb}$ finally. This is why it is called binary. We give examples in App.~\ref{app:notbinary} which involve three kinds of boxing differential operators simultaneously. Thus binary DCI integrals are generalized ladders which can be evaluated to SVHPLs that satisfy both the extended Steinmann constraints and boxing differential equations with at most two kinds of differential operators involved. We name this function space as the binary Steinmann SVHPL space. 

The logic of our survey into integrated $f$-graphs is that we first study the function space which we have good control of, then we search for those $f$-graphs that are related to this function space and finally the $f$-graphs give us back some information about the periods related to this function space. Therefore, we will first discuss this function space before discussing the possible forms of binary DCI integrals which will be evaluated to binary Steinmann SVHPLs and their connection to $f$-graphs.
\subsection{Binary Steinmann SVHPL function space}\label{sec:binaryDCI integrals}
We study the explicit forms of binary Steinmann SVHPLs and their number counting in this section. 
First of all, for all the DCI integrals integrals, reordering the external legs is equivalent to performing a conformal transformation to the original defined $z,\bar{z}$. We already used this trick in \eqref{eq:ztrans} to simplify the differential equations. However, the periods will not change under the conformal transformation of $z,\bar{z}$ since the integration measure of \eqref{eq:pn}
\[ \int_{\mathcal{C}}\frac{\mathrm{d}\mathtt{z}\mathrm{d}\bar{\mathtt{z}}}{2\pi i}\frac{(\mathtt{z}-\bar{\mathtt{z}})}{\mathtt{z}\bar{\mathtt{z}}(1-\mathtt{z})(1-\bar{\mathtt{z}})} \]
is invariant under the conformal transformation. Thus, we can choose a convenient ordering of external legs or a canonical definition of $z,\bar{z}$ by $u,v$ and then all its conformal transformations will not be discussed any more. The extended Steinmann relations have the simplest form under the following definition of $\mz,\mzb$ by $u,v$ (the definitions of $u,v$ by $x_{ij}^{2}$ remain unchanged):
\begin{equation}
    u=\frac{\mz\mzb}{(1-\mz)(1-\mzb)}, \, v=\frac{1}{(1-\mz)(1-\mzb)}.
\end{equation}
This definition of $\mz,\mzb$ is also consistent with the transformation performed in \eqref{eq:ztrans} and we distinguish $\mz$ from $z$ by different fonts as in the last section. 

By using $\mz,\mzb$, the extended Steinmann conditions state that there can not be consecutive $\mz-1$ or $\mzb-1$ in the symbols~\cite{He:2021mme}. This can be further translated into the requirement that there are no consecutive '1's in the words for SVHPL according to the symbol formula (Definition 2.7) of SVHPL in \cite{Schnetz:2013hqa}. In the following, we will denote the SVHPL directly by its words. $\mathcal{L}_{w}\equiv\mathrm{I}_{z,w,0} $. For example,
\begin{equation}
    \mathcal{L}_{0010}(z)\equiv\mathrm{I}_{z,0,0,1,0,0}.
\end{equation}
Note that one can easily counts such SVHPL functions in general and subject to constraints such as extended Steinmann relations. In the following table, we enumerate the total number of SVHPL functions ($2^w$ for weight $w$), parity-odd case (see below), as well as those after imposing extended Steinmann conditions, which are counted by the Fibonacci sequence: 
\begin{table}[htbp]
    \centering
    \begin{tabular}{c|c|c|c|c|c|c|c|c|c|c|c|c}
    \hline\hline
       $w$  & 1 & 2 & 3 & 4 & 5 & 6 & 7 & 8 & 9 & 10 & 11 & 12 \\
       \hline 
       $N_w$ & 2 & 4 & 8 & 16 & 32 & 64 & 128 & 256 & 512 & 1024 & 2048 & 4096\\
       \hline
       $N^{\rm odd}_w$ & 0 & 1 & 2 & 6 & 12 & 28 & 56 & 120 & 240 & 496 & 992 & 2016\\
    \hline
        $N^{\rm ES}_w$ &1 & 1 & 2 & 3 & 5 & 8 & 13 & 21 & 34 & 55 & 89 & 144\\
        \hline\hline
    \end{tabular}
    \caption{The total number of SVHPL, $N_w$, that of parity-odd SVHPL, $N_w^{\rm odd}$ and that of extended-Steinmann-satisfying SVHPL, $N_w^{\rm ES}$, at weight $w$.}
    \label{tab:nDCI integrals}
\end{table}

Since dual conformal integrals satisfying the boxing differential equation have the same leading singularity $1/(\mz-\mzb)$ and these integrals should be invariant under complex conjugate $\mz\leftrightarrow \mzb$, the binary Steinmann SVHPLs must only involve parity-odd SVHPLs. We denote them as
\begin{equation}
    \tilde{B}_{w}(\mz)=(-1)^{|w|}\left(\mathcal{L}_{w}(\mz)-\mathcal{L}_{\tilde{w}}(\mz)+\sum\zeta_{i}\times\text{ lower-weight SVHPLs}\right),
\end{equation}
where $w$ denotes the word of SVHPL and $\tilde{w}$ is the reverse order of $w$. $|w|$ is the length of word $w$. For example, if $w=0010$, $\tilde{w}=0100$ and their length are $4$. The coefficient $(-1)^{|w|}$ is just a normalization factor to incorporate the minus signs in boxing differential equations.  In the symbol level, $\mathcal{L}_{w}(\mz)-\mathcal{L}_{\tilde{w}}(\mz)$ is already parity-odd by definition because $\mathcal{S}(\mathcal{L}_{w}(\mzb))=\mathcal{S}(\mathcal{L}_{\tilde{w}}(\mz))$ where $\mathcal{S}(f)$ denotes the symbol of function $f$. The lower-weight part can be fixed by solving differential equations and keeping $\tilde{B}_{w}(\mz)$ parity-odd in the function level.
Finally, since the boxing differential equation is either $\mz\mzb\partial_{\mz}\partial_{\mzb}\tilde{B}_{w}(\mz)=-\tilde{B}_{v}(\mz)$ or $(1-\mz)(1-\mzb)\partial_{\mz}\partial_{\mzb}\tilde{B}_{w}(\mz)=-\tilde{B}_{v}(\mz)$, the inverse boxing in the function level, which is defined as solving the boxing differential equations, requires that $w=0v0$ or $w=1v1$. That is, the inverse boxing prepends and appends one '0' or '1' to the original words in the same time. 

In summary, the set of binary Steinmann SVHPLs can be represented by such a series of binary words consisting of '0' and '1':
\begin{enumerate}
    \item They are of even length and centrally symmetric except for the innermost two letters which are '01'. For example, $w=0010$ represents the two-loop ladder and $w=01001010$ represents the four-loop zigzag DCI integrals. This is a consequence of boxing differential equation involving at most two kinds of differential operators.
    \item There are no two consecutive '1's in the words and this is the consequence of extended Steinmann conditions.
\end{enumerate}
The numbers of such words as a function of loop numbers form the Fibonacci series, $N(L)$ where $L$ is the loop number. This can be easily proved by showing that $N(L)=N(L-1)+N(L-2)$. The boxing differential equation defines a natural map from words of length $2L$ (loop $L$) to $2(L-1)$ (loop $L-1$). This map is not injective for words like $\_0v0\_$ and it is two to one. So we only need to count the number of such words which is determined by the number of $v$, that is, $N(L-2)$. Then we have $N(L)=N(L-1)+N(L-2)$. The number of them for the first few loops has been listed in Table~\ref{tab:nbDCI integrals}\footnote{Note that Tab.~\ref{tab:nbDCI integrals} is different from the last row of Tab.~\ref{tab:nDCI integrals}. It is the number growing with the loop, not the weight of SVHPLs. The binary Steinmann DCI integrals form a rather small subset of all Steinmann-satisfying SVHPLs.}.
\begin{table}[htbp]
    \centering
    \begin{tabular}{c|c|c|c|c|c|c|c|c|c|c}
    \hline\hline
       $L$  & 1 & 2 & 3 & 4 & 5 & 6 & 7 & 8 & 9 & 10\\
       \hline 
        $N(L)$ & 1 & 1 & 2 & 3 & 5 & 8 & 13 & 21 & 34 & 55\\
        \hline\hline
    \end{tabular}
    \caption{The number of binary dual conformal integrals $N(L)$ defined in this section. $L$ is the loop number. It is a Fibonacci sequence in loop order $L$ (as opposed to that in weight $w=2L$ for general ES-satisfying SVHPL functions).}
    \label{tab:nbDCI integrals}
\end{table}

Since the words for binary Steinmann SVHPLs are centrally symmetric except for the innermost two letters, we further abbreviate them by taking the right half the whole words. For example, $0010$ is abbreviated to $10$. This abbreviation will always take the form $10...$ and we need to keep in mind that the left part of the whole word is the mirror of this abbreviation except that its rightmost is always '0' instead of '1'. Hereafter, we will denote the binary Steinmann SVHPLs $B_{w}(\mz)$ by this abbreviation. For example,
\begin{equation}\label{eq:abbrev}
    B_{10}(\mz)=\tilde{B}_{0010}(\mz), \, B_{1010}(\mz)=\tilde{B}_{01001010}(\mz).
\end{equation}
In this convention, the series of zigzag DCI integrals will be $B_{10},B_{101},B_{1010},B_{10101},B_{101010},\ldots$ and the first few order results can be found in Table~\ref{tab:zigzagDCI integrals}. Another special series is the ladder DCI integrals and they are $B_{10},B_{100},B_{1000},B_{10000},\ldots$. For later convenience, we will sometimes use the following notation:
\begin{equation}
    B_{n_1,n_2,\ldots,n_{r}}(\mz) \equiv B_{10^{n_1-1}10^{n_2-1}\cdots 10^{n_{r}-1}}(\mz),
\end{equation}
where $n_{i}\ge 2, \forall\, 1\le i< r$ and $n_{r}\ge 1$. Here we emphasize that $w$ in $B_{w}(\mz)$ is \textit{not} the real word for the SVHPL and its length is \textit{half} the transcendental weight of $B_{w}(\mz)$. It corresponds to the loop number. We define the whole set of binary Steinmann SVHPLs as $\mathcal{B}$. It is also useful to define $\hat{\mathcal{B}}$ which is the subset of $\mathcal{B}$, consisting of binary Steinmann SVHPLs whose words always end with '0'. Both the ladder series and zigzag series belong to $\hat{\mathcal{B}}$.

\subsection{A canonical series of binary DCI integrals from inverse boxing}\label{sec:canonical}
In this section, we find the dual conformal integral representations for the integral family $\mathcal{B}$. We note that many different integrals can be evaluated to the same SVHPL. Especially, there are magic identities~\cite{Drummond:2006rz} (or called twist identities~\cite{Schnetz:2008mp,Borinsky:2021gkd} in a graphical view) among different dual conformal integrals. Since the word $w$ of the binary Steinmann SVHPL $B_{w}$ indicates how it is solved by boxing differential equations, we can construct a canonical series of DCI integrals integrands corresponding to the word $w$. The rule is simple,
\begin{enumerate}
    \item We always start from a box in Fig.~\ref{fig:box} with external points $x_1,x_2,x_3,x_4$ in order, which corresponds to the first letter '1' in $w$. The first letter of $w$ is always '1'.
    \item When coming across letter '1' in the remaining word, we always multiply a factor $u=\frac{x_{12}^{2}x_{34}^{2}}{x_{13}^{2}x_{24}^{2}}$ to the current diagram and apply the inverse boxing to the vertex $x_4$ afterwards. When coming across letter '0', we always apply inverse boxing to vertex $x_1$. The inverse boxing to an external vertex $x_i$ in the integrand level is defined to be the reverse process of what we present in \eqref{eq:boxingdiag}.
\end{enumerate}
In such a way, we can construct a unique integrand for a given $B_{w}$. The canonical series constructed in this way will recover what we have presented in \eqref{eq:antiprismDCI integrals} for zigzag DCI integrals. Here we also draw the corresponding series in loop momentum space, additionally including those odd-loop ones,
\begin{equation}\label{eq:zigzagloopmomen}
    \begin{aligned}
        &\vcenter{\hbox{\vbox{\hbox to 3cm{\hfill \includegraphics[scale=0.3]{fig/msB10.pdf} \hfill}\hbox to 3cm{\hfill \scriptsize $B_{10}(\mz)$ \hfill}}}},\vcenter{\hbox{\vbox{\hbox to 3cm{\hfill \includegraphics[scale=0.3]{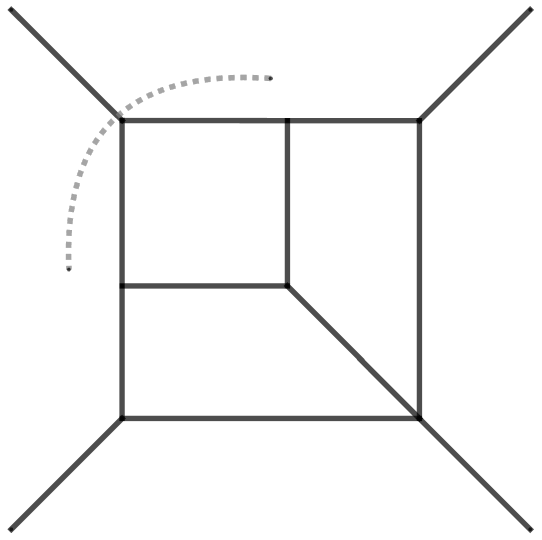} \hfill}\hbox to 3cm{\hfill \scriptsize $B_{101}(\mz)$ \hfill}}}}, \vcenter{\hbox{\vbox{\hbox to 3cm{\hfill \includegraphics[scale=0.3]{fig/msB1010.pdf} \hfill}\hbox to 3cm{\hfill \scriptsize $B_{1010}(\mz)$ \hfill}}}},\vcenter{\hbox{\vbox{\hbox to 3cm{\hfill \includegraphics[scale=0.3]{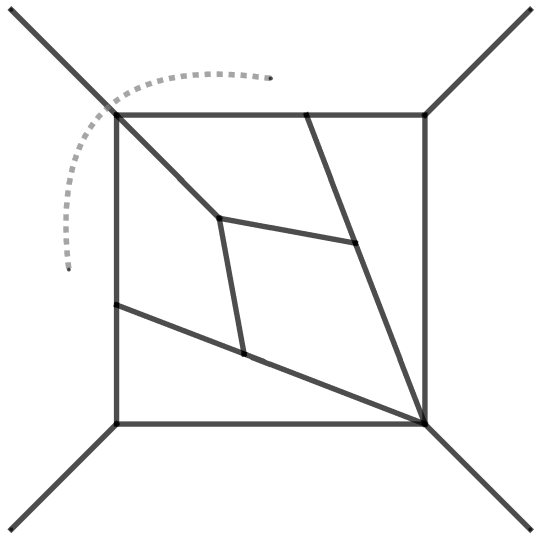} \hfill}\hbox to 3cm{\hfill \scriptsize $B_{10101}(\mz)$ \hfill}}}}, \\
        &\vcenter{\hbox{\vbox{\hbox to 3cm{\hfill \includegraphics[scale=0.3]{fig/msB101010.pdf} \hfill}\hbox to 3cm{\hfill \scriptsize $B_{101010}(\mz)$ \hfill}}}}, \, \vcenter{\hbox{\vbox{\hbox to 3cm{\hfill \includegraphics[scale=0.3]{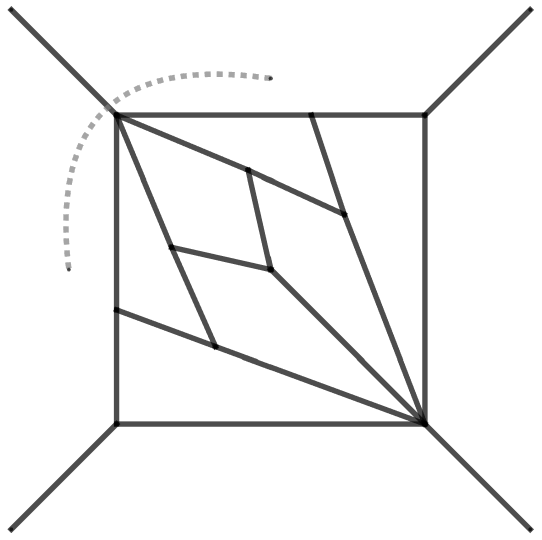} \hfill}\hbox to 3cm{\hfill \scriptsize $B_{1010101}(\mz)$ \hfill}}}},\, \vcenter{\hbox{\vbox{\hbox to 3cm{\hfill \includegraphics[scale=0.3]{fig/msB10101010.pdf} \hfill}\hbox to 3cm{\hfill \scriptsize $B_{10101010}(\mz)$ \hfill}}}},\vcenter{\hbox{\vbox{\hbox to 3cm{\hfill \includegraphics[scale=0.3]{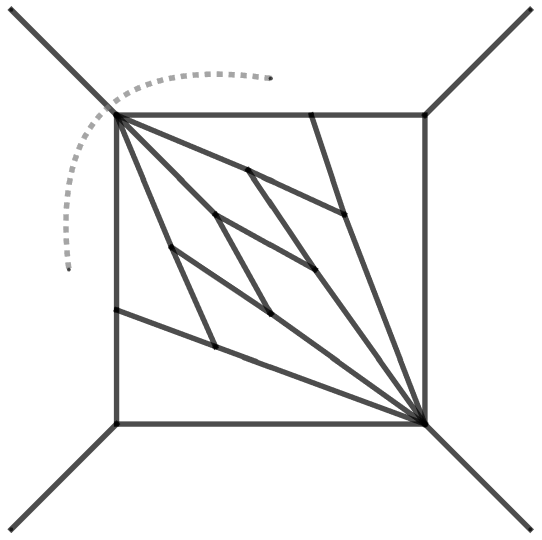} \hfill}\hbox to 3cm{\hfill \scriptsize $B_{101010101}(\mz)$ \hfill}}}},\ldots
    \end{aligned}
\end{equation}
Dashed lines denote the numerators. The zigzag path is from northeast to southwest along the diagonal, then the points on the zigzag path are alternatively joined to the two corners on the northwest and southeast.
Let us show another construction process for the binary Steinmann SVHPL $B_{1001010}$ in dual space:
\begin{equation}\label{eq:b1001010}
        \begin{aligned}
        &\vcenter{\hbox{\vbox{\hbox to 3cm{\hfill \includegraphics[scale=0.35]{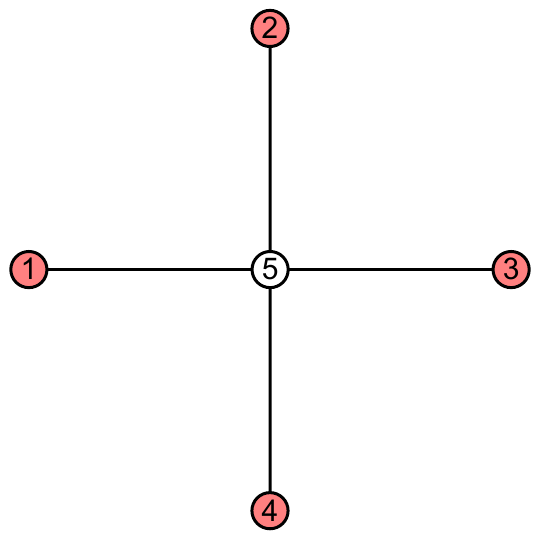} \hfill}\hbox to 3cm{\hfill \scriptsize $B_{1}(\mz)$ \hfill}}}}\rightarrow \vcenter{\hbox{\vbox{\hbox to 3cm{\hfill \includegraphics[scale=0.35]{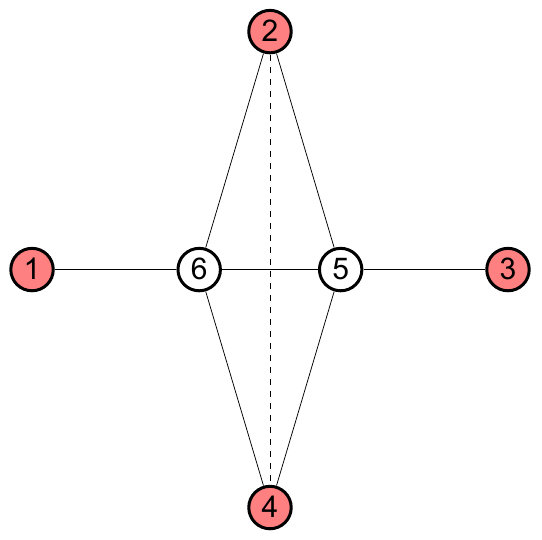} \hfill}\hbox to 3cm{\hfill \scriptsize $B_{10}(\mz)$ \hfill}}}}
        \rightarrow\vcenter{\hbox{\vbox{\hbox to 3cm{\hfill \includegraphics[scale=0.35]{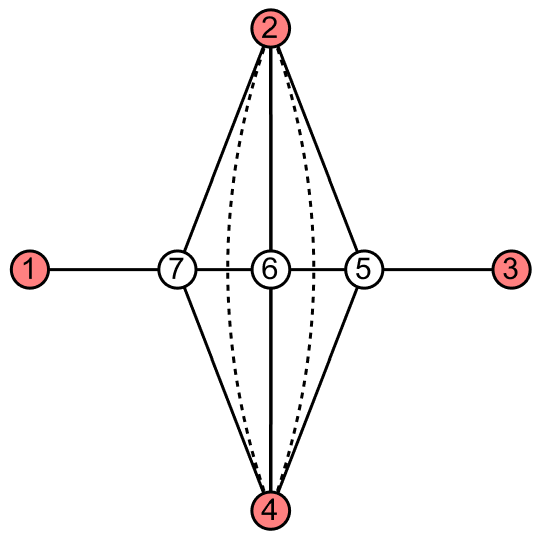} \hfill}\hbox to 3cm{\hfill \scriptsize $B_{100}(\mz)$ \hfill}}}}\rightarrow\\
        &\vcenter{\hbox{\vbox{\hbox to 3cm{\hfill \includegraphics[scale=0.35]{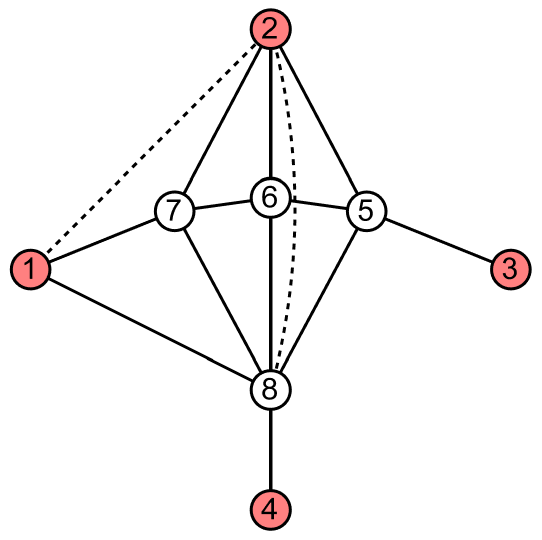} \hfill}\hbox to 3cm{\hfill \scriptsize $B_{1001}(\mz)$ \hfill}}}}\rightarrow \vcenter{\hbox{\vbox{\hbox to 3cm{\hfill \includegraphics[scale=0.35]{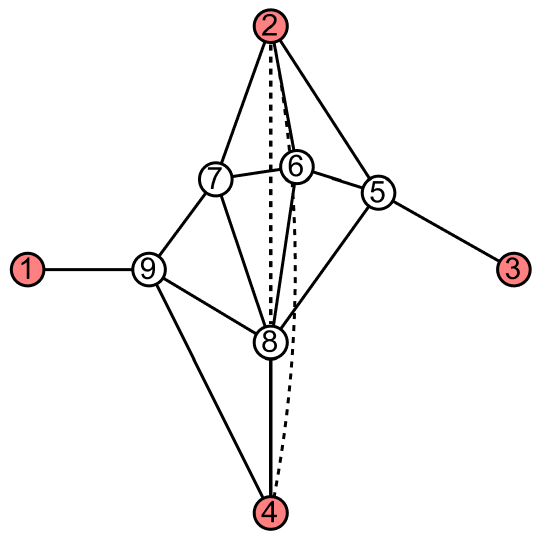} \hfill}\hbox to 3cm{\hfill \scriptsize $B_{10010}(\mz)$ \hfill}}}}\rightarrow \vcenter{\hbox{\vbox{\hbox to 3cm{\hfill \includegraphics[scale=0.35]{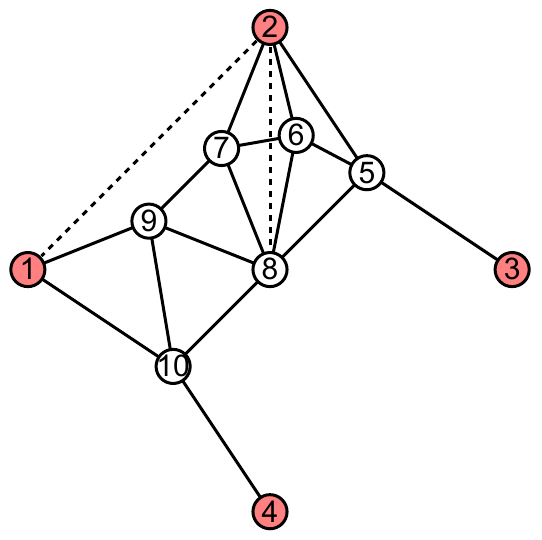} \hfill}\hbox to 3cm{\hfill \scriptsize $B_{100101}(\mz)$ \hfill}}}}\rightarrow \vcenter{\hbox{\vbox{\hbox to 3cm{\hfill \includegraphics[scale=0.35]{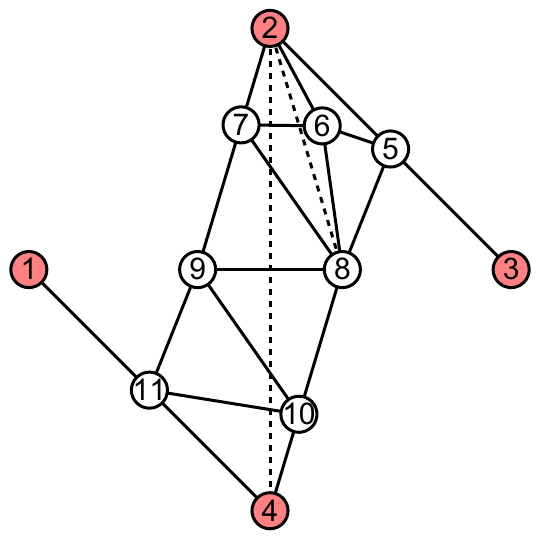} \hfill}\hbox to 3cm{\hfill \scriptsize $B_{1001010}(\mz)$ \hfill}}}}.
        \end{aligned}
\end{equation}
We notice that the canonical integrand of $B_{1001010}$ can be obtained from that of zigzag $B_{101010}$ by directly dividing a factor $x_{62}^{2}x_{65}^{2}x_{67}^{2}x_{68}^{2}/(x_{57}^{2}x_{28}^{2})$. This is actually the inverse of square rule defined in \cite{Bourjaily:2016evz}. However, not all of them can be related to zigzag in this way. For example, $B_{1010010}$ is not related to $B_{101010}$ by square rule. 

We summarize some general features of the canonical integrand constructed in this way and \eqref{eq:b1001010} serves as an example to demonstrate these features. For canonical binary DCI integrals whose words end with '0', there is at least one dashed line connecting $2$ and $4$. $1$ is attached to a single inner point in this case. For canonical binary DCI integrals whose words end with '1', there is one dashed line connecting $1$ and $2$. $4$ is attached to a single inner point in this case. $3$ is always attached to a single inner point in all cases. These can be easily derived from the rules. The diagrams of these canonical integrands in dual coordinate space is straightforward since they are constructed in dual coordinate space. We can also compare the diagrams of them in momentum space. For example, $B_{1010010}$, $B_{1010100}$ and $B_{1001010}$ in the loop momentum space are
\begin{equation}\label{eq:1010100}
    \begin{aligned}
        \vcenter{\hbox{\vbox{\hbox to 3cm{\hfill \includegraphics[scale=0.35]{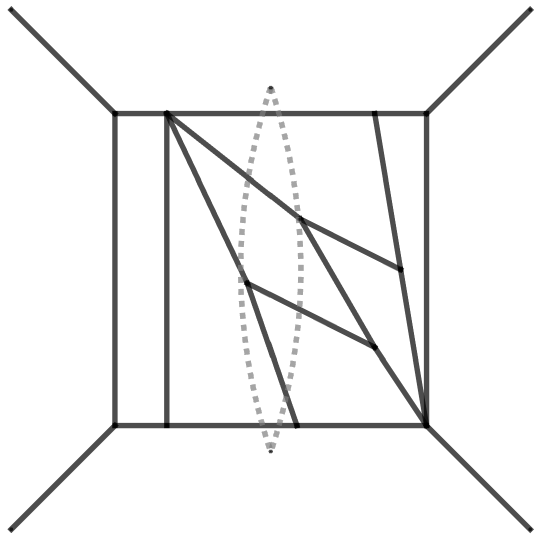} \hfill}\hbox to 3cm{\hfill \scriptsize $B_{1010100}(\mz)$ \hfill}}}}, \quad \vcenter{\hbox{\vbox{\hbox to 3cm{\hfill \includegraphics[scale=0.35]{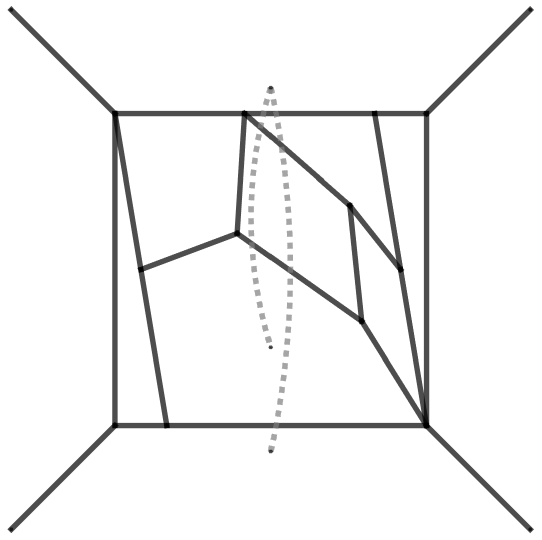} \hfill}\hbox to 3cm{\hfill \scriptsize $B_{1010010}(\mz)$ \hfill}}}},\quad  \vcenter{\hbox{\vbox{\hbox to 3cm{\hfill \includegraphics[scale=0.35]{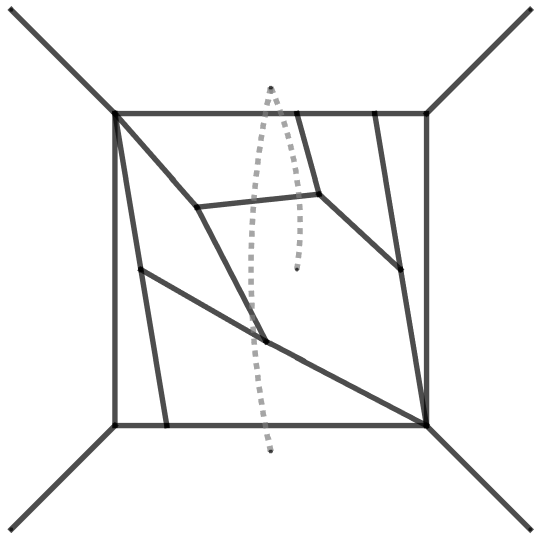} \hfill}\hbox to 3cm{\hfill \scriptsize $B_{1001010}(\mz)$ \hfill}}}}.
    \end{aligned}
\end{equation}
They take very different forms. We remind the readers that above diagrams of binary DCI integrals belong to the canonical series we construct, they are not the only representative of corresponding binary Steinmann SVHPL.  For example, in six loops, except for the diagram we presented in \eqref{eq:zigzagloopmomen} there is another one which can evaluate to $B_{101010}(\mz)$:
\begin{equation}
    \vcenter{\hbox{\vbox{\hbox to 3cm{\hfill \includegraphics[scale=0.35]{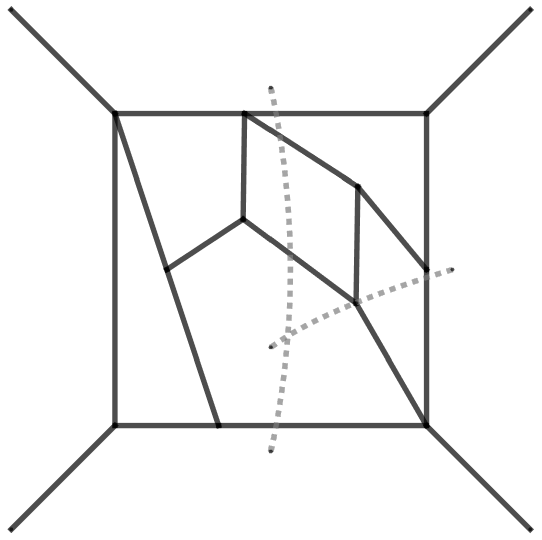} \hfill}\hbox to 3cm{\hfill \scriptsize $B_{101010}(\mz)$ \hfill}}}}
\end{equation}
Careful reader may notice that it takes a very similar form with $B_{1010010}(\mz)$ in \eqref{eq:1010100}. Actually, we can perform boxing differential operator to $B_{1010010}(\mz)$ in \eqref{eq:1010100} from the right and derive the following integrand for $B_{101001}(\mz)$\footnote{The integrand derived in this way is not our canonical form any more, since to derive the canonical form, the boxing must be performed to $x_{1}$, it is on the left hand side in above diagram.}:
\begin{equation}
    \vcenter{\hbox{\vbox{\hbox to 3cm{\hfill \includegraphics[scale=0.35]{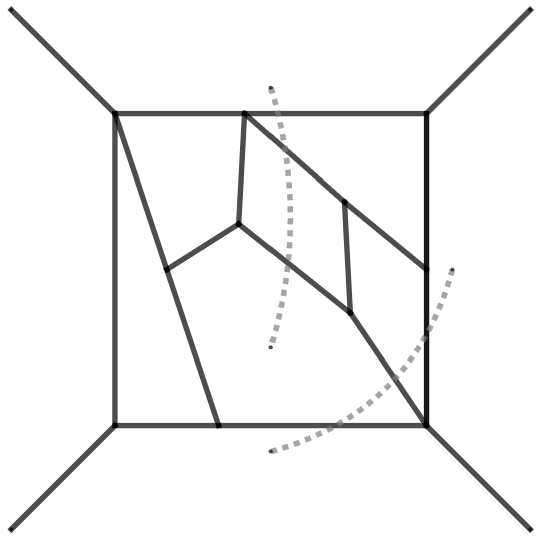} \hfill}\hbox to 3cm{\hfill \scriptsize $B_{101001}(\mz)$ \hfill}}}}
\end{equation}
The two integrands only differ from each other by the numerators, however, they will be evaluated to two totally different SVHPL functions.

The canonical DCI integrals integrands we constructed can be straightforwardly transformed to $f$-graphs. There are some benefits to study the periods of these DCI integrals through $f$-graphs and we will discuss all these things in detail in the next section.

\section{Periods, binary DCI integrals and \texorpdfstring{$f$-graphs}{f-graphs}}\label{sec:periods}
Inspired by the above relation between antiprism $f$-graphs and the zigzag periods studied in Sec.~\ref{sec:antiprism}, we extend our study of the relationships among $f$-graphs, binary DCI integrals and periods. The maps from $f$-graph or DCI integrals to periods are not injective. Thus, the representation of periods with the $f$-graphs or DCI integrals is redundant in general. In what follows, we will mostly concentrate on the relationship between \textit{planar} $f$-graphs\footnote{The $f$-graphs we will discuss are all planar if we do not mention the non-planar case explicitly.} and binary DCI integrals. Their relations reveal more structure of the map from themselves to periods.

On the one hand, all binary DCI integrals can be generated from some $f$-graph. The relation between zigzag DCI integrals and antiprism $f$-graphs is an example. On the other hand, only a small subset of $f$-graphs can be related to binary DCI integrals. So we are studying some special cases whose periods will always be multiple zeta values with uniform transcendental weights. The whole $f$-graphs form a much larger set. However, even the relations between these most special $f$-graphs and binary DCI integrals reveal some interesting properties. We will elaborate on these in the following.

\subsection{Periods of binary Steinmann SVHPLs}
Compared to $f$-graphs or binary DCI integrals, the maps from the function space $\mathcal{B}$ to periods are more faithful. That is, there are fewer degeneracies from the binary Steinmann SVHPLs to periods\footnote{As a contrast, there are many more different $f$-graphs or DCI integrals evaluated to the same period.}. Therefore, in this section, we first study the periods of the binary Steinmann SVHPLs defined in the last section. We will relate these binary Steinmann SVHPLs to binary DCI integrals and $f$-graphs later, and the periods of these binary Steinmann SVHPLs are the periods of corresponding binary DCI integrals and $f$-graphs. In general, we can define the periods of binary Steinmann SVHPLs in the following way:
\begin{equation}
    P_{n_1,n_2,\ldots,n_{r}}\equiv P_{10^{n_1-1}10^{n_2-1}\ldots 10^{n_{r}-1}}\equiv P(B_{10^{n_1-1}10^{n_2-1}\ldots 10^{n_{r}-1}}(\mz)). 
\end{equation}
where $P(f(\mz))$ denotes the period of some integrals evaluated to $f(\mz)/(\mz-\mzb)$~\cite{Wen:2022oky,Borinsky:2022lds},
\begin{equation}
    P(f(\mz))=\int_{\mathcal{C}}\frac{\mathrm{d}\mz\mathrm{d}\mzb}{2\pi i}\frac{(\mz-\mzb)^2}{\mz\mzb(1-\mz)(1-\mzb)}\frac{f(\mz)}{\mz-\mzb}.
\end{equation}
and $0^{n-1}$ means a string with $n\!-\!1$ '0's for $n\ge 0$. $0^{0}$ is understood as the null string. When $n=0$, $10^{0-1}$ is understood as deleting this whole part from the string. That is, $P_{n_1,\ldots,n_{k-1},0,n_{k+1},\ldots,n_{r}}\equiv P_{n_1,\ldots,n_{k-1},n_{k+1},\ldots,n_{r}}$. $L=\sum_{i=1}^{r}n_{i}$ is the loop number of corresponding Feynman diagrams. For later convenience, we also define the period of some integrals evaluated to $f(\mz)/(\mz-\mzb)^2$ as
\begin{equation}\label{eq:defp2}
    P^{\prime}(f(\mz))=\int_{\mathcal{C}}\frac{\mathrm{d}\mz\mathrm{d}\mzb}{2\pi i}\frac{(\mz-\mzb)^2}{\mz\mzb(1-\mz)(1-\mzb)}\frac{f(\mz)}{(\mz-\mzb)^2}.
\end{equation}

There are two special series in the binary DCI integrals family. They are the ladders and zigzags. The periods of both have been calculated to arbitrary loop orders~\cite{Broadhurst:1985vq,Broadhurst:1995km,Brown:2012ia,Derkachov:2023xqq}:
\begin{equation}
    \begin{aligned}
        &P_{ladder}(L)=P_{L}=C^{L+1}_{2L+2}\zeta_{2L+1}=\frac{(2L+2)!}{[(L+1)!]^2}\zeta_{2L+1}. \\
        &P_{zigzag}(L)=P_{{\scriptsize \underbrace{2,2,\ldots,2}_{\left\lfloor L/2\right\rfloor}},L\!\bmod\! 2}=\frac{4}{L+2}C^{L+1}_{2L+2}\left(1-\frac{1-(-1)^{L}}{2^{2L+1}}\right)\zeta_{2L+1}.
    \end{aligned}
\end{equation}
For example, $P_{2,1}=\frac{441}{8}\zeta_{7}=Z_{5}$ in Tab.~\ref{tab:zn}. We have calculated all periods of binary Steinmann SVHPLs up to 10 loops by using \texttt{HyperlogProcedures} and found some interesting phenomena.

First, only very special binary Steinmann SVHPLs can be evaluated to a single zeta value. Ladders and zigzags are such examples. We also find two special binary Steinmann SVHPLs which are not ladders or zigzags and their periods are proportional to a single zeta value. One is in 7 loop:
\begin{equation}
    P_{2,3,2}\equiv P_{1010010}\equiv P(B_{1010010}(\mz))=6006\zeta_{15}.
\end{equation}
The expression of $B_{1010010}(\mz)$ takes the form
\begin{equation}
    \begin{aligned}
        &B_{1010010}(\mz)=-\mathcal{L}_{01001001010010}+\mathcal{L}_{01001010010010}+4\zeta_{3}(\mathcal{L}_{01001001010}-\mathcal{L}_{01001010010}) \\
        &+8\zeta_{3}^{2}(\mathcal{L}_{01001010}-\mathcal{L}_{01001001})-6\zeta_{7}\mathcal{L}_{0100101}+40\zeta_{3}\zeta_{5}\mathcal{L}_{010010}+\zeta_{9}(168\mathcal{L}_{01010}-78\mathcal{L}_{01001})\\
        &+(4240\zeta_{11}-80f_{3,3,5}-80f_{3,5,3}+160f_{5,3,3})\mathcal{L}_{010}-60(\zeta_{5}\zeta_{7}+3\zeta_{3}\zeta_{9})\mathcal{L}_{01},
    \end{aligned}
\end{equation}
where 
\begin{equation}\label{eq:f533}
    -80f_{3,3,5}-80f_{3,5,3}+160f_{5,3,3}=-48\zeta_{5,3,3}-40\zeta_{3}^{2}\zeta_{5}-360\pi^{2}\zeta_{9}-\frac{8}{5}\pi^{4}\zeta_{7}+\frac{8}{63}\pi^{6}\zeta_{5}
\end{equation}
and we have suppressed $\mz$ dependence of $\mathcal{L}$. Another special binary one is in 10 loop which evaluates to
\begin{equation}
    P_{2,3,3,2}\equiv P_{1010010010}=251940\zeta_{21}.
\end{equation}
The expression of $B_{1010010010}$ is too lengthy to list here. Though we have only a few data points for now~\footnote{The next one in this series is $P_{2,3,3,3,2}$ which is of weight 27, still within the reach of \texttt{HyperlogProcedures}. However, we failed to simplify the expression due to memory constraints of the laptop used.}, it is reasonable to make the following conjecture.
\begin{conj}\label{conj:singlezeta}
    $P_{2,{\scriptsize \underbrace{3,\ldots,3}_{\text{m}}},2}$ evaluates to a single zeta value $\zeta_{6m+9}$ with some integer coefficient for arbitrary $m\in\mathbb{Z}$. $m=0,1,2$ have been explicitly checked.
\end{conj}

Second, we find that binary Steinmann SVHPLs whose orders of $10$, $100$ and $100...$ are reverse of each other will give the same periods. For example, $P_{2,3}=P_{3,2}$, $P_{2,2,3}=P_{3,2,2}$ and $P_{2,3,4}=P_{4,3,2}$. We note two points. First, this relation applies only to binary Steinmann SVHPLs whose words consist of patterns like $10$, $100$, $100\ldots$, without any isolated $1$; equivalently, $n_i \ge 2$ for all $i=1,2,\ldots,r$ in $P_{n_1,n_2,\ldots,n_r}$. Second, this is not an antipode in the usual MPL sense, since the subscript in
$P_{n_1,n_2,\ldots,n_r} \equiv P_{10^{n_1-1}10^{n_2-1}\cdots10^{n_r-1}}$
is merely an abbreviation (as in \eqref{eq:abbrev}), not the actual word of the corresponding SVHPL. When the indices are reversed, the blocks $10$, $100$, $100\ldots$ are reversed as whole units — the substring $10$ does not become $01$. Then we can formulate the statement as
\begin{prop}\label{conj:reverse}
    $P_{n_1,n_2,\ldots,n_{r-1},n_{r}}=P_{n_{r},n_{r-1},\ldots,n_{2},n_{1}}$ holds for arbitrary integers $n_1,\ldots,n_{r}\ge 2$. 
\end{prop} 
\noindent This has been explicitly checked up to $L=10$ loops. It is actually a corollary of the relation between binary Steinmann SVHPLs and $f$-graphs, and we will prove this later in the next section. 

\begin{figure}[htbp]
    \centering
    \includegraphics[width=0.85\linewidth]{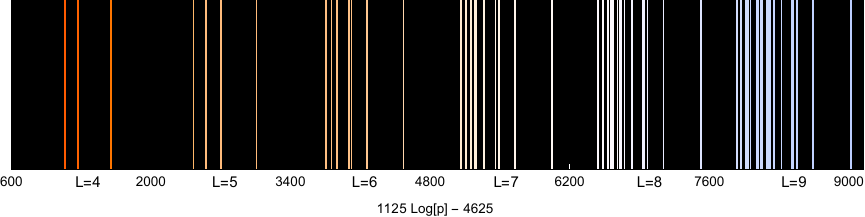}
    \caption{The comparison plot of magnitudes of periods of binary Steinmann SVHPLs. The magnitudes has been rescaled by a monotonic function $1125\log x-4625$. We can see that the periods of different loops are separated. In each loop, the largest value corresponds to the ladder and the smallest value corresponds to the zigzag.}
    \label{fig:periodspectrum}
\end{figure}

Third, although the magnitudes of MZVs are not important and their motivic version is mostly studied which manifests their algebraic properties, the magnitudes of \textit{single-valued} MZVs may still be interesting since they may be related to the anomalous dimensions in $N=4$ SYM and thus correspond to energy of string configurations according to AdS/CFT correspondence~\cite{Leurent:2013mr}. We find that the magnitudes of all binary DCI integrals periods with a given loop order $L$ lie within the closed interval formed by $P_{ladder}(L)$ and $P_{zigzag}(L)$. This can be understood qualitatively, since ladders are binary Steinmann SVHPLs with the least depth and zigzags are binary Steinmann SVHPLs with the greatest depth. Here, the depth can be understood as the number of '1's in the words of corresponding highest-weight SVHPLs. Usually, for a given weight, multiple zeta values tend to have smaller magnitudes as their depth increases, in a certain range of depth\footnote{This is a rather rough statement. It can be understood from the sum theorem of multiple zeta value (conjectured in \cite{hoffman1992} and proven by Granville \cite{Granville1997AnalyticNT} and Zagier independently): $\sum_{\{(n_1,n_2,\ldots,n_{r})|n_1+\ldots+n_r=L,n_r>1\}}\zeta_{n_1,n_2,\ldots,n_r}=\zeta_{L}$ for given $r$. Then with the increase of $r$, when the number of partition increase, their average magnitudes will decrease. When $r$ cross some threshold, the number of partition will decrease and then the process reverses. In the other end, we will have $\zeta_{{\tiny \underbrace{1,1,\ldots,1}_{L-2}},2}=\zeta_{L}$ \cite{hoffman1992}. So this statement roughly hold in a certain range of $r$. }.  Our second conjecture about periods of binary Steinmann SVHPLs is
\begin{conj}
    Given the loop number $L=\sum_{i=1}^{r}n_{i}$, $P_{{\scriptsize \underbrace{2,2,\ldots,2}_{\left\lfloor L/2\right\rfloor}},L\!\bmod\! 2}\le P_{n_1,n_2,\ldots,n_{r}}\le P_{L}$.
\end{conj}
\noindent This conjecture is also checked up to ten loops. And we depict the spectrum formed by periods of binary Steinmann SVHPLs up to nine loops in Fig.~\ref{fig:periodspectrum}.
This gives us a rough idea of how these periods grow with weight and how they decrease with depth.

Finally, by calculating all the periods of binary Steinmann SVHPLs, we can identify their basis in the lower-loop cases. It is summarized in Table.~\ref{tab:periodbasis}.
\begin{table}[htbp]
    \centering
    \begin{tabular}{c|c|c}
    \hline\hline
        loop & basis of periods of $\mathcal{B}$ & \#  \\
        \hline
        2 & \thead{$f_{5}$} & 1\\
        3 & \thead{$f_{7}$} & 1\\
        4 & \thead{$f_{9},{ f_{3,3,3}}$} & 2 \\
        5 & \thead{$f_{11},f_{3,3,5}+f_{3,5,3}, {f_{5,3,3}}$} & 3 \\
        6 & \thead{$f_{13},f_{5,5,3}+f_{5,3,5}, f_{3,3,7}+f_{3,7,3},{f_{3,5,5}, f_{7,3,3}}$} & 5\\
        7 & \thead{$f_{15},f_{7,3,5}\!+\!f_{7,5,3},\!f_{5,3,7}\!+\!f_{5,7,3},f_{3,5,7}\!+\!f_{3,7,5},$ 
        {$f_{5,5,5},f_{3,3,9}+f_{3,9,3},f_{9,3,3},f_{3,3,3,3,3}$}} & 8 \\
        8 & \thead{$f_{17},f_{7,3,7}+f_{7,7,3},f_{5,5,7}+f_{5,7,5},f_{3,5,9}+f_{3,9,5},f_{5,3,9}+f_{5,9,3},f_{9,3,5}+f_{9,5,3},$\\ $f_{3,3,11}+f_{3,11,3},f_{3,3,3,3,5}\!+\!f_{3,3,3,5,3}\!-\!2 f_{3,5,3,3,3},$ \\ { $f_{3,7,7},f_{7,5,5},f_{11,3,3},f_{3,3,5,3,3}+3f_{3,5,3,3,3},f_{5,3,3,3,3}$} } & 13\\
        9 & \thead{$f_{19},f_{5,7,7},f_{9,5,5},f_{13,3,3},f_{7,3,3,3,3},f_{3,7,9}+f_{3,9,7},f_{3,5,11}+f_{3,11,5},f_{3,3,13}+f_{3,13,3},$\\ $f_{5,5,9}+f_{5,9,5},f_{5,3,11}+f_{5,11,3},f_{7,5,7}+f_{7,7,5},f_{7,3,9}+f_{7,9,3},f_{9,3,7}+f_{9,7,3},f_{11,3,5}+f_{11,5,3},$ \\ $f_{3,3,3,5,5}+f_{3,5,5,3,3},f_{3,3,7,3,3}+3 f_{3,7,3,3,3},f_{5,3,5,3,3}+3f_{5,5,3,3,3},$\\$f_{3,3,5,3,5}\!+\!f_{3,3,5,5,3}\!-\!2 f_{3,5,5,3,3},f_{3,5,3,3,5}\!+\!f_{3,5,3,5,3}\!+\!2 f_{3,5,5,3,3},f_{3,3,3,3,7}\!+\!f_{3,3,3,7,3}\!-\!2
   f_{3,7,3,3,3}$ \\ $3(f_{5,3,3,3,5}+f_{5,3,3,5,3})+2 f_{5,3,5,3,3}$ } & 21 \\
   10 & \thead{$f_{21},f_{3,9,9},f_{7,7,7},f_{11,5,5},f_{15,3,3},f_{9,3,3,3,3},f_{3,3,3,3,3,3,3},f_{3,7,11}\!+\!f_{3,11,7},f_{3,5,13}\!+\!f_{3,13,5},$\\ $f_{3,3,15}\!+\!f_{3,15,3},f_{5,7,9}\!+\!f_{5
   ,9,7},f_{5,5,11}\!+\!f_{5,11,5},f_{5,3,13}\!+\!f_{5,13,3},f_{7,5,9}\!+\!f_{7,9,5},f_{7,3,11}\!+\!f_{7,11,3},$ \\ $f_{9,5,7}\!+\!f_{9,7,5},f_{9,3,9}\!+\!f_{9,9,3},f_{11,3,7}\!+\!f_{11,7,3},f_{13,3,5}\!+\!f_{13,5,3},f_{5,3,3,5,5}\!+\!f_{5,5,5,3,3},$\\
   $f_{3,3,9,3,3}\!\!+\!\!3 f_{3,9,3,3,3},f_{5,3,7,3,3}\!\!+\!\!3 f_{5,7,3,3,3},f_{7,3,5,3,3}\!\!+\!\!3 f_{7,5,3,3,3},f_{3,3,5,3,7}\!\!+\!\!f_{3,3,5,7,3}\!\!-\!\!2
   f_{3,7,5,3,3},$ \\ $f_{3,3,3,3,9}\!\!+\!\!f_{3,3,3,9,3}\!\!-\!\!2 f_{3,9,3,3,3},f_{5,3,5,3,5}\!\!+\!\!f_{5,3,5,5,3}\!\!-\!\!2 f_{5,5,5,3,3},f_{5,5,3,3,5}\!\!+\!\!f_{5,5,3,5,3}\!\!+\!\!2
   f_{5,5,5,3,3},$ \\ $f_{5,3,3,3,7}\!\!+\!\!f_{5,3,3,7,3}\!\!-\!\!2 f_{5,7,3,3,3},f_{7,3,3,3,5}\!\!+\!\!f_{7,3,3,5,3}\!\!-\!\!2
   f_{7,5,3,3,3},$ \\ $f_{3,5,7,3,3}\!\!+\!\!f_{3,7,3,3,5}\!\!+\!\!f_{3,7,3,5,3}\!\!+\!\!f_{3,7,5,3,3}, f_{3,5,5,3,5}\!\!+\!\!f_{3,5,5,5,3}\!\!-\!\!2f_{3,3,5,5,5},$ \\ $3(f_{3,5,5,5,3}\!+\!f_{3,5,5,3,5})\!+\!2f_{3,5,3,5,5},f_{3,3,3,5,7}\!+\!f_{3,3,3,7,5}\!-\!f_{3,7,3,3,5}\!-\!f_{3,7,3,5,3},$ \\ $f_{3,5,3,3,7}\!\!+\!\!f_{3,5,3,7,3}\!\!-\!\!f_{3,7,3,3,5}\!\!-\!\!f_{3,7,3,5,3},f_{3,3,7,3,5}\!\!+\!\!f_{3,3,7,5,3}\!\!+\!\!2(
   f_{3,7,3,3,5}\!\!+\!\!f_{3,7,3,5,3}\!\!+\!\!f_{3,7,5,3,3})$} & 35 \\
        \hline\hline
    \end{tabular}
    \caption{Basis for periods of binary Steinmann SVHPLs. Independent periods for $\mathcal{B}$ saturate the space for single-valued versions of  motivic multiple zeta values \cite{Brown:2013gia} up to ten loops. Some patterns will repeat across different loops like $f_{3,3,5,3,3}+3f_{3,5,3,3,3}$. }
    \label{tab:periodbasis}
\end{table}
We also provide the explicit results of these periods up to 7 loops in Tab.~\ref{tab:periodsupto7}\footnote{We provide the complete results up to ten loops in the electronic supplemental materials \texttt{binarytable.nb}.}.
\begin{table}[htbp]
    \centering
    \begin{tabular}{c|c|c}
    \hline\hline
       Loop & $B_{\bullet}$ & period \\
       \hline
       \multirow{3}{*}{4}  & $B_{1000}$ & $252f_{9}$ \\
         & $B_{1001}$ & $\frac{1567}{9}f_{9}+48f_{3,3,3}$ \\
         & $B_{1010}$ &  $168f_{9}$ \\
         \hline 
       \multirow{4}{*}{5} & $B_{10000}$ & $924 f_{11}$ \\
       & $B_{10001}$ & $-\frac{3557 f_{11}}{12}+100 (f_{3,3,5}+f_{3,5,3})+40 f_{5,3,3}$\\
       & $B_{10100},B_{10010}$ & $4240 f_{11}-80 (f_{3,3,5}+ f_{3,5,3})+160 f_{5,3,3}$\\
       & $B_{10101}$ & $\frac{33759 f_{11}}{64}$\\
       \hline
       \multirow{7}{*}{6} & $B_{100000}$ & $3432 f_{13}$ \\
       & $B_{100001}$ & $\frac{12054 f_{13}}{25}+280 (f_{3,3,7}+ f_{3,7,3})+148(f_{5,3,5}+f_{5,5,3})+200 f_{3,5,5}$\\
       & $B_{100010},B_{101000}$ & $278 f_{13}-560 (f_{3,3,7}+ f_{3,7,3})+400(f_{5,3,5}+f_{5,5,3})+800 f_{3,5,5}$\\
       & $B_{100100}$ & $27488 f_{13}-1600 f_{3,5,5}+800 (f_{5,3,5}+f_{5,5,3})$\\
       & $B_{100101}$ & \thead{$\displaystyle \frac{252579011 f_{13}}{22400}-\frac{891}{4} (f_{3,3,7}+f_{3,7,3})-270 f_{3,5,5}$ \\ $\displaystyle +357(f_{5,3,5}+f_{5,5,3})+\frac{873}{2}f_{7,3,3}$}\\
       & $B_{101001}$ & \thead{$\displaystyle -\frac{2131719 f_{13}}{22400}+\frac{675}{4} (f_{3,3,7}+f_{3,7,3})-243 (f_{5,3,5}+ f_{5,5,3})+\frac{675}{2}f_{7,3,3}$}\\
       & $B_{101010}$ & $1716 f_{13}$\\
       \hline
       \multirow{10}{*}{7} & $B_{1000000}$ & $12870 f_{15}$ \\
       & $B_{1000001}$ & \thead{$\displaystyle-\frac{41511511 f_{15}}{600}+1008 (f_{3,3,9}+f_{3,9,3})+392 (f_{3,5,7}+f_{3,7,5})$ \\ $\displaystyle+420 (f_{5,3,7}+ f_{5,7,3})+320f_{5,5,5}+84 (f_{7,3,5}+f_{7,5,3})$}\\
       & $B_{1000010},B_{1010000}$ & \thead{$\displaystyle\frac{21687511 f_{15}}{120}-2016 (f_{3,3,9}+ f_{3,9,3})+1680 (f_{3,5,7}+f_{3,7,5})$\\ $\displaystyle+1920 f_{5,5,5}+672 (f_{7,3,5}+f_{7,5,3})$} \\
       & $B_{1000100},B_{1001000}$ & \thead{$-\frac{310483 f_{15}}{8}-2800 (f_{3,5,7}+ f_{3,7,5})$ $+1400(f_{5,3,7}+f_{5,7,3}+f_{7,3,5}+f_{7,5,3})$}\\
       & $B_{1000101}$ & \thead{$\frac{59568283583 f_{15}}{332640}-\frac{20753}{18}(f_{3,3,9}+f_{3,9,3})+\frac{581}{2}(f_{3,5,7}+f_{3,7,5})+\frac{4319}{9} f_{9,3,3}$\\ $+880 f_{5,5,5}+280 (f_{5,3,7}+f_{5,7,3})+\frac{2877}{2} (f_{7,3,5}+f_{7,5,3})-768 f_{3,3,3,3,3}$}\\
       & $B_{1001001}$ & \thead{$-\frac{1443935413 f_{15}}{22176}+\frac{20443}{18}(f_{3,3,9}+f_{3,9,3})-\frac{9783}{4}(f_{3,5,7}+f_{3,7,5})+\frac{7907}{9} f_{9,3,3}$\\ $+1080f_{5,5,5}+\frac{1935}{4}(f_{5,3,7}+f_{5,7,3})+\frac{297}{2}(f_{7,3,5}+f_{7,5,3})+384f_{3,3,3,3,3}$}\\
       & $B_{1010001}$ & \thead{$-\frac{222210137 f_{15}}{1575}-\frac{1684}{9} (f_{3,3,9}+f_{3,9,3})+1310(f_{3,5,7}+f_{3,7,5})-\frac{4408}{9} f_{9,3,3}$\\$+750(f_{5,3,7}+f_{5,7,3})-1040 (f_{7,3,5}+f_{7,5,3})-1920f_{3,3,3,3,3}$}\\
       & $B_{1010100},B_{1001010}$ & \thead{$\frac{4349824 f_{15}}{45}-1344 (f_{3,3,9}+f_{3,9,3})+1120(f_{5,3,7}+f_{5,7,3})$\\ $-1920 f_{5,5,5}+1568 (f_{7,3,5}+f_{7,5,3})$}\\
       & ${\color{orange} B_{1010010}}$ & $6006 f_{15}$\\
       & $B_{1010101}$ & $\frac{11713845 f_{15}}{2048}$\\
    \hline\hline
    \end{tabular}
    \caption{Explicit results of periods of binary Steinmann SVHPLs up to 7 loops. $B_{1010010}$ (which is colored orange) is a special binary Steinmann SVHPL whose period is a single zeta value as those of ladder and zigzag integrals. These binary Steinmann SVHPLs are ordered by the magnitude of their periods in the decreasing order.}
    \label{tab:periodsupto7}
\end{table}
Here $f$ denotes the $f$-alphabets for multiple zeta values~\cite{Brown:2011ik}. It is especially useful to apply the $f$-alphabets when checking the independence of multiple zeta values, and the results are usually more concise using the $f$-alphabets. In the following, we will express multiple zeta values with $f$-alphabets except for some special cases where it is more straightforward to use zeta values, such as $f_{i}=\zeta_{i}$\footnote{The $f$-alphabets can be easily converted to multiple zeta values by \texttt{Convert(\_,zeta)} of \texttt{HyperlogProcedures}.}. The periods of $\mathcal{B}$ already saturate the space of a single-valued version of motivic multiple zeta values~\cite{Brown:2013gia} up to ten loops, indicating that the period of any SVHPL with weight less than 22 can be expressed as a combination (products) of periods of binary DCI integrals. However, we must note here that this will not generally be true for arbitrary loops, because the dimensions of single-valued versions of motivic multiple zeta values \cite{Brown:2013gia}, $\mathrm{dim}\mathcal{H}_{N}^{sv}$, will be greater than the number of different periods generated by binary Steinmann SVHPLs after considering the equivalent relation in \textbf{Proposition}~\ref{conj:reverse} when the loop number is greater than 13. We do not know whether the space of single-valued version of motivic multiple zeta values can still be saturated if we slightly relax the restriction of binary Steinmann SVHPLs to Steinmann SVHPLs which can be solved by boxing differential equations. That is, three different kinds of boxing differential operator can appear simultaneously in the solving sequence which we call the ternary cases. It would be interesting if this turns out to be the case, since that would indicate the special role that the boxing differential operator (which is a second-order differential operator) play in constructing a basis for motivic single-valued MZVs.

\subsection{Identities of periods between different DCI integrals from \texorpdfstring{$f$-graphs}{f-graphs}}
As reviewed in Sec.~\ref{sec:antiprism}, $f$-graphs can generate dual conformal integrals and thus SVHPLs by taking different four cycles in the diagram. The periods of these different DCI integrals and SVHPLs are the same, since they all equal to the periods of corresponding $f$-graphs which is defined as \cite{Wen:2022oky,Brown:2023zbr}
\begin{equation}
    P(f)=\frac{1}{(\pi^2)^{n-3}}\int\frac{\mathrm{d}x_1\mathrm{d}x_2\ldots \mathrm{d}x_{n}}{\mathrm{Vol}(SO(2,4))}f(x_1,\ldots,x_{n}),
\end{equation}
where $f(x_1,\ldots,x_{n})$ is the integrand of corresponding $f$-graph and $\mathrm{Vol}(SO(2,4))$ means modding out conformal transformations by fixing three points to $0,1,\infty$. The period remains the same regardless of which four points in a $f$-graph have been chosen as external points. This simple fact generates a lot of identities between periods of different DCI integrals, including \textbf{Proposition}~\ref{conj:singlezeta}. 
Exploring these identities serves as an important step in understanding the map between binary Steinmann SVHPLs and $f$-graphs.

Before going into details about these identities, we note that not all $f$-graphs can be related to binary DCI integrals and the function space $\mathcal{B}$ of binary Steinmann SVHPLs. We leave such examples and why they are not related to binary DCI integrals to App.~\ref{app:notbinary}. In the following discussion, we will focus on those $f$-graphs which are related to binary DCI integrals and $\mathcal{B}$. Then, two questions arise naturally. First, can a single $f$-graph generate different kinds of binary DCI integrals (different binary Steinmann SVHPLs)? Second, if an $f$-graph generates more than one kind of binary DCI integrals, is there any relation between these different DCI integrals? These two questions are closely related to each other and we will answer them one by one.

A single $f$-graph can generate different kinds of binary DCI integrals (different binary Steinmann SVHPLs). For example, the five-loop $f$-graph $f^{(5)}_{5}$ can generate the following two binary DCI integrals,
\begin{equation}\label{eq:f5l5}
    f_{5}^{(5)}=\vcenter{\hbox{\includegraphics[scale=0.4]{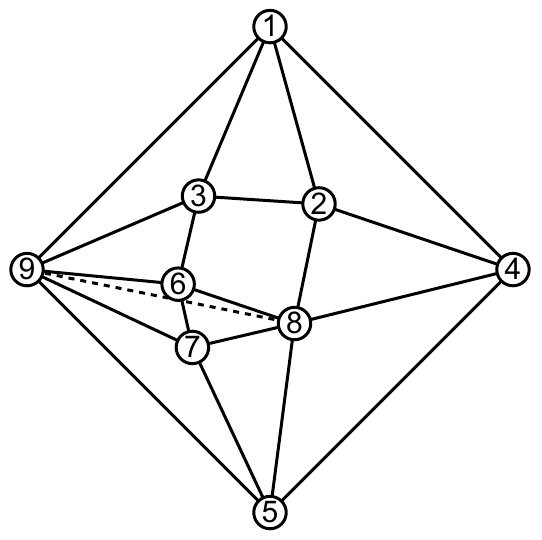}}}\Rightarrow\left\{\begin{array}{c}
         \vcenter{\hbox{\includegraphics[scale=0.4]{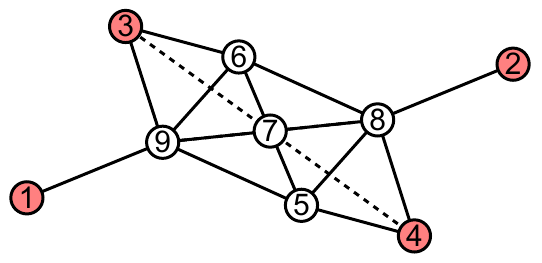}}}=\frac{1}{\mz-\mzb}B_{10010}(\mz)\\
         \vcenter{\hbox{\includegraphics[scale=0.4]{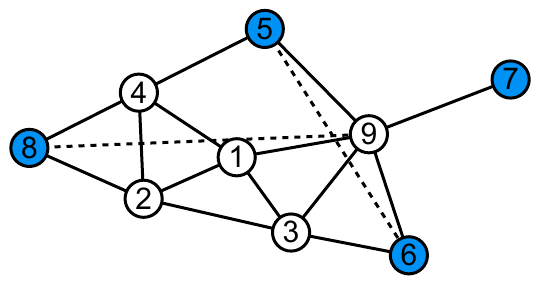}}}=\frac{1}{\mz-\mzb}B_{10100}(\mz)
    \end{array}\right. .
\end{equation}
We can notice that these two binary DCI integrals are evaluated to two binary Steinmann SVHPLs with the order of $10$ and $100$ reversed in their words. This is not a coincidence. Actually, we can prove the following statement,
\begin{prop}\label{prop:reverse}
    If a binary DCI integrals which is evaluated to $B_{n_1,n_2,\ldots,n_{r}}(\mz)$ with $n_i\ge 2$ can be generated from a weight-preserving $f$-graph by taking a four-cycle, then so will the binary DCI integrals which is evaluated to $B_{n_{r},n_{r-1},\ldots,n_1}(\mz)$ from the same $f$-graph.
\end{prop}
The proof, given in App.~\ref{app:proof}, follows from the reversibility of the boxing operation on the diagram. A direct corollary of this statement is the \textbf{Proposition} \ref{conj:reverse}, since the periods of $f$-graphs are independent of how the ``external'' points are chosen. Thus $P_{n_1,n_2,\ldots,n_{r}}=P_{n_r,n_{r-1},\ldots,n_1}$. When $\{n_1,n_2,\ldots,n_{r}\}=\{n_r,n_{r-1},\ldots,n_1\}$, these two kinds of binary DCI integrals become one and the identity holds trivially.

There are a few comments here that further reveal the map between binary Steinmann SVHPLs and $f$-graphs. First, the above proposition and corresponding proof do not hold for binary DCI integrals which are evaluated to $B_{n_1,n_2,\ldots,1}(\mz)$. Actually, all the DCI integrals obtained by taking four-cycles are binary Steinmann SVHPLs whose words end with '0', belonging to the smaller set $\hat{\mathcal{B}}$. This can be argued as follows. We can apply the boxing differential operators to the binary DCI integrals obtained by taking a four-cycle of planar $f$-graph where $1,2,3,4$ lie in order. Supposing that the external point on which we act the boxing differential operator is $x_{4}$, $x_{4}$ must be attached to a single inner point without any other attachments. Then $x_1$ and $x_{3}$ must be connected with each other by at least one dashed line (numerator), because when taking the four-cycle we have multiplied a factor $\xi_{4}=x_{12}^{2}x_{23}^{2}x_{34}^{2}x_{14}^{2}x_{13}^{2}x_{24}^{2}$\footnote{It is actually $x_{12}^{2}x_{23}^{2}x_{34}^{2}x_{14}^{2}x_{13}^{4}x_{24}^{4}$. However, we will always factor out an $x_{13}^{2}x_{24}^{2}$. That is, in the diagram we keep external points with degree 1 rather than 0. See Sec.~\ref{sec:antiprism} for the related conventions.} and for planar $f$-graphs we cannot have a substructure like $x_{12}^{2}x_{23}^{2}x_{34}^{2}x_{14}^{2}x_{13}^{2}x_{24}^{2}$ in denominator which is against planarity in the diagram. If we already have $x_{24}^{2}$ in the denominator in the integrand (there must be one $x_{24}^{2}$ in the denominator for $x_{4}$ to be attached to a single inner point without any other attachment after multiplying $\xi_{4}$), we cannot further have $x_{13}^{2}$ in the denominator. Therefore, multiplying $\xi_{4}$ will introduce a $x_{13}^{2}$ in numerator. This indicates that the boxing differential operator $\frac{x_{14}^{2}x_{34}^{2}}{x_{13}^{2}}\square_{4}$ will not introduce a denominator $x_{13}^{2}$ to the diagram since it will at least be canceled by $x_{13}^{2}$ we multiplied in $\xi_{4}$ if there is no additional numerator containing $x_{13}^{2}$. In this case, there is no additional normalization factor as in \eqref{eq:boxingdiag}, which indicates the last word of corresponding SVHPL is '0'. How $B_{n_1,n_2,\ldots,1}(\mz)$ will be related to $f$-graphs is discussed in App.~\ref{app:Bxxx1}. It is less relevant to the following discussions.

Second, \textbf{Proposition}~\ref{prop:reverse} does not apply for the weight-dropping $f$-graphs, since their boxing differential equation will not end at a box but some p-integrals as discussed in App.~\ref{app:weight}. However, the same fact that periods of $f$-graphs are independent of how the ``external points'' are chosen still gives very strong constraints on the result of these p-integrals. For example, in eight loops, there is such a weight-dropping $f$-graph, which can generate two different kinds of binary DCI integrals:
\begin{equation}
    f_{2669}^{(8)}=\vcenter{\hbox{\includegraphics[scale=0.35]{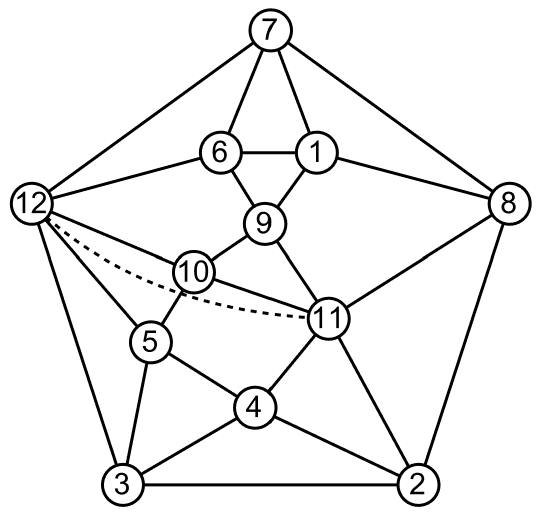}}}\Rightarrow\left\{\begin{array}{c}
         \vcenter{\hbox{\includegraphics[scale=0.25]{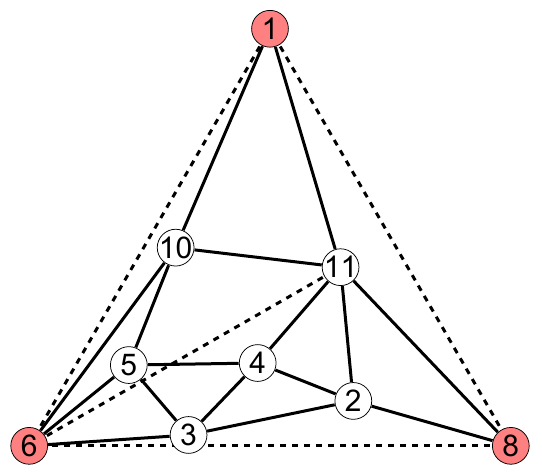}}}\times B_{10}(\mz)\\
         \vcenter{\hbox{\includegraphics[scale=0.25]{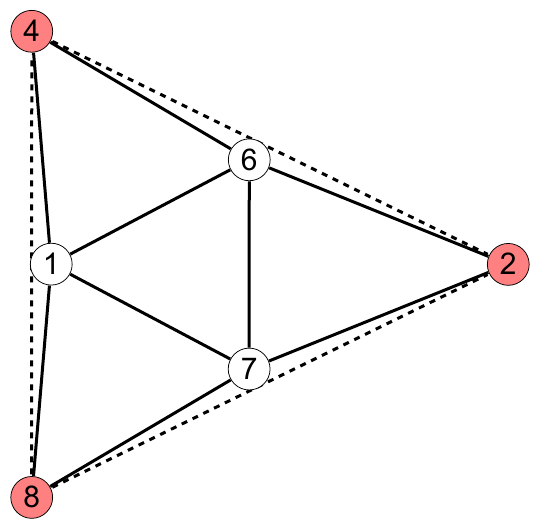}}}\times B_{10010}(\mz)
    \end{array}\right. .
\end{equation}
Then we can derive
\begin{equation}
    \vcenter{\hbox{\includegraphics[scale=0.3]{fig/f2669l8dci1.pdf}}}=P(B_{10010}(\mz))=4240\zeta_{11}-80(f_{3,3,5}+f_{3,5,3}-2f_{5,3,3}),
\end{equation}
since
\begin{equation}
    \vcenter{\hbox{\includegraphics[scale=0.25]{fig/f2669l8dci2.pdf}}}=P(B_{10}(\mz))=20\zeta_{5}
\end{equation}
Conformal diagrams with only three external points will always be evaluated to some number. 

Third, the above proposition relies on an assumption that we can find an $f$-graph that generates the binary DCI integrals $B_{n_1,n_2,\ldots,n_{r}}(\mz)$ with $n_i\ge 2$. This can be proven constructively. We can directly promote the canonical series of binary DCI integrals constructed in Sec.~\ref{sec:canonical} to a canonical series of $f$-graphs. For the $L$-loop canonical binary DCI integrals whose corresponding word end with '0', supposing that the external legs are named $1,2,3,4$ in order, we can multiply the integrand with $1/(x_{12}^{2}x_{23}^{2}x_{34}^{2}x_{14}^{2}x_{13}^{2}x_{24}^{2})$, which brings it to an $L$-loop $f$-graph. Then for arbitrary binary Steinmann SVHPL whose words end with '0', we can indeed find an $f$-graph that generates it. This finishes the proof of \textbf{Proposition} \ref{conj:reverse}. We have checked up to ten loops that all such $f$-graphs have nonzero coefficients as part of ansatz for four-point correlators in $\mathcal{N}=4$ SYM and the absolute values of their coefficients are all Catalan numbers. They form a canonical series of $f$-graphs which we denote as $\mathcal{F}$. The canonical $f$-graphs can be taken as standard representatives of the binary Steinmann SVHPLs.

Then the remaining question is, can the same weight-preserving $f$-graph generate binary Steinmann SVHPLs other than the above two? The existence of any other binary Steinmann SVHPLs will imply additional identities between periods of these SVHPLs and we have checked up to ten loops that the answer is no. We believe this to be true for arbitrary loops but do not know how to prove this in general. However, the same $f$-graph can indeed generate other SVHPLs which are not binary Steinmann SVHPLs. For example, the four-loop antiprism $f$-graph can generate $2\times 2$ fishnet DCI integrals\footnote{All fishnet integrals can be calculated as determinants of ladders~\cite{Basso:2017jwq,Coronado:2018cxj}.} and $B_{1010}(\mz)$ in the same time,
\begin{equation}\label{eq:fishnet22}
    f_{2}^{(4)}=\vcenter{\hbox{\includegraphics[scale=0.4]{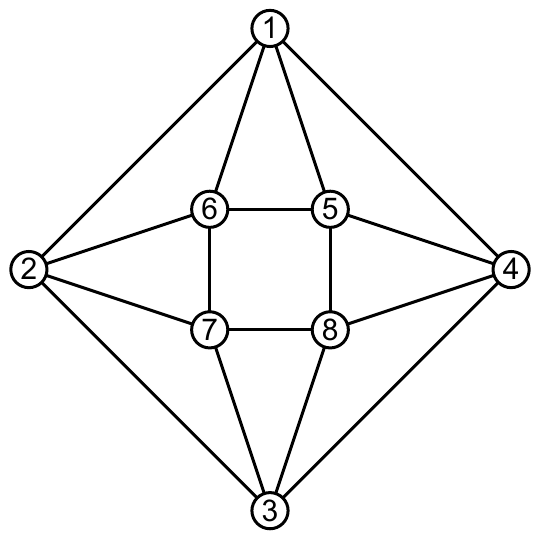}}}\Rightarrow\left\{\begin{array}{l}
         \vcenter{\hbox{\includegraphics[scale=0.35]{fig/apdci8.pdf}}}=\frac{1}{\mz-\mzb}B_{1010}(\mz)\\
         \vcenter{\hbox{\includegraphics[scale=0.3]{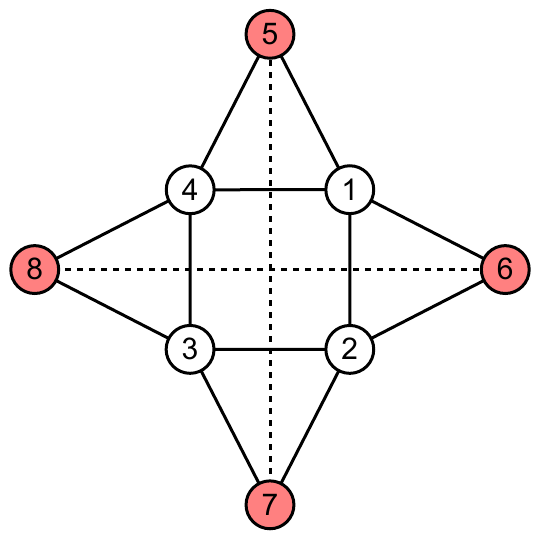}}}\frac{B_{1}(\mz)B_{3}(\mz)-\frac{1}{3}B_{2}^{2}(\mz)}{(\mz-\mzb)^2}
    \end{array}\right. .
\end{equation}
In this way, we can calculate the periods of many DCI integrals which do not satisfy boxing differential equation. We mainly give three typical examples here. One is the $2\times n$ fishnet integral family. They can be generated from such $f$-graphs that also generate the DCI integrals which are evaluated to zigzag SVHPLs. The first non-trivial example of this family, $2\times 2$ fishnet, has been given in \eqref{eq:fishnet22}. $2\times 3$ fishnet can be generated from the following 6-loop f-graph
\begin{equation}\label{eq:f31l6}
    f_{31}^{(6)}=\vcenter{\hbox{\includegraphics[scale=0.4]{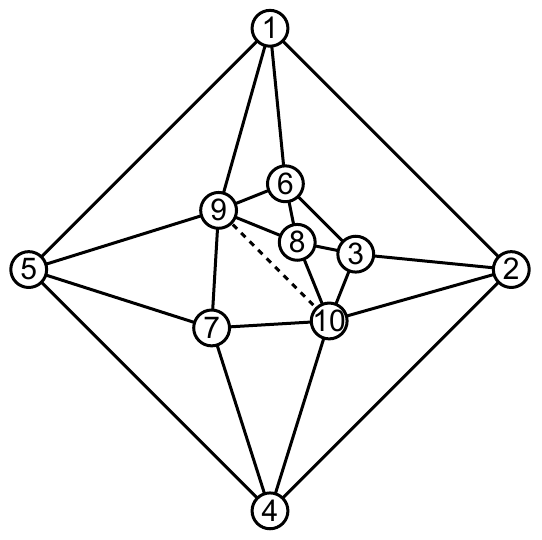}}}\Rightarrow\left\{\begin{array}{l}
         \vcenter{\hbox{\includegraphics[scale=0.35]{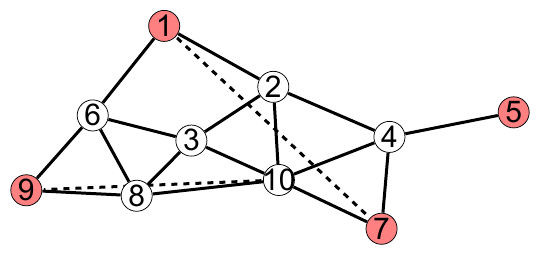}}}= \frac{1}{\mz-\mzb}B_{101010}(\mz)\\
         \vcenter{\hbox{\includegraphics[scale=0.3]{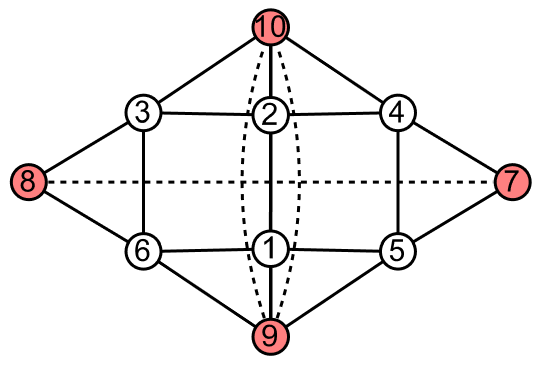}}}=\frac{B_{2}(\mz)B_{4}(\mz)-\frac{1}{2}B_{3}^{2}(\mz)}{(\mz-\mzb)^2}
    \end{array}\right. .
\end{equation}
and $2\times 4$ fishnet is generated from
\begin{equation}\label{eq:f2524l8}
    f_{2524}^{(8)}=\vcenter{\hbox{\includegraphics[scale=0.4]{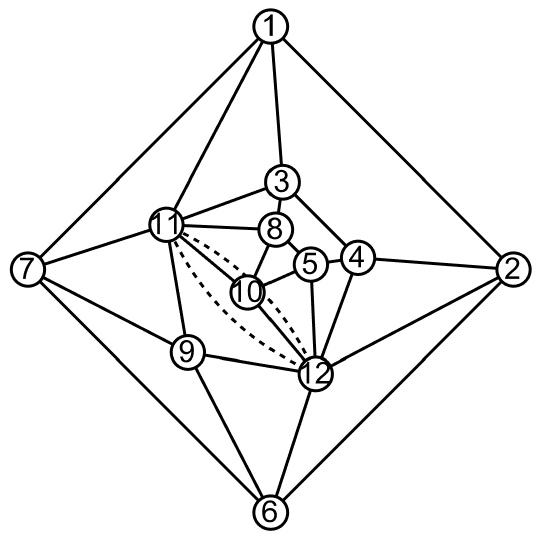}}}\Rightarrow\left\{\begin{array}{l}
         \vcenter{\hbox{\includegraphics[scale=0.35]{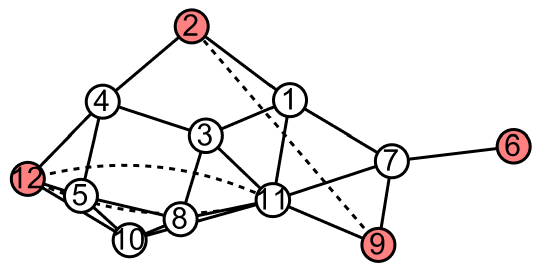}}}= \frac{1}{\mz-\mzb}B_{10101010}(\mz)\\
         \vcenter{\hbox{\includegraphics[scale=0.3]{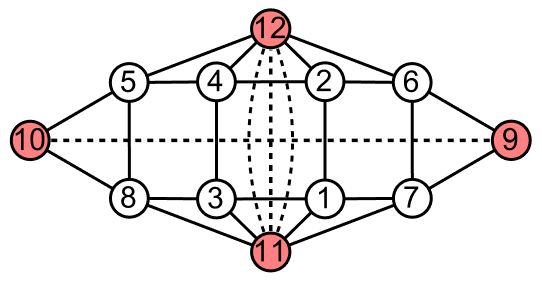}}}=\frac{B_{3}(\mz)B_{5}(\mz)-\frac{3}{5}B_{4}^{2}(\mz)}{(\mz-\mzb)^2}
    \end{array}\right. .
\end{equation}
The above series can be directly generated to higher loops. It is interesting that the periods of $2\times n$ fishnet integrals\footnote{The leading singularities of $2\times n$ fishnets are $1/(\mz-\mzb)^2$. When calculating their periods, we should use definition \eqref{eq:defp2}. For example, for $2\times 2$ fishnet $P(f_{5}^{(5)})=\int_{\mathcal{C}}\frac{\mathrm{d}\mz\mathrm{d}\mzb}{2\pi i}\frac{(\mz-\mzb)^2}{\mz\mzb(1-\mz)(1-\mzb)}\frac{B_{1}(\mz)B_{3}(\mz)-\frac{1}{3}B_{2}^{2}(\mz)}{(\mz-\mzb)^2}$.  } equal to those of zigzag DCI integrals. The second is the ladder family. Considering the canonical $f$-graph of ladders, we can take the four cycle to split the diagram into two parts and each of them is a ladder. Then we will derive the following identity which has been proved in \cite{Dorigoni:2021guq},
\begin{equation}
    P(f^{(L)}(\mz))=P^{\prime}(f^{(L-M)}(\mz)\times f^{(M)}(\mz)),
\end{equation}
where $f^{(L)}(\mz)$ is the $L$-loop ladder. This can be proved in the following way. Let us consider different choices of four cycles in Fig.~\ref{fig:ladderfgraph} as \eqref{eq:ladderperiodidg}; the period of this $f$-graph is independent of this choice.
\begin{equation}\label{eq:ladderperiodidg}
    P\left(\vcenter{\hbox{\includegraphics[scale=0.35]{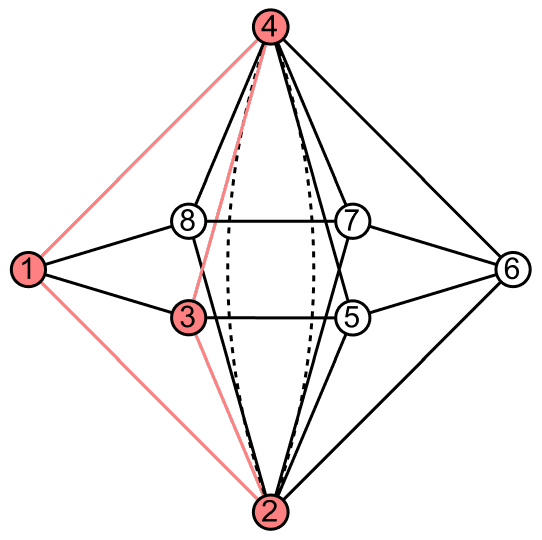}}}\right)=P\left(\vcenter{\hbox{\includegraphics[scale=0.35]{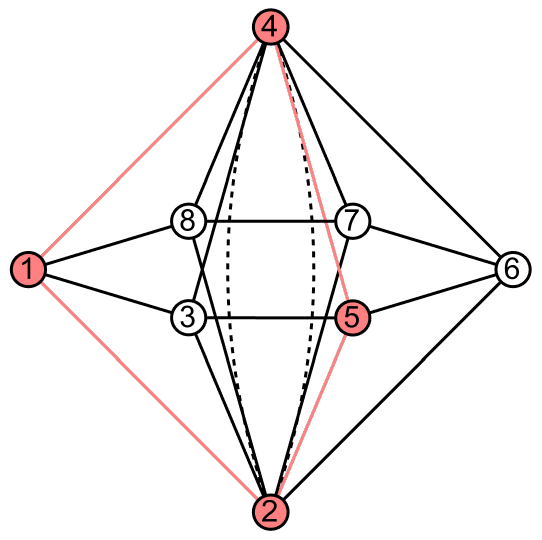}}}\right)
\end{equation}
where vertices colored red are chosen as external points. It can be directly translated into
\begin{equation}
    P(f^{(4)}(\mz))=P^{\prime}(f^{(1)}(\mz)\times f^{(3)}(\mz)).
\end{equation}
The readers can easily generate above case to arbitrary $M,L-M\ge 1$ by different partition of the graph (which changes $M$) and upgrade the hexagon in the middle plane to $n$-gon (which changes $L$). 
The last one is a non-trivial 8-loop example which is used to show that some very nontrivial dual conformal integrals can have simple periods,
\begin{equation}
    f_{1976}^{(8)}=\vcenter{\hbox{\includegraphics[scale=0.4]{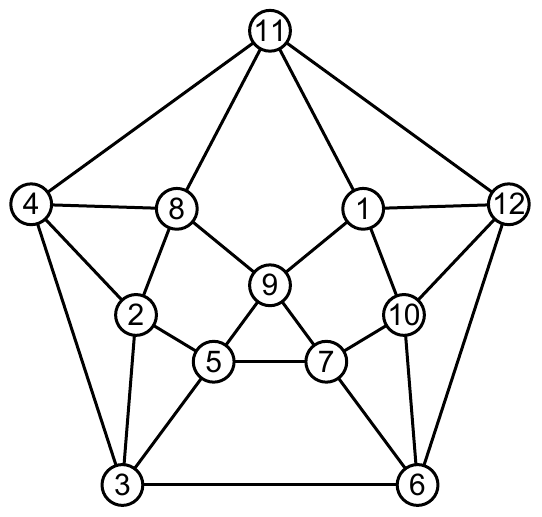}}}\Rightarrow\left\{\begin{array}{l}
         \vcenter{\hbox{\includegraphics[scale=0.35]{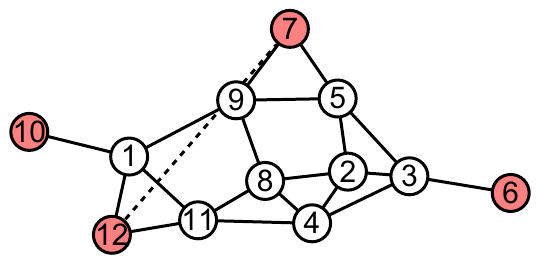}}}\\
         \vcenter{\hbox{\includegraphics[scale=0.3]{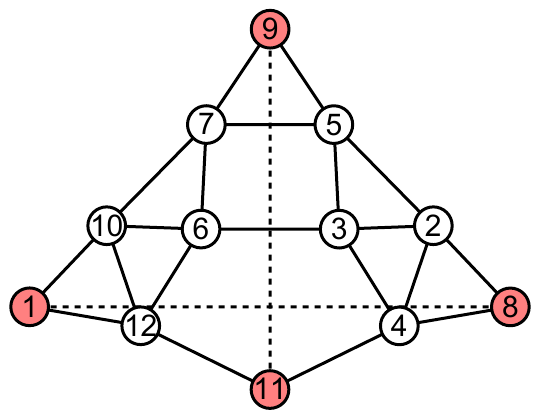}}}
    \end{array}\right. .
\end{equation}
Its period has been calculated in \eqref{eq:f1976l8period} by studying boxing differential equation of the DCI integrals in the first row. We can thus derive the period of the non-trivial DCI integrals in the second row.

At last, we present the planar $f$-graphs that have nonzero coefficients and can be related to binary Steinmann SVHPLs by taking four cycles up to nine loops in the ancillary \texttt{Mathematica} notebook \texttt{binarytable.nb} and a data file \texttt{fmap.m}. The function \texttt{fmap[]} in the notebook can work in two direction: given a binary word $w$ for binary Steinmann SVHPLs, \texttt{fmap[w]} gives the $f$-graphs related and given the number $a$ and loop $l$ for an $f$-graph $f^{(l)}_{a}$\footnote{This only works for those $f$-graphs that can be solved recursively by boxing differential equations. The number $a$ is labeled as in \cite{Bourjaily:2016evz}.}, \texttt{fmap[a,l]} gives the binary words and periods related to $f^{(l)}_{a}$.

\section{Conclusions}\label{sec:conclusion}

In this paper we have started the study of infinite families of all-loop ``binary" DCI integrals originated from $f$-graphs contributing to four-point amplitudes/correlators, which evaluate to certain SVHPL functions satisfying extended Steinmann relations. In this function space, we identified two ``inverse-boxing" operations which allow us to build such families of DCI integrals recursively (starting from the one-loop box integral), which label any such integral with a string of $0$ and $1$ with no consecutive $1$'s, and they include all-loop ladder integrals and the so-called ``zigzag" integrals (as the unique DCI integral from even-loop antiprism $f$-graphs and odd-loop counterparts) as two extreme cases.  In addition to studying these functions, we have also studied their periods which evaluate to single-valued MZVs contributing to integrated correlators. Both calculations are made possible due to the graphical function method proposed in~\cite{Borinsky:2022lds,Schnetz:2013hqa} and in particular the \texttt{HyperlogProcedures}. Based on such explicit data, we have also explored interesting relations among $f$-graphs/DCI integrals, binary Steinmann SVHPLs and their periods: for example, we have studied periods of all these binary DCI integrals up to $10$ loops, which exhibit very interesting structures and properties, some of which are either conjectured or proven to hold up to arbitrary loops. We have seen that the $f$-graphs themselves serve as a useful tool to study all these new structures of periods and functions, especially that one can discover relations among various periods and functions utilizing graphical properties of the $f$-graphs. We believe that these binary DCI integrals, which incorporate all-loop ladders and zigzag DCIs as two extreme cases, have only scratched the surface of the landscape of solvable infinite families of (conformal) Feynman integrals, with the notable example of fishnet integrals~\cite{Basso:2017jwq} and those given by octagons~\cite{Caron-Huot:2021usw}. 

We also note the double triangle rule in \cite{Schnetz:2008mp,Brown:2009rc} and its possible relation to the role that the boxing differential equation plays. The double triangle rule in \cite{Schnetz:2008mp} concerns the transcendental weight of an integral. It mainly states that the result of this integral has a weight drop if and only if the double triangle reduction of it has a weight drop. The boxing differential equation can generate double triangle rule in specific cases:
\begin{equation}\label{eq:boxingbasic}
        \begin{aligned}
        \vcenter{\hbox{\includegraphics[scale=0.35]{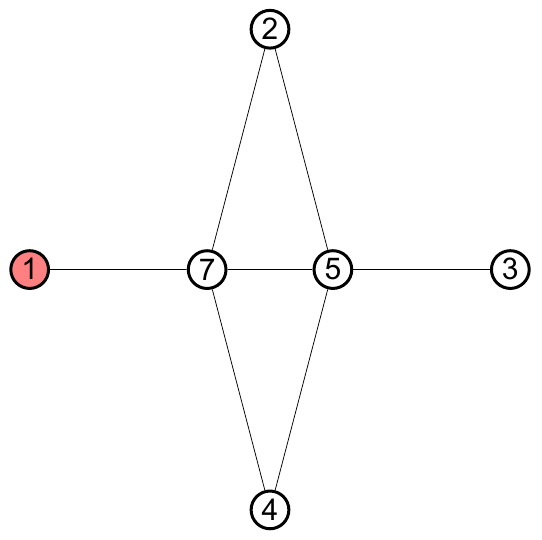}}}\xrightarrow{\frac{x_{12}^2x_{14}^{2}}{x_{24}^2}\square_{1}}\vcenter{\hbox{\includegraphics[scale=0.35]{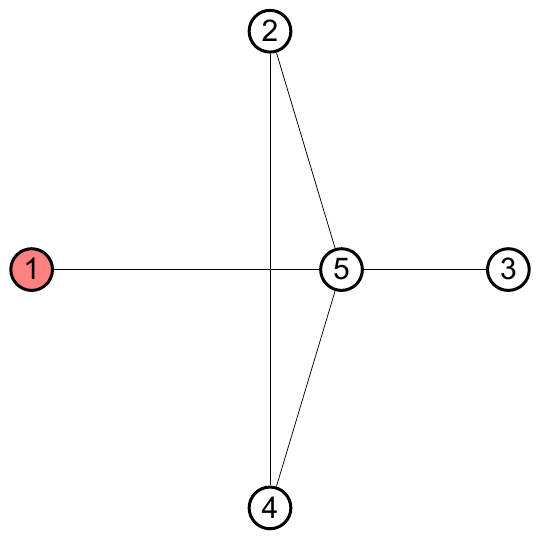}}}
        \end{aligned}
\end{equation}
where vertex $1$ colored pink can not be connected to any other inner points, while the other points without color can be connected to any other points except for vertex $1$. At the function level, the boxing operator $z\bar{z}\partial_{z}\partial_{\bar{z}}$ will reduce the transcendental weight of the original integral by 2. Therefore, if the original integral has a weight drop, it happens for the reduced integral by applying double triangle rule and vice versa\footnote{It is interesting that there is also a double triangle rule for the $f$-graphs~\cite{Bourjaily:2025iad,He:2024cej}. It reduces $L$-loop $f$-graphs to $(L-1)$-loop ones and this results in linear relations between the coefficients of them. Although in pure graphical view, the double triangle rule presented in \cite{Schnetz:2008mp} is a special case of the double triangle rule in \cite{He:2024cej} (which allows more edges attached to vertices), they have totally different applications. It can be argued in the same way by boxing differential equation that the double triangle rule in \cite{He:2024cej} also indicates the transcendental weight properties of corresponding integrals between $f$-graphs and their double triangle reduction if the DCI integrals generated are still related to (generalized) SVHPLs.}.

There are many directions that can be further explored, and here we only mention a few of them. For those Steinmann-satisfying SVHPLs which can be solved by boxing differential equations, we can directly generate the binary Steinmann SVHPLs to ternary Steinmann SVHPLs, which involve all the three kinds of differential operators. Such ternary Steinmann SVHPLs involve a larger set of $f$-graphs. There are more identities involving periods of such ternary Steinmann SVHPLs, which can be predicted by corresponding $f$-graphs. Another direction is to further explore those $f$-graphs that cannot be solved recursively by boxing differential equations. Since these $f$-graphs contribute to amplitudes/correlators of $\mathcal{N}=4$ SYM, it would be highly desirable to compute them both as SVMPL functions and periods for perturbative study of half-BPS and integrated correlators. For the latter, we would like to accumulate more higher-loop data and analyze possible all-loop patterns for certain classes of $f$-graphs, as well as understanding more types of relations among these periods. 

It would also be interesting to study further the implications of (extended) Steinmann relations and the associated function spaces, which has played an important role in our study. Numerous other properties in this space can also be explored, {\it e.g.} recently a surprising {\it antipodal self-duality}~\cite{Dixon:2025zwj} has been discovered for all squared fishnet integrals~\cite{Basso:2017jwq} and other infinite families of polynomials of ladder integrals. It would be highly desirable to relate those functions to $f$-graphs, and even to see if there is any such duality for more generalized space {\it e.g.} possible polynomials of generalized ladder functions. Another important point is that unlike ladder integrals, it is an open question if one could resum some of these infinite families to all loops, such as the odd-loop or even-loop zigzag DCI integrals, as a function of the coupling (see~\cite{Arkani-Hamed:2021iya} for a recent study from a different perspective). 

In general for $f$-graphs/DCI integrals, it would be interesting to develop a systematic bootstrap method for corresponding (SV) multiple-polylogarithmic functions, which generally involve non-trivial {\it leading singularities} and new symbol letters~\cite{He:2025vqt} (for the correlator elliptic MPL functions already start to appear at four loops~\cite{He:2025rza}). In this regard, a direct look into their periods may give us hints at better organization of these integrals {\it e.g.} such that each of them has uniform, maximal transcendentality (weight $2\ell$), which may simplify such bootstrap program for higher loop integrals. Last but not least, having computed some of these DCI integrals at higher loops (or even all loops) should have applications for computing other physical quantities (amplitudes/correlators) which can be expressed in terms of these integrals, and we leave such applications to future investigations. 
\acknowledgments
It is our pleasure to thank Lance Dixon, Zhenjie Li and Jiahao Liu for inspiring discussions, and Jacob Bourjaily, Canxin Shi and Yichao Tang for collaborations on generating the $f$-graph data. The work of SH is supported by the National Natural Science Foundation of China under Grant No. 12225510, 12447101, 12247103, and by the New Cornerstone Science Foundation through the XPLORER PRIZE.

\clearpage
\appendix
\section{Relations between integrated antiprism and zigzag periods}\label{app:fourier}
In Sec.~\ref{sec:antiprism}, we prove the periods of antiprism diagrams to be zigzag periods in an algebraic way. Here we prove in a diagrammatical way that they are equivalent. This relation between integrals in dual coordinate space and momentum space is called planar duality and studied by Fourier transformation in \cite{Schnetz:2008mp}. In \cite{Schnetz:2008mp}, the overall divergence of diagrams in momentum space is tackled in the same way as dual coordinate space. Here we show that traditional ways of dealing with the divergence in momentum space will also result in the same result. Let us take the eight-point antiprism $f$-graph as an example:
\begin{equation}\label{eq:integratedd2}
        \vcenter{\hbox{\includegraphics[scale=0.35]{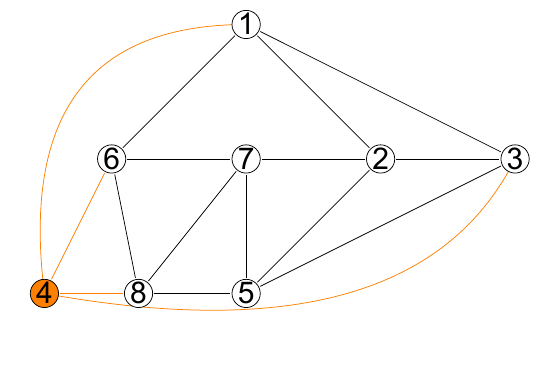}}}\xrightarrow{x_4\to \infty} \vcenter{\hbox{\includegraphics[scale=0.35]{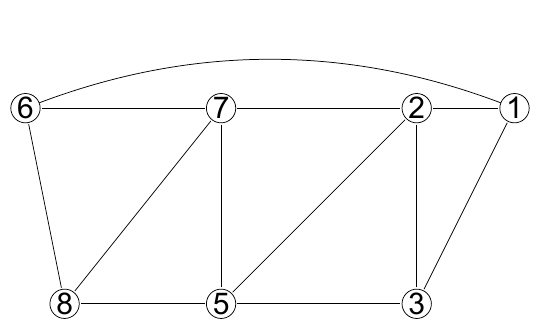}}}
    \end{equation}
    The left-hand-side of \eqref{eq:integratedd2} is the integrand of the $f$-graph. To calculate the integrated antiprism, we need to integrated out every vertex of this diagram. However, we also need to fix three points to mod out the symmetry group $SO(2,4)$. Here we fix $x_4$ to $\infty$, so the orange lines can be removed. Now the right hand side is very close to the vacuum $\phi_4$ zigzag diagram. They take the same form, but have different meanings as stated in Tab.~\ref{tab:zigzag}.
    \begin{table}[htbp]
        \centering
        \begin{tabular}{c|c}
            \hline\hline
            \thead{$\vcenter{\hbox{\includegraphics[scale=0.3]{fig/d2int1fig.pdf}}}$ \\ as integrated correlator} & \thead{$\vcenter{\hbox{\includegraphics[scale=0.3]{fig/d2int1fig.pdf}}}$  \\ as $\phi_4$ diagram} \\
            \hline
            dual coordinate space & loop momentum space \\
            \hline
            \thead{fix two points to be $0$ and $1$.\\ a way will be fixing $x_6$ to 0 and $x_1$ to 1.} & \thead{introduce a scale to extract the UV divergence\\  standard way will be introducing a mass or cutoff\\  to line 61\\} \\
            \hline
            $x_4$ has been fixed to infinity & \thead{the external momenta attached to degree-3 vertices \\have been set to 0 because we only need the divergence}\\
            \hline\hline   
        \end{tabular}
        \caption{The same diagram has different meanings for integrated correlator and $\phi_4$ diagram. They are related by Fourier transformation.}\label{tab:zigzag}
    \end{table}

    Now we show by Fourier transformation that they are equivalent to each other. We start from the same diagram (the right-hand-side of \eqref{eq:integratedd2}) as a $\phi_{4}$ vacuum diagram. We need to extract its divergence. The standard way is to introduce a mass to line 61~\cite{Baikov:2010hf} and this integral can be calculated as
    \begin{equation}
        \begin{aligned}
            \vcenter{\hbox{\includegraphics[scale=0.35]{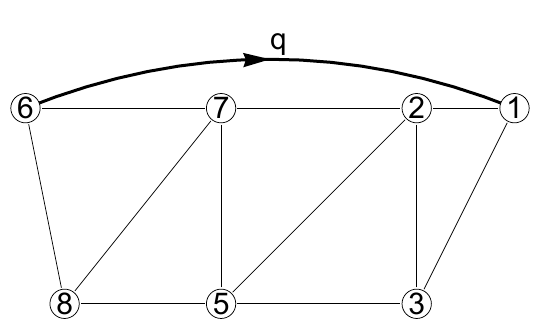}}}=&\int\frac{\mathrm{d}^{d}q}{i\pi^{d/2}}\frac{1}{-q^2+m^2}\times \vcenter{\hbox{\includegraphics[scale=0.4]{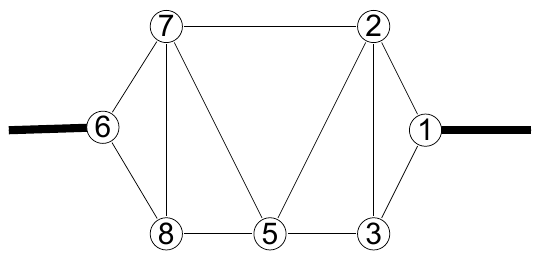}}} \\
            =&\int\frac{\mathrm{d}^{d}q}{i\pi^{d/2}}\frac{1}{-q^2+m^2}\frac{1}{(-q^2)}Z_{6} +\mathcal{O}(1) \\
            =&\frac{1}{\epsilon}Z_{6}+\mathcal{O}(1)
        \end{aligned}
    \end{equation}
    where $Z_{6}$ is the zigzag period we defined before, it is $168\zeta_{9}$. However, we can see that $Z_6$ has nothing to do with the integration of $q$. It is actually related to the self-energy integral. $\int\frac{\mathrm{d}^{d}q}{i\pi^{d/2}}\frac{1}{-q^2+m^2}$ only specifies a way how we glue the two external legs of the self-energy diagram together.
    Thus we can glue in another way (this is equivalent to that we specify another way to extract the coefficients of the divergence):
    \begin{equation}\label{eq:anotherglue}
        \begin{aligned}
            \vcenter{\hbox{\includegraphics[scale=0.35]{fig/zigzag6.pdf}}}=&\Gamma(-\epsilon)\int\frac{\mathrm{d}^{d}q}{i\pi^{d/2}}\frac{e^{2iq\cdot\mathbf{1}}}{(-q^2)^{-\epsilon}}\times \vcenter{\hbox{\includegraphics[scale=0.4]{fig/d2sefig.pdf}}} \\
            =&\Gamma(-\epsilon)\int\frac{\mathrm{d}^{d}q}{i\pi^{d/2}}\frac{e^{2iq\cdot\mathbf{1}}}{(-q^2)^{1-\epsilon}}Z_{6} +\mathcal{O}(1) \\
            =&\frac{1}{\epsilon}Z_{6}+\mathcal{O}(1)
        \end{aligned}
    \end{equation}
    where we have used the Fourier transformation
    \begin{equation}
        \frac{1}{(x^2)^{\alpha}}=\frac{\Gamma(2-\alpha-\epsilon)}{\Gamma(\alpha)}\int\frac{\mathrm{d}^{d}q}{\pi^{d/2}}\frac{e^{2iq\cdot x}}{(q^2)^{2-\alpha-\epsilon}}.
    \end{equation}
    The missing minus sign and imaginary unit $i$ is due to the difference between Minkowski and Euclidean metric. Now we perform the Fourier transformation to
    \begin{equation}
       I= \int\frac{\mathrm{d}^{d}q}{i\pi^{d/2}}\frac{e^{2iq\cdot\mathbf{1}}}{(-q^2)^{-\epsilon}}\times \vcenter{\hbox{\includegraphics[scale=0.4]{fig/d2sefig.pdf}}}
    \end{equation}
    where $\mathbf{1}^2=1$. That is, we change all the propagator of loop momenta to an expression of coordinates:
    \begin{equation}
        \frac{1}{(k^2)^{\alpha}}=\frac{\Gamma(2-\alpha-\epsilon)}{\Gamma(\alpha)}\int\frac{\mathrm{d}^{d}x}{\pi^{d/2}}\frac{e^{2ik\cdot x}}{(x^2)^{2-\alpha-\epsilon}}.
    \end{equation}
    Since the divergence has been extracted into $\Gamma(-\epsilon)$ before in \eqref{eq:anotherglue}, now we can set the dimension back to 4 ($\alpha=1,\epsilon=0$). Then after applying
    \begin{equation}
        \int\mathrm{d}^{4}k e^{2ik\cdot x}=\pi^{4}\delta^{(4)}(x).
    \end{equation}
    we arrive at the following integral in coordinate space
    \begin{equation}\label{eq:xcoordint}
       I= \int\frac{d^4x_{1}d^4x_{2}d^4x_{3}d^4x_{5}d^4x_{6}d^4x_{7}d^4x_{8}}{(\pi^2)^{5}}\frac{\delta^{4}(x_1\!-\!x_6\!-\!\mathbf{1})\delta^{4}(x_6)}{x_{12}^{2}x_{13}^{2}x_{23}^{2}x_{35}^{2}x_{25}^{2}x_{27}^{2}x_{57}^{2}x_{58}^{2}x_{78}^{2}x_{67}^{2}x_{68}^{2}}
    \end{equation}
    Due to the momentum conservation in momentum space or the translation invariance in coordinate space, we need to choose a root \cite{Chetyrkin:1980pr}, that is, the coordinate set to 0. We have chosen it to be $x_{6}$. And the integration of $q$ gives another delta function in \eqref{eq:xcoordint}. Above is exactly the integrated 8-point antiprism by fixing $x_6=0,x_1=1,x_4=\infty$. Though the above proof is all about an 8-point example, it can be generalized to arbitrary antiprism in a straightforward way. Thus we have shown in another way the integrated antiprism is directly related to the zigzag periods in the calculation of the anomalous dimension of $\phi_{4}$ theory. This relation can also be generally applied to other cases beyond the antiprism as long as there are no sub-divergences in the corresponding momentum space integral.

\section{Examples of \texorpdfstring{$f$-graphs}{f-graphs} and periods not related to binary DCI integrals}\label{app:notbinary}
First, if an $f$-graph is related to binary DCI integrals, its period will be of uniform weight since its period is the period of binary Steinmann SVHPLs. Then if the period of an $f$-graph is not of uniform weight, it is definitely not related to binary DCI integrals. Such graphs arise starting from five loops. Let us list the two of them:
\begin{equation}\label{eq:fgraphnDCI integrals}
   f_{1}^{(5)}= \vcenter{\hbox{\includegraphics[scale=0.4]{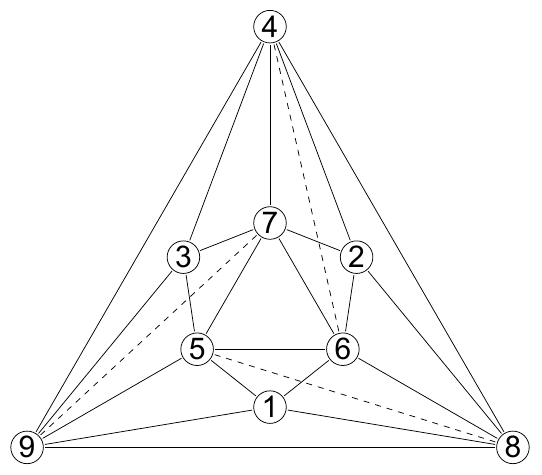}}},\,f_{2}^{(5)}= \vcenter{\hbox{\includegraphics[scale=0.4]{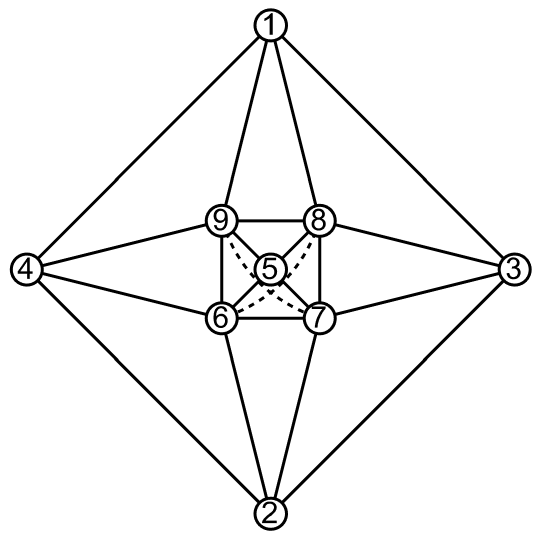}}}. 
\end{equation}
Their periods have been calculated in \cite{Wen:2022oky}\footnote{Note that the label of these diagrams in this paper is different from \cite{Wen:2022oky}.} and are not of uniform transcendental weight. Thus, they have no chance to generate binary DCI integrals which are always of uniform weight. This can be verified by taking all possible four cycles existing in these two diagrams, and we will find that the DCI integrals generated can not be reduced by boxing differential equations \textit{recursively} since they lack external points that are attached to a single inner point without any other attachments. For example, there is one such DCI integrals generated from $f_{2}^{(5)}$:
\begin{equation}\label{eq:f5DCI integrals}
  \vcenter{\hbox{\includegraphics[scale=0.5]{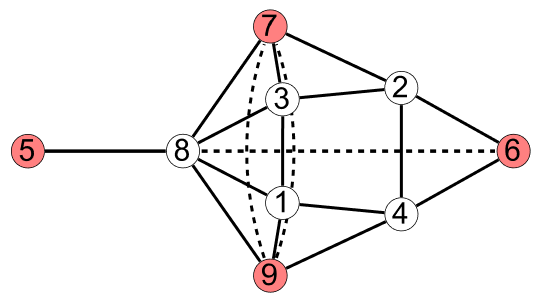}}}\xrightarrow{\frac{x_{57}^2x_{59}^{2}}{x_{79}^2}\square_{5}}\vcenter{\hbox{\includegraphics[scale=0.35]{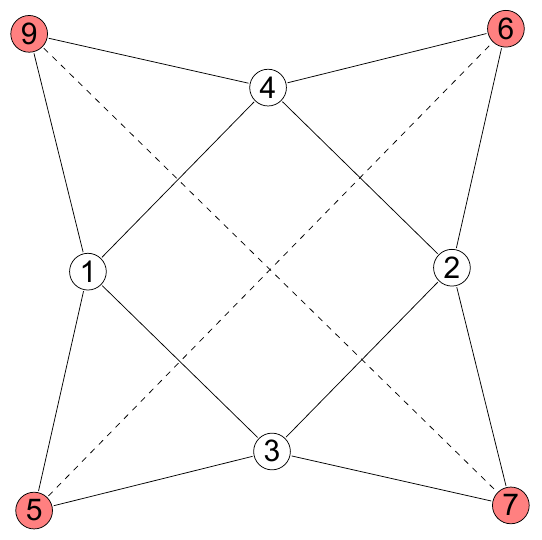}}}. 
\end{equation}
The external points are colored red. We can perform the boxing differential operator to external point $x_5$ once and the resulting integral does not satisfy additional boxing differential equations. DCI integrals generated from $f$-graphs such as $f_{1}^{(5)}$ and $f_{2}^{(5)}$ are usually not of uniform weight, this is the reason why the corresponding periods are not of uniform weight too.

Second, if the period of an $f$-graph is of uniform weight, it is still not necessarily related to a binary DCI integrals. We take three counterexamples at 8 loops for example:
\begin{equation}\label{eq:speciall8}
  f_{1976}^{(8)}=\vcenter{\hbox{\includegraphics[scale=0.35]{fig/f1976l8.pdf}}},f_{2521}^{(8)}=\vcenter{\hbox{\includegraphics[scale=0.35]{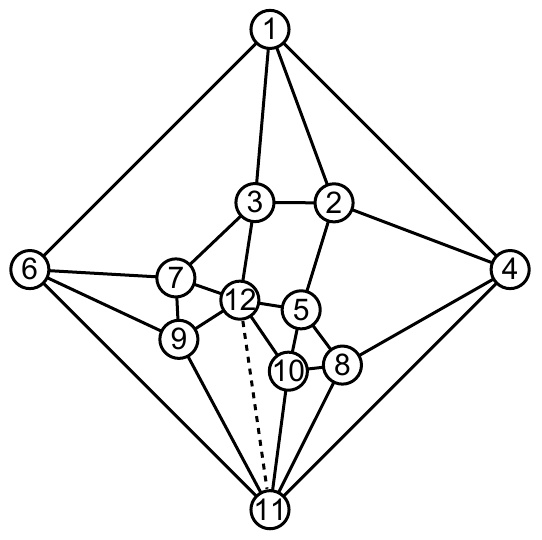}}},f_{2528}^{(8)}=\vcenter{\hbox{\includegraphics[scale=0.35]{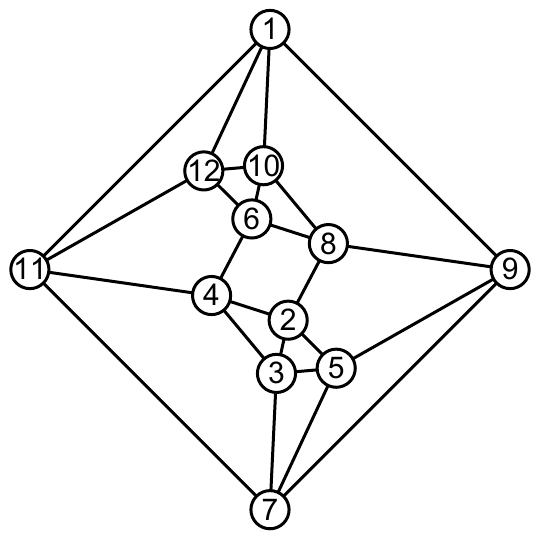}}}. 
\end{equation}
The periods of them are the same,
\begin{equation}\label{eq:periodid}
    P(f_{1976}^{(8)})=P(f_{2521}^{(8)})=P(f_{2528}^{(8)}),
\end{equation}
because the DCI integrals generated from them share the same set of boxing differential equations\footnote{The DCI integrals from them certainly have different integrands. However, they will be evaluated to the same function because they satisfy the same set of boxing differential equations. The identification of boxing sequences is fast and straightforward.} and they can be solved recursively.
However, these DCI integrals generated are not binary DCI integrals because they involve three different kinds of boxing differential operator in the solving sequence, although they are still Steinmann-satisfying SVHPLs. Let us display the boxing sequence of one DCI integrals generated by $f_{1976}^{(8)}$,
\begin{equation}\label{eq:boxingdiag3}
        \begin{aligned}
        &\mathcal{I}_{1976}^{(8)}=\vcenter{\hbox{\includegraphics[scale=0.4]{fig/I1976l8b1.pdf}}}\xrightarrow{\frac{x_{7,10}^2x_{10,12}^{2}}{x_{7,12}^2}\square_{10}}\vcenter{\hbox{\includegraphics[scale=0.35]{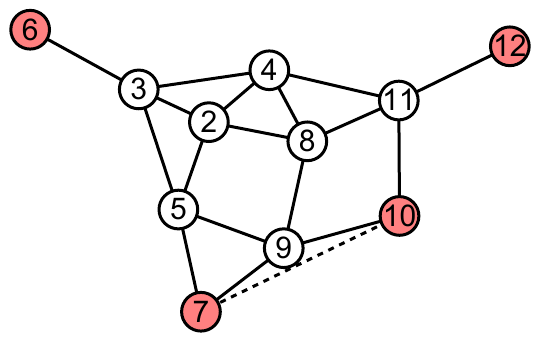}}}\xrightarrow{\frac{x_{6,12}^2x_{10,12}^{2}}{x_{6,10}^2}\square_{12}}\\
        &{\color{lightblue}\frac{x_{6,12}^{2}x_{7,10}^{2}}{x_{6,10}^{2}x_{7,12}^{2}}}\times\vcenter{\hbox{\includegraphics[scale=0.35]{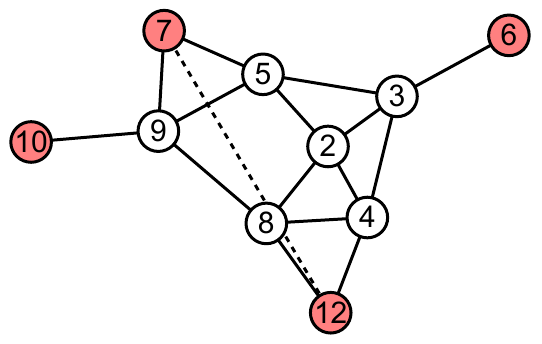}}}\xrightarrow{\frac{x_{7,10}^2x_{10,12}^{2}}{x_{7,12}^2}\square_{10}}\vcenter{\hbox{\includegraphics[scale=0.35]{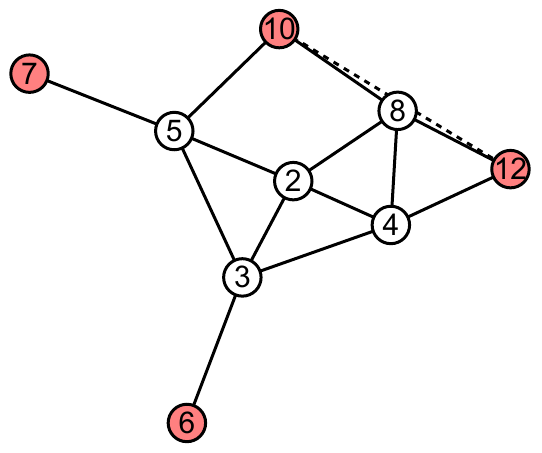}}}\xrightarrow{\frac{x_{6,7}^2x_{6,12}^{2}}{x_{7,12}^2}\square_{6}} \\
        &{\color{orange}\frac{x_{6,7}^{2}x_{10,12}^{2}}{x_{6,10}^{2}x_{7,12}^{2}}}\times\vcenter{\hbox{\includegraphics[scale=0.35]{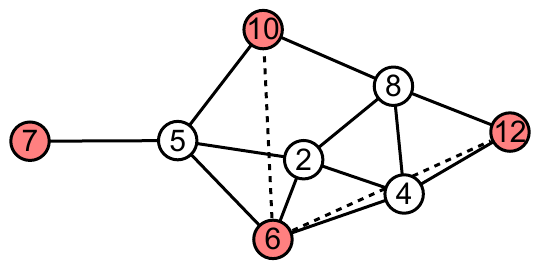}}}\xrightarrow{\frac{x_{7,10}^2x_{6,7}^{2}}{x_{6,10}^2}\square_{7}}\vcenter{\hbox{\includegraphics[scale=0.35]{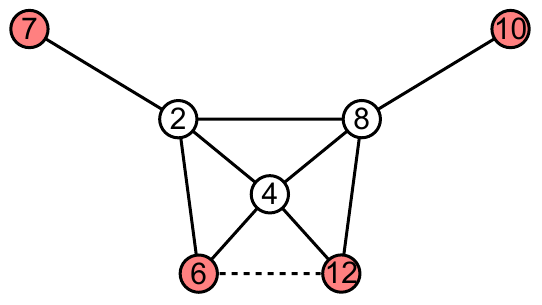}}}\xrightarrow{\frac{x_{7,10}^2x_{10,12}^{2}}{x_{7,12}^2}\square_{10}} \\
        &{\color{lightblue}\frac{x_{6,12}^{2}x_{7,10}^{2}}{x_{6,10}^{2}x_{7,12}^{2}}}\times\vcenter{\hbox{\includegraphics[scale=0.35]{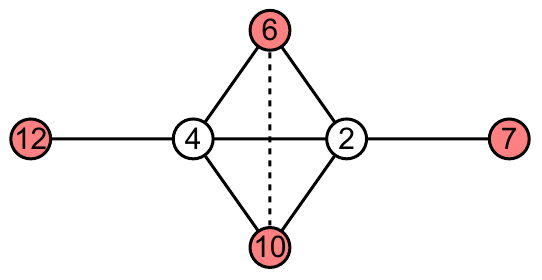}}}.
        \end{aligned}
\end{equation}
Notice that there are two different normalization factors $v=\frac{x_{6,7}^{2}x_{10,12}^{2}}{x_{6,10}^{2}x_{7,12}^{2}}$ and $u=\frac{x_{6,12}^{2}x_{7,10}^{2}}{x_{6,10}^{2}x_{7,12}^{2}}$ appearing alternatively. This means that this boxing sequence involves three different kinds of operators $z\bar{z}\partial_{z}\partial_{\bar{z}}$, $(1-z)(1-\bar{z})\partial_{z}\partial_{\bar{z}}$ and $z\bar{z}(1-z)(1-\bar{z})\partial_{z}\partial_{\bar{z}}$ simultaneously. This is not ``binary'' any more. They belong to a larger function space which we will call ternary Steinmann SVHPL space. We may notice that such phenomena should appear at 6 loops already. However, the corresponding 6-loop $f$-graph has coefficient 0 in $\mathcal{N}=4$ SYM\footnote{As a contrast, the coefficients of above three 8-loop $f$-graphs are $1,\frac{1}{2},\frac{1}{2}$.}. Although for the purpose of studying periods, we actually do not care whether the coefficient is 0, we still give a more complicated example to show that it indeed also occurs in the calculation of physical observables. We can solve this boxing sequence and obtain its result, which is
\begin{equation}
    \begin{aligned}
        &\mathcal{I}_{1976}^{(8)}(\mz)=-4 f_{5} \mathrm{I}_{z ,0,1,0,1,0,1,0,1,0,0,1,0}+4 f_{3} \mathrm{I}_{z ,0,1,0,1,0,1,0,1,0,0,1,0,1,0}+4 f_{5} \mathrm{I}_{z ,0,1,0,0,0,1,0,1,0,0,1,0}\\
        &-4 f_{3} \mathrm{I}_{z ,0,1,0,0,0,1,0,1,0,0,1,0,1,0}-12 f_{7} \mathrm{I}_{z ,0,1,0,1,0,1,0,0,1,0}+4 f_{5} \mathrm{I}_{z ,0,1,0,1,0,1,0,0,1,0,1,0}\\
        &-4 f_{3} \mathrm{I}_{z ,0,1,0,1,0,1,0,0,1,0,1,0,1,0}+12 f_{7} \mathrm{I}_{z ,0,1,0,0,0,1,0,0,1,0}-4 f_{5} \mathrm{I}_{z ,0,1,0,0,0,1,0,0,1,0,1,0}\\
        &+4 f_{3} \mathrm{I}_{z ,0,1,0,0,0,1,0,0,1,0,1,0,1,0}+24\left( f_{3,5}+ f_{5,3}\right) \mathrm{I}_{z ,0,1,0,0,0,1,0,1,0}\\
        &-\left(13466 f_{13}+560 f_{3,3,7}-1600 f_{3,5,5}+560 f_{3,7,3}\right) \mathrm{I}_{z ,0,1,0,0}+180 \mathrm{I}_{z ,0,1,0,0,0,1,0,0} f_{9}\\
        &\!\!-\!\!180 \mathrm{I}_{z ,0,1,0,1,0,1,0,0} f_{9}\!-\!12 \mathrm{I}_{z ,0,1,0,0,0,1,0,1,0,0} f_{7}\!+\!12 \mathrm{I}_{z ,0,1,0,1,0,1,0,1,0,0} f_{7}\!+\!20 \mathrm{I}_{z ,0,1,0,0,0,1,0,0,0,1,0,0} f_{5}\\
        &+20 \mathrm{I}_{z ,0,1,0,1,0,1,0,1,0,1,0,0} f_{5}-4 \mathrm{I}_{z ,0,1,0,0,0,1,0,0,1,0,0,1,0,0} f_{3}+4 \mathrm{I}_{z ,0,1,0,0,0,1,0,1,0,0,0,1,0,0} f_{3}\\
        &-20 f_{5} \mathrm{I}_{z ,0,1,0,0,0,1,0,1,0,1,0,0}-20 f_{5} \mathrm{I}_{z ,0,1,0,1,0,1,0,0,0,1,0,0}+4 f_{3} \mathrm{I}_{z ,0,1,0,1,0,1,0,0,1,0,0,1,0,0}\\
        &-4 f_{3} \mathrm{I}_{z ,0,1,0,1,0,1,0,1,0,0,0,1,0,0}-24\left( f_{3,5}+ f_{5,3}\right) \mathrm{I}_{z ,0,1,0,1,0,1,0,1,0}\\
        &-\left(-168 f_{7,7}+360 f_{9,5}+360 f_{5,9}-594 f_{3,11}-594 f_{11,3}\right) \mathrm{I}_{z ,0,1,0}\\
        &+\left(140 f_{3,7}-160 f_{5,5}+140 f_{7,3}\right) \mathrm{I}_{z ,0,1,0,0,0,1,0}-\left(140 f_{3,7}-160 f_{5,5}+140 f_{7,3}\right) \mathrm{I}_{z ,0,1,0,1,0,1,0}\\
        &+\mathrm{I}_{z ,0,1,0,1,0,1,0,0,1,0,1,0,1,0,1,0,0}+\mathrm{I}_{z ,0,1,0,1,0,1,0,1,0,0,1,0,0,0,1,0,0}-\mathrm{I}_{z ,0,1,0,1,0,1,0,1,0,0,1,0,1,0,1,0,0}\\
        &-\mathrm{I}_{z ,0,1,0,0,0,1,0,1,0,0,1,0,0,0,1,0,0}-\mathrm{I}_{z ,0,1,0,0,0,1,0,0,1,0,1,0,1,0,1,0,0}+\mathrm{I}_{z ,0,1,0,0,0,1,0,0,1,0,1,0,0,0,1,0,0}\\
        &+\mathrm{I}_{z ,0,1,0,0,0,1,0,1,0,0,1,0,1,0,1,0,0}-\mathrm{I}_{z ,0,1,0,1,0,1,0,0,1,0,1,0,0,0,1,0,0}.
    \end{aligned}
\end{equation}
Then we can calculate its period to be
\begin{equation}\label{eq:f1976l8period}
\begin{aligned}
    &P(f_{1976}^{(8)})=\frac{93566592}{5}f_{17}-4800 f_{7,5,5}-\frac{3200}{3} (f_{9,3,5}+ f_{9,5,3})+152512 f_{11,3,3}\!\\
    &-\!47808 (f_{3,3,11}+f_{3,11,3})+\frac{40000}{3} (f_{3,5,9}+
   f_{3,9,5})+2560( f_{3,3,3,5,3}+f_{3,3,3,3,5}-2 f_{3,5,3,3,3}).
\end{aligned}
\end{equation}
It is of uniform weight and not one of those periods of binary DCI integrals. We will refer to such periods as ternary type.

\section{Weight-preserving \texorpdfstring{$f$-graphs}{f-graphs} and weight-dropping \texorpdfstring{$f$-graphs}{f-graphs}}\label{app:weight}
 $f$-graphs which are related to binary DCI integrals and whose periods are of uniform weight can be further divided into two types. One is the weight-preserving type and the other is the weight-dropping type. This notation follows directly from studies of the weight drop of Feynman integrals in $\phi^{4}$ theory~\cite{Brown:2009rc,Brown:2009ta,Schnetz:2008mp}. There are well-studied operations that can explain the weight drop in $\phi^{4}$ theory~\cite{Brown:2009rc}. However, here we will study these two types of $f$-graphs by boxing differential equations. The first reason is that $f$-graphs in $\mathcal{N}=4$ SYM generally have nontrivial numerators which are different from $\phi^{4}$. The second reason is that since the $f$-graphs we will study can all be solved by boxing differential equations, it is natural to understand the difference in this way.
 
 DCI integrals generated from these two types of $f$-graphs can be solved recursively by boxing differential equations. The difference is the end point of this recursion. Applying the boxing differential operator recursively, the weight-preserving type will end at the box integrals we give in Fig.~\ref{fig:box}, thus the original integral is directly the binary DCI integrals by definition whose transcendental weight is $2L$ where $L$ is the loop number or number of inner points. However, the weight-dropping type will end at a transcendental number which is calculated from an equivalent ``two-point'' diagram. They are also called p-integrals in the literature~\cite{Baikov:2010hf}. The weights of these binary DCI integrals are lower than $2L$. The weight-preserving type is easy to understand and an example is the antiprism $f$-graph we analyzed in Sec.~\ref{sec:antiprism}. The weight-dropping types arise starting from five loops as well. And in five-loop cases, there is only one such example,
\begin{equation}\label{eq:fgraphweightdrop}
   f_{7}^{(5)}= \vcenter{\hbox{\includegraphics[scale=0.35]{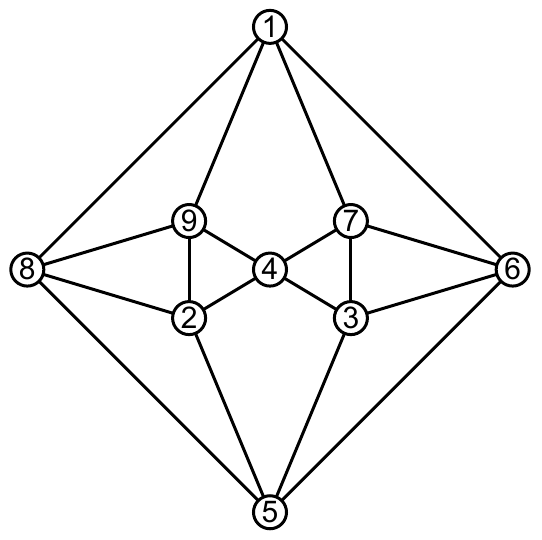}}}.
\end{equation}
Let us take a look at the dual conformal integrals taken from it and its boxing sequence:
\begin{equation}\label{eq:boxingdiag2}
        \begin{aligned}
        &\vcenter{\hbox{\includegraphics[scale=0.35]{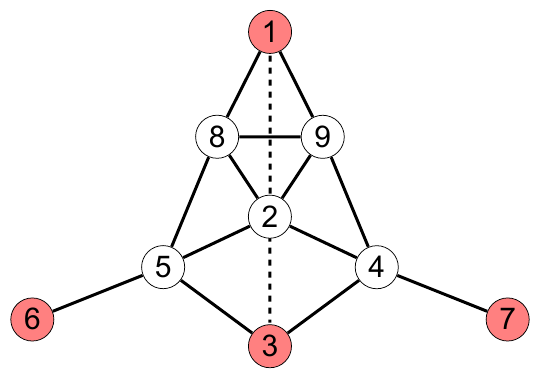}}}\xrightarrow{\frac{x_{16}^2x_{36}^{2}}{x_{13}^2}\square_{6}}\vcenter{\hbox{\includegraphics[scale=0.35]{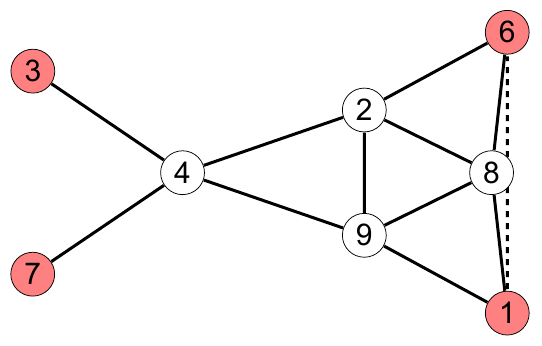}}}\xrightarrow{\frac{x_{36}^2x_{37}^{2}}{x_{67}^2}\square_{3}}\vcenter{\hbox{\includegraphics[scale=0.35]{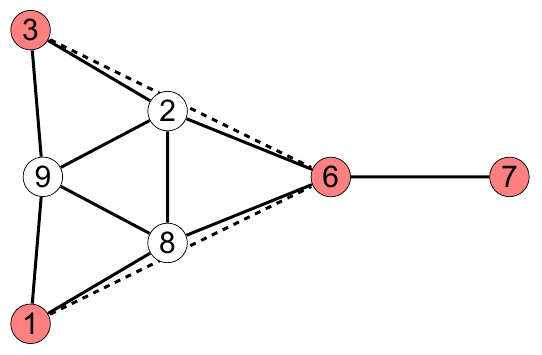}}} \\
        &=\frac{1}{x_{13}^2x_{67}^2}\times\vcenter{\hbox{\includegraphics[scale=0.3]{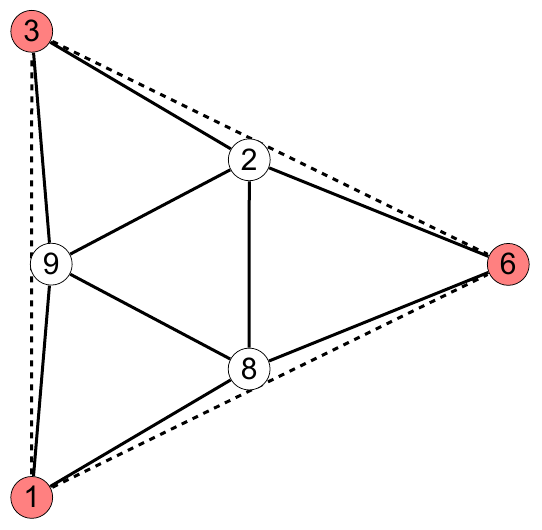}}}
        \end{aligned}
\end{equation}
The last integral in the second line depends only on three external points. Since this is a conformal integral, we can always take one point to infinity. This integral depends only on two external points, and it is $20\zeta_{5}$. The inverse boxing of a constant number corresponds to the box integral. So this DCI integrals generated from $f_{7}^{(5)}$ is $20\zeta_{5}B_{10}(\mz)$ where $B_{10}(\mz)$ is the two loop ladder in our convention. It can be immediately derived that $P(f_{7}^{(5)})=(20\zeta_{5})^2$, one weight less than the standard five-loop periods which are of weight 11. Such weight-dropping phenomena are related to the appearance of DCI integrals depending only on three external points. They are usually evaluated to be multiple zeta values in lower loops. Besides the three-loop example, there are also higher-loop ones such as
\begin{equation}
  \vcenter{\hbox{\includegraphics[scale=0.35]{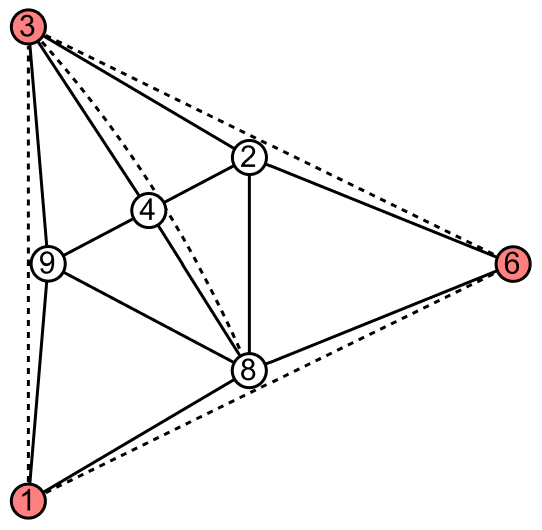}}} = 70\zeta_{7}.
\end{equation}
We remind the readers here that, for general $f$-graphs which we do not study here, these ``two-point'' diagrams originating from the recursive solving of boxing differential equations can be very complicated, and they may even not be of uniform transcendental weight. For example, in eight-loop case, there is an $f$-graph,
\begin{equation}
  f_{75}^{(8)}=\vcenter{\hbox{\includegraphics[scale=0.35]{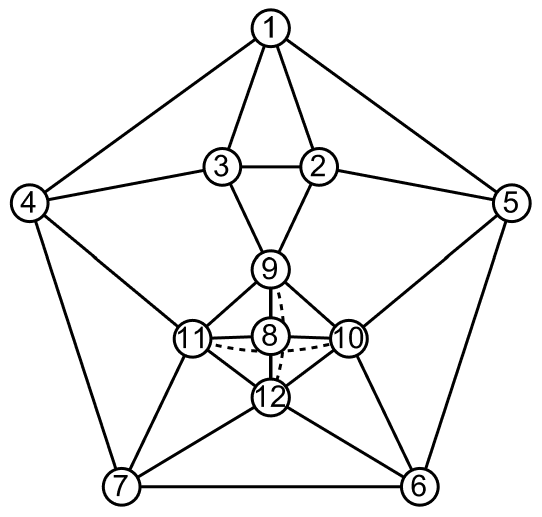}}}
\end{equation}
Its period can be calculated as
\begin{equation}
\begin{aligned}
    P(f_{75}^{(8)})=&P(B_{10}(\mz))\times  \vcenter{\hbox{\includegraphics[scale=0.35]{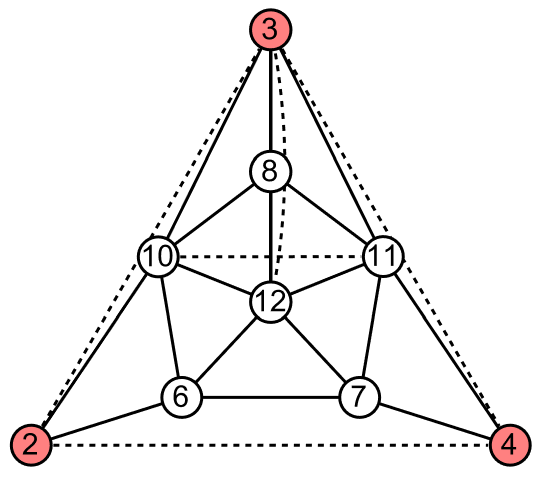}}} \\
    =&20\zeta_{5}\times\left(-29300\zeta_{11}+800\zeta_{5}^{2}+640(f_{3,3,5}+f_{3,5,3}-2f_{5,3,3})\right),
\end{aligned}
\end{equation}
where $f_{3,3,5}+f_{3,5,3}-2f_{5,3,3}$ has been presented in \eqref{eq:f533}. $P(f_{75}^{(8)})$ is not of uniform weight because of the weight-$10$ sub-part $800\zeta_{5}^{2}$.
We will not consider such cases but focus on $f$-graphs whose periods are of uniform weight as stated earlier. 

As we have discussed earlier, if an $f$-graph is weight-preserving, then it can be solved recursively and ends at a box integral. Their results are SVHPLs, thus finite, and there is no divergence or ambiguity. However, in the weight-dropping case, subtlety related to divergence arises when discussing the boxing differential equations. The boxing for weight-dropping $f$-graphs will end at a p-integral. If this p-integral is divergent, then the original DCI integrals generated from $f$-graph is actually divergent. An even crazy thing is that, a DCI integrals generated from $f$-graph may be divergent in four dimensions but if we still apply the boxing differential equation formally, it may well end at a finite p-integral! We will give such an example later. For simplicity, we will restrict ourselves to finite $f$-graphs. However, there may still be some interesting observations about the leading divergent terms of those divergent $f$-graphs.

Let us first study weight-dropping $f$-graphs which will contribute to the four-point correlators. They have nonzero coefficients. The periods of these ``two-point'' integrals appearing in the weight-dropping $f$-graphs are $P_{n_1,n_2,\ldots,n_r}$ with $n_{i}\ge 2$, just like those binary DCI integrals. The periods of corresponding $f$-graphs are products of them. For example, in five loops, we have the last integral in \eqref{eq:boxingdiag2} equal to $P_{2}$ and the period of corresponding $f$-graph is $P_{2}^{2}$. In six loops, ``two-point'' integrals evaluated to $P_{2}$ and $P_{3}$ appear and the period of corresponding weight-dropping $f$-graph is $P_{2}P_{3}$. In seven loops, we find ``two-point'' integrals which are $P_{4},P_{2,2},P_{3},P_{2}$ after recursively solving differential equations. And corresponding $f$-graphs will be evaluated to $P_{4}P_{2},P_{3}P_{3},P_{2,2}P_{2}$. In eight loops, such weight-dropping $f$-graphs are evaluated to $P_{4}P_{3},P_{2,2}P_{3},P_{5}P_{2},P_{2,3}P_{2},P_{2}^{3}$. In nine loops, they are $P_{6}P_{2},P_{2,4}P_{2},P_{3,3}P_{2},P_{2,2,2}P_{2},P_{5}P_{3},P_{2,3}P_{3},P_{4}P_{4},P_{2,2}P_{4},P_{2,2}P_{2,2},P_{2}^{2}P_{3}$. This indicates that the periods of weight-dropping types can all be represented as products of periods from lower-loop weight-preserving types. The weight-dropping phenomena in lower loops will repeat in higher loops. For example, in eight loops there is $P_{5}P_{2}$ which is a product of five-loop period and two-loop period. However, we know that five-loop period can have a weight drop to $P_{2}^{2}$. Then there will be another period $P_{2}^{3}$ in eight loops. $P_{2}^{2}P_{3}$ in nine loops is another example. These weight-dropping $f$-graphs with nonzero coefficients are all finite\footnote{We note here that there are divergent $f$-graphs that have nonzero coefficients as stated in \cite{Bourjaily:2015bpz}. However, up to 9 loops, we have found that none of them can be solved by boxing differential equations and thus they do not belong to the objects we are discussing.}. We remind the readers that here we restrict ourselves to those $f$-graphs that are related to binary DCI integrals. If we relax this condition to all those $f$-graphs that can be solved by boxing differential equations, the ternary-type p-integrals will also appear starting from 9 loops, which correspond to the ternary DCI integrals discussed in App.~\ref{app:notbinary}. 

Although weight-dropping $f$-graphs with zero coefficients are not related to physical objects, their periods are still interesting. Weight-dropping $f$-graphs with zero coefficients are usually related to divergences. Such divergences are studied in \cite{Bourjaily:2015bpz} and classified as $k$-divergences. To understand what has happened here, we give a seven-loop example. The following $f$-graph
\begin{equation}
  f_{206}^{(7)}=\vcenter{\hbox{\includegraphics[scale=0.35]{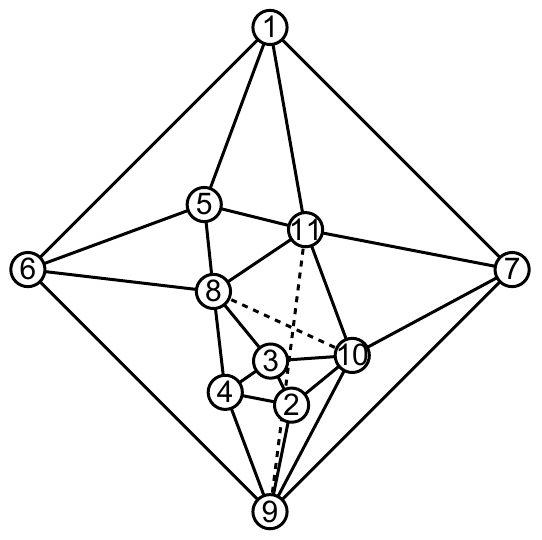}}}
\end{equation}
involves the $k=4$ divergences and has coefficient 0. It can generate a DCI integrals integral $\mathcal{I}_{206}^{(7)}$ by taking four cycle of $1,5,8,11$. The boxing sequence of this integral is (formally)
\begin{equation}\label{eq:boxingdiagf206}
        \begin{aligned}
        &\mathcal{I}_{206}^{(7)}=\vcenter{\hbox{\includegraphics[scale=0.35]{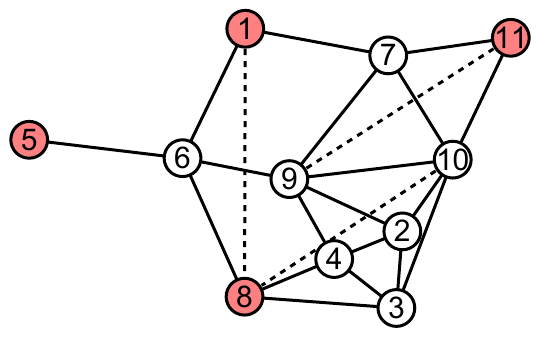}}}\xrightarrow{\frac{x_{15}^2x_{58}^{2}}{x_{18}^2}\square_{5}}\vcenter{\hbox{\includegraphics[scale=0.35]{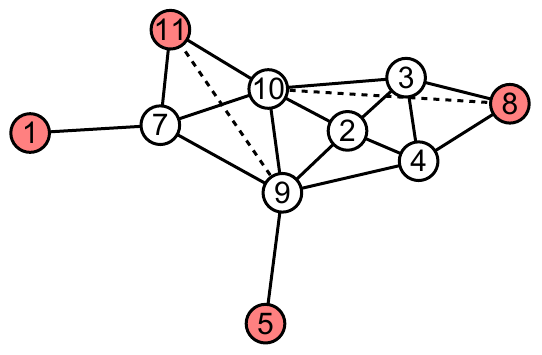}}}\xrightarrow{\frac{x_{15}^{2}x_{1,11}^2}{x_{5,11}^2}\square_{1}} \\ 
        &\frac{x_{15}^{2}x_{8,11}^2}{x_{18}^{2}x_{5,11}^{2}}\times\vcenter{\hbox{\includegraphics[scale=0.35]{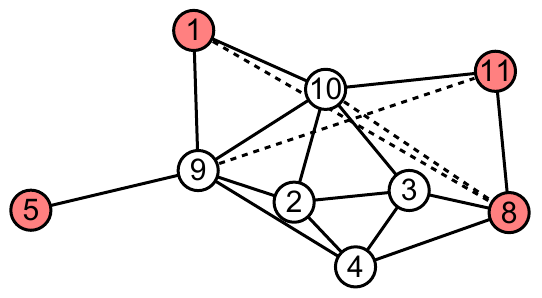}}} \xrightarrow{\frac{x_{15}^2x_{58}^{2}}{x_{18}^2}\square_{5}}\vcenter{\hbox{\includegraphics[scale=0.35]{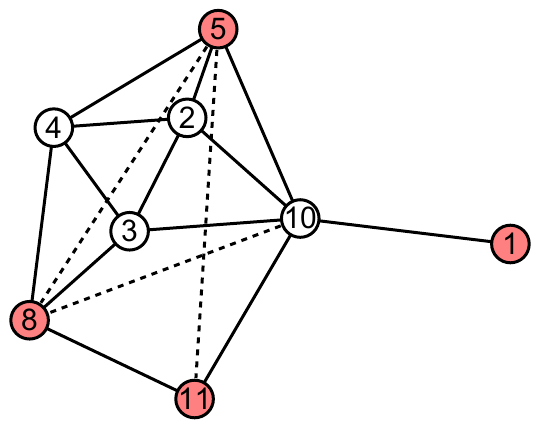}}}\xrightarrow{\frac{x_{15}^2x_{1,11}^{2}}{x_{5,11}^2}\square_{1}}\vcenter{\hbox{\includegraphics[scale=0.35]{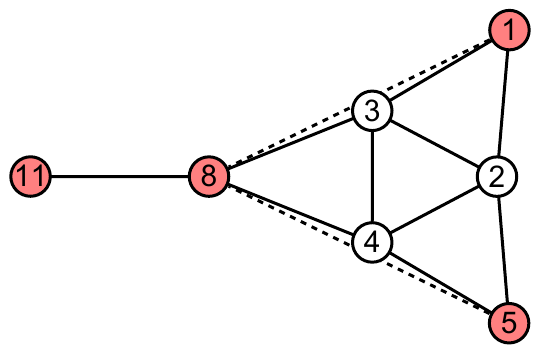}}} \\
        &=\frac{1}{x_{18}^2x_{5,11}^2}\times\frac{x_{18}^2x_{5,11}^2}{x_{15}^{2}x_{8,11}^2}\vcenter{\hbox{\includegraphics[scale=0.3]{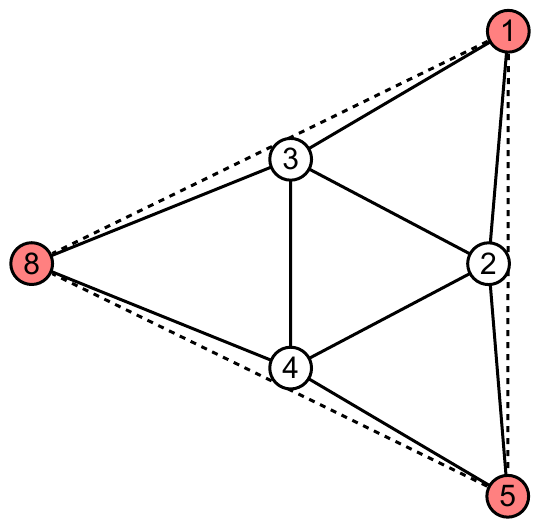}}}.
        \end{aligned}
\end{equation}
As we can see, the second integral in the second row has a superficial divergence if we first fix $x_5\to \infty$. This indicates that all the upriver integrals in this sequence have subdivergences. However, we observe that the leading divergence of the first integral in the second row is proportional to the result of the integral in the last row which is $20\zeta_{5}$. This indicates that the leading divergence of $f^{(7)}_{206}$ is proportional to $20\zeta_{5}B_{10}$. If we define the periods of $f^{(7)}_{206}$ as the periods of its leading divergence, then this period is proportional to $(20\zeta_{5})^2$. Let us show how these can be derived from the boxing differential equation. The last boxing differential equation is 
\begin{equation}\label{eq:diffdoublepole}
    -\mz\mzb\partial_{\mz}\partial_{\mzb}f(\mz)=\frac{1}{\frac{\mz\mzb}{(1-\mz)(1-\mzb)}}\frac{\mz-\mzb}{(1-\mz)(1-\mzb)}C=\frac{\mz-\mzb}{\mz\mzb}C,
\end{equation}
where $C=20\zeta_{5}$. We recall that here $u\equiv \frac{\mz-\mzb}{(1-\mz)(1-\mzb)}=\frac{x_{15}^{2}x_{8,11}^2}{x_{18}^{2}x_{5,11}^{2}}$. And we can compare it with the boxing differential equation for box $B_{1}(\mz)$:
\begin{equation}\label{eq:diffbox}
    -\mz\mzb\partial_{\mz}\partial_{\mzb}B_{1}(\mz)=\frac{\mz-\mzb}{(1-\mz)(1-\mzb)}.
\end{equation}
A huge difference between \eqref{eq:diffdoublepole} and \eqref{eq:diffbox} is that the former involves the double pole of $\mz,\mzb$ around 0 when we move the $\mz\mzb$ from the left hand side to the right. We can ``regularize'' such double pole by introducing a regulator $(\mz\mzb)^{\epsilon}$ to the right hand side:
\begin{equation}\label{eq:diffreg}
    -\mz\mzb\partial_{\mz}\partial_{\mzb}f(\mz)=C\frac{\mz-\mzb}{\mz\mzb}(\mz\mzb)^{\epsilon}.
\end{equation}
Single-valued solution for the above differential equation with the pole ``regularized'' would be 
\begin{equation}
    f(\mz)=\frac{C}{\epsilon(1-\epsilon)}\frac{\mz-\mzb}{\mz\mzb}(\mz\mzb)^{\epsilon}.
\end{equation}
$f(\mz)$ is the eigenfunction of $-\mz\mzb\partial_{\mz}\partial_{\mzb}$ with eigenvalue $\epsilon(1-\epsilon)$. So when we recursively solve \eqref{eq:diffreg}, each recursion introduces a factor $1/\epsilon/(1-\epsilon)$ until the right hand side is normalized by $u$ again. For example, the first integral in the second row multiplied by its normalization factor $u$ is
\begin{equation}
    \frac{\mz\mzb}{(1-\mz)(1-\mzb)}\times \frac{C}{\epsilon^2(1-\epsilon)^2}\frac{\mz-\mzb}{\mz\mzb}(\mz\mzb)^{\epsilon}=\frac{C}{\epsilon^2(1-\epsilon)^2}\frac{\mz-\mzb}{(1-\mz)(1-\mzb)}(\mz\mzb)^{\epsilon}
\end{equation}
Taking $\epsilon\to 0$, its leading divergent part is $\frac{\mz\mzb}{(1-\mz)(1-\mzb)}$ and the differential equation goes back to \eqref{eq:diffbox} and the remaining solution is regular like finite integrals. A direct consequence of above regularization is that the boxing sequence is chopped at the second row and we can directly replace the first integral in the second row with $C=20\zeta_{5}$. If we define the period of divergent $f$-graphs as their leading divergent parts regularized in the above way, then the period of $f^{(7)}_{206}$ is $(20\zeta_{5})^{2}$, which agrees with the result of \texttt{HyperlogProcedures}. We can easily extend the above example to more general cases. The periods of them have an interesting correspondence with the finite weight-dropping $f$-graphs type we present in the last paragraph, due to the chopped boxing sequence. For example, in seven loops, it is $P_{2}P_{2}$. It can be derived from $P_{2,2}P_{2}$ by removing the first or last index from $P_{2,2}$. In the same way, we will have $P_{2}P_{3}$ and $P_{2}P_{2}$ in eight loops, corresponding to $P_{2,3}P_{2}$. In nine loops, we have found $P_{2}P_{2},P_{3}P_{2},P_{4}P_{2},P_{2,2}P_{2},P_{3}P_{3}$, corresponding to $P_{2,4}P_{2},P_{3,3}P_{2},P_{2,2}P_{2},P_{2,3}P_{3},P_{2,2}P_{4},P_{2,2}P_{2,2}$. However, we find up to nine loops that such kind of divergent $f$-graphs will all have zero coefficients. We conjecture that all these $f$-graphs are related to k=4 divergences and thus will not contribute to correlators for all loops.

\section{Proof of proposition 2}\label{app:proof}
Here we demonstrate the proof of Proposition~\ref{prop:reverse} step by step. First, we consider a planar $f$-graph which corresponds to some bianry DCI integral $B_{n_1,n_2,\ldots,n_{r}}(\mz)$ with $n_{i}\ge 2$ (e.g. $B_{2,3}(\mz)\equiv B_{10100}(\mz)$). We can choose four points which form a four cycle without inner points as external points\footnote{We require no inner points because we do not want to split the $f$-graph into two parts, in which case we will get products of DCI integrals instead of a single one.}. They are named $x_1,x_2,x_3,x_4$ with a chosen order. Then the boxing operations starting from $x_1,x_2,x_3,x_4$ can be translated into a set of operations on the diagrams of DCI integrals as in \eqref{eq:boxingdiag}. They can also be expressed as operations on corresponding $f$-graph. Let us show this process using examples in \eqref{eq:f5l5}:
\begin{equation}\label{eq:b10010}
        \begin{aligned}
        &\vcenter{\hbox{\vbox{\hbox to 3.5cm{\hfill \includegraphics[scale=0.4]{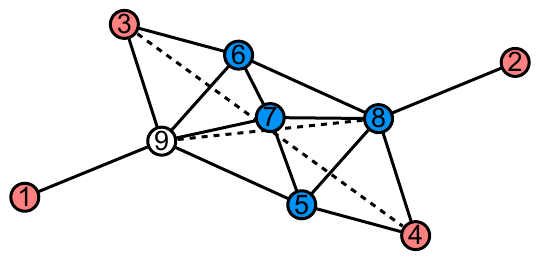} \hfill}\hbox to 3.5cm{\hfill \scriptsize $B_{10010}(\mz)$ \hfill}}}}\xrightarrow[\text{remove 0}]{\frac{x_{13}^{2}x_{14}^{2}}{x_{34}^2}\square_{1}} \vcenter{\hbox{\vbox{\hbox to 3.8cm{\hfill \includegraphics[scale=0.35]{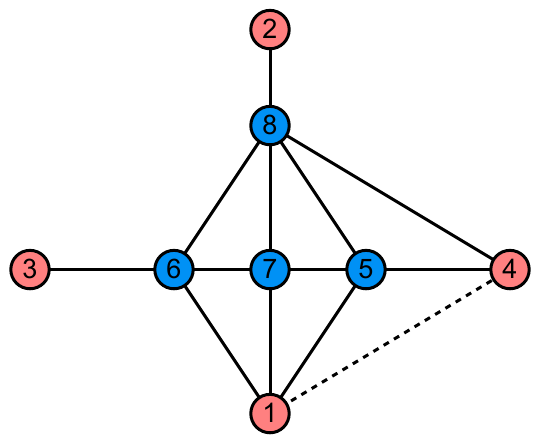} \hfill}\hbox to 3.8cm{\hfill \scriptsize $B_{1001}(\mz)$ \hfill}}}}
        \xrightarrow[\text{remove 1}]{{\color{orange}\frac{x_{12}^{2}x_{34}^{2}}{x_{14}^{2}x_{23}^{2}}\times}\frac{x_{23}^{2}x_{13}^{2}}{x_{12}^2}\square_{3}}\vcenter{\hbox{\vbox{\hbox to 3cm{\hfill \includegraphics[scale=0.35]{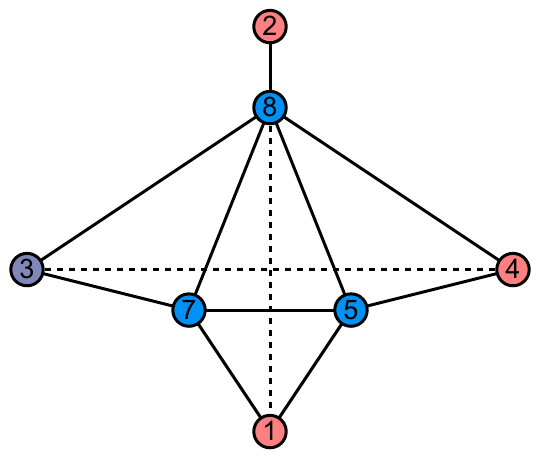} \hfill}\hbox to 3cm{\hfill \scriptsize $B_{100}(\mz)$ \hfill}}}}\\
        &\xrightarrow[\text{remove 0}]{\frac{x_{23}^{2}x_{24}^{2}}{x_{34}^2}\square_{2}}\vcenter{\hbox{\vbox{\hbox to 3.5cm{\hfill \includegraphics[scale=0.35]{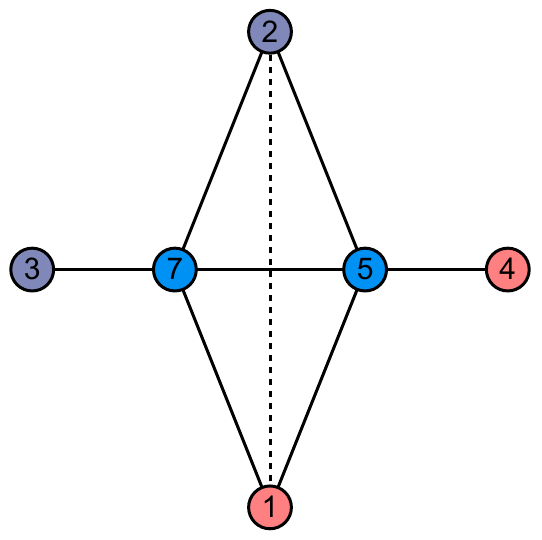} \hfill}\hbox to 3.5cm{\hfill \scriptsize $B_{10}(\mz)$ \hfill}}}}\xrightarrow[\text{remove 0}]{\frac{x_{24}^{2}x_{14}^{2}}{x_{12}^2}\square_{4}}\vcenter{\hbox{\vbox{\hbox to 3.5cm{\hfill \includegraphics[scale=0.35]{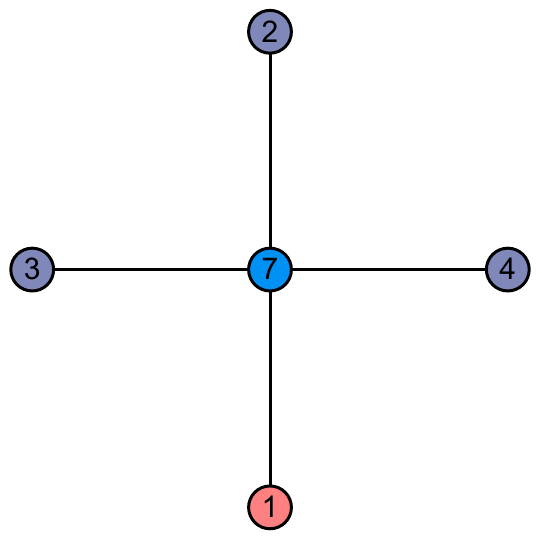} \hfill}\hbox to 3.5cm{\hfill \scriptsize $B_{1}(\mz)$ \hfill}}}}\xrightarrow[\text{remove 1}]{\frac{x_{13}^{2}x_{14}^{2}}{x_{34}^2}\square_{1}} \!\!\vcenter{\hbox{\vbox{\hbox to 3cm{\hfill \includegraphics[scale=0.3]{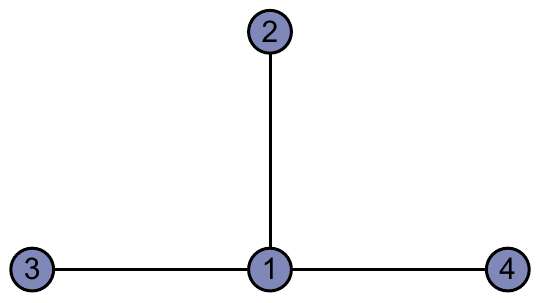} \hfill}\hbox to 3cm{\hfill \scriptsize $\frac{\mz\mzb}{(1-\mz)(1-\mzb)}$ \hfill}}}}
        \end{aligned}
\end{equation}
The boxing operator is acted on the start of an arrow and results in the expression in its end. In the second step, we perform a re-normalization (colored orange) for the third graph which is why `1' instead of `0' is removed. Above operation can be represented as a flow in the original $f$-graph:
\begin{equation}\label{eq:I5f5path}
    \begin{aligned}
        &\vcenter{\hbox{\includegraphics[scale=0.4]{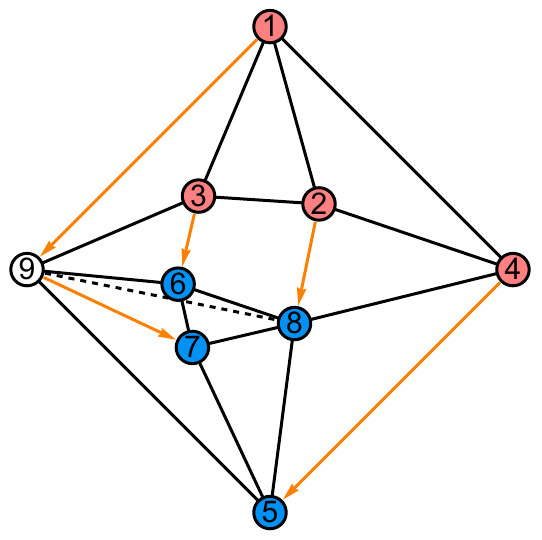}}}\!\xrightarrow[\text{remove 0}]{\frac{x_{13}^2x_{14}^{2}}{x_{34}^2}}\!\vcenter{\hbox{\includegraphics[scale=0.4]{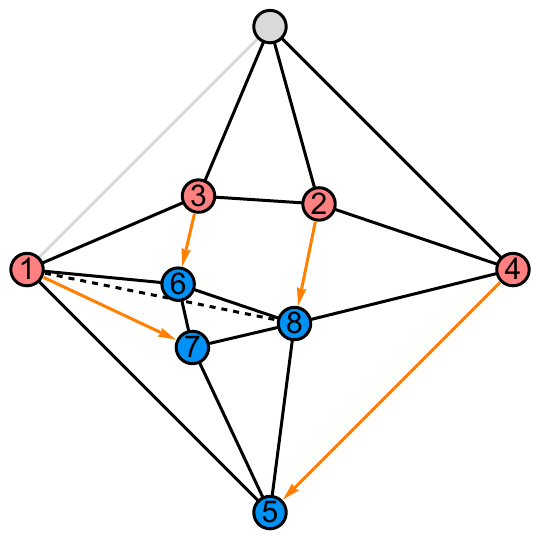}}}\xrightarrow[\text{remove 1}]{{\color{orange}\frac{x_{12}^{2}x_{34}^{2}}{x_{14}^{2}x_{23}^{2}}\times}\frac{x_{23}^{2}x_{13}^{2}}{x_{12}^2}} \vcenter{\hbox{\includegraphics[scale=0.4]{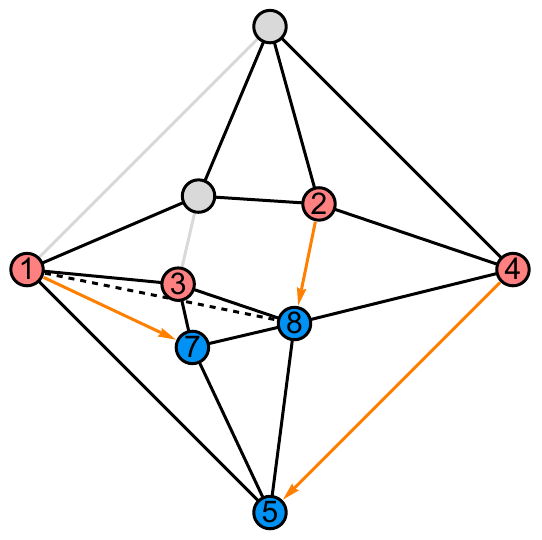}}} \\
        &\!\!\xrightarrow[\text{remove 0}]{\frac{x_{23}^{2}x_{24}^{2}}{x_{34}^2}}\!\!\vcenter{\hbox{\includegraphics[scale=0.4]{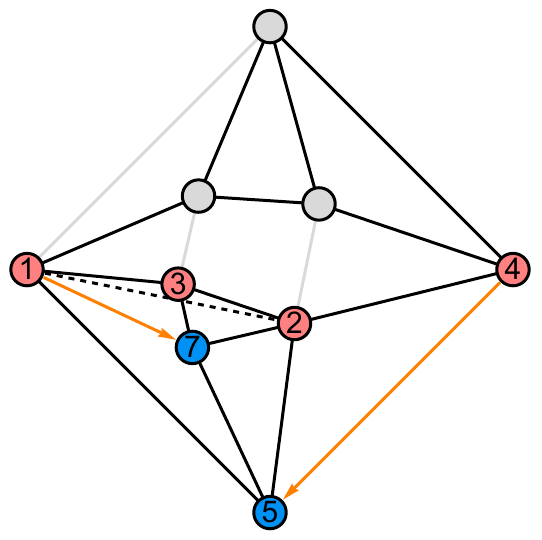}}}\!\!\xrightarrow[\text{remove 0}]{\frac{x_{24}^{2}x_{14}^{2}}{x_{12}^2}}\!\!\vcenter{\hbox{\includegraphics[scale=0.4]{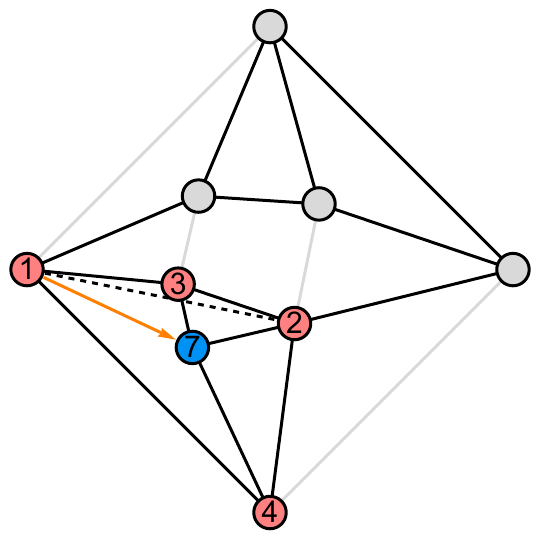}}} \!\!\xrightarrow{\frac{x_{13}^{2}x_{14}^{2}}{x_{34}^2}} \!\!\vcenter{\hbox{\includegraphics[scale=0.4]{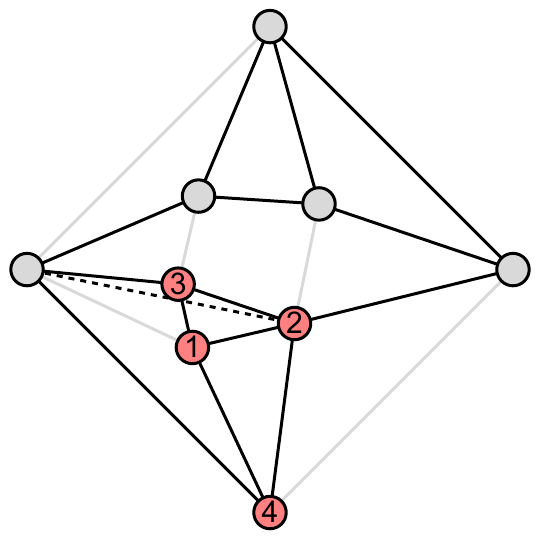}}}
    \end{aligned}
\end{equation}
where the light gray lines should be pinched and in each step the factor on the arrow should be multiplied. These factors come from boxing or re-normalization and only consist of the original four external points $x_{1},x_{2},x_{3},x_{4}$. In this example, $x_{1},x_{2},x_{3},x_{4}$ are finally identified with $x_{7},x_{8},x_{6},x_{5}$ in the $f$-graph. In the general case, the weight-preserving condition \footnote{The weight-preserving condition here is the same as the requirement that the binary DCI integrals evaluated to $B_{n_1,n_2,\ldots,n_{r}}(\mz)$ can be solved recursively until it ends as a box and we elaborate on the concept of weight-preserving and weight-dropping $f$-graph in App.~\ref{app:weight}.} in Proposition~\ref{prop:reverse} guarantees that the original four points we start with will end with and only with four points after a series of boxing operations. We can always name the final four points as $x_{5},x_{6},x_{7},x_{8}$. 

\begin{figure}[htbp]
    \centering
    \includegraphics[scale=0.45]{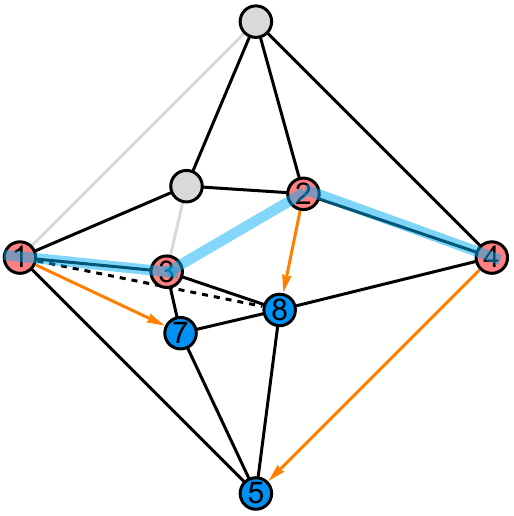}
    \hspace{10pt}
    \includegraphics[scale=0.45]{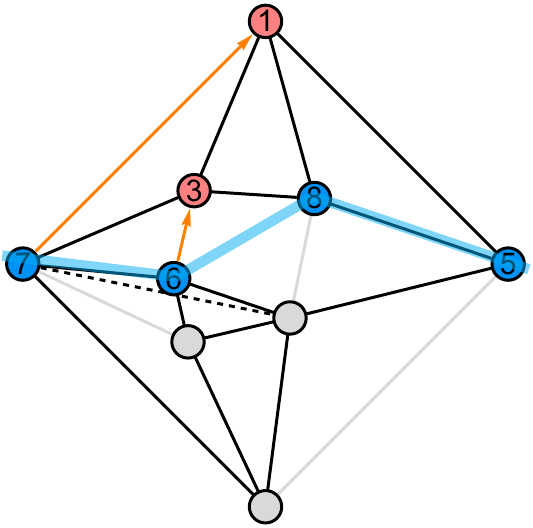}
    \caption{We mark the ``squall lines'' of the third graph in two contrary boxing sequence \eqref{eq:I5f5path} and \eqref{eq:I5f5pathrev} out with thick blue lines. They coincide with each other and there is a correspondence between $x_{1},x_{2},x_{3},x_{4}$ and $x_{7},x_{8},x_{6},x_{5}$.}
    \label{fig:squallline}
\end{figure}

Now we can directly reverse above process by starting from the four cycle formed by $x_{5},x_{6},x_{7},x_{8}$. The boxing direction is reversed and we can directly replace $x_{1},x_{2},x_{3},x_{4}$ with $x_{7},x_{8},x_{6},x_{5}$ in each step since they are identified in each pinch step. We depict the reverse process of above example:
\begin{equation}\label{eq:I5f5pathrev}
    \begin{aligned}
        &\vcenter{\hbox{\includegraphics[scale=0.4]{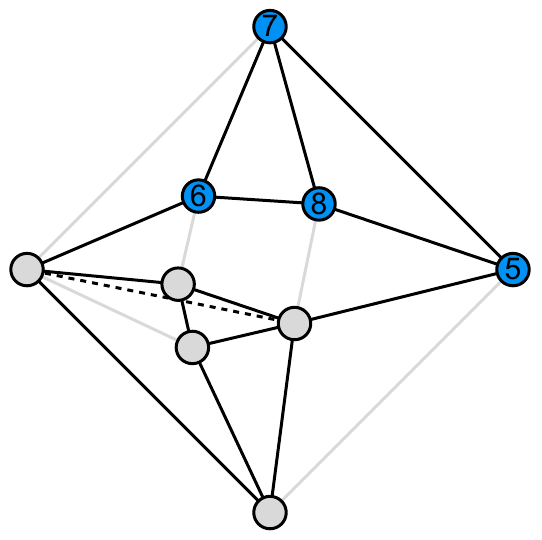}}}\!\xleftarrow{\frac{x_{67}^{2}x_{57}^{2}}{x_{56}^2}}\!\vcenter{\hbox{\includegraphics[scale=0.4]{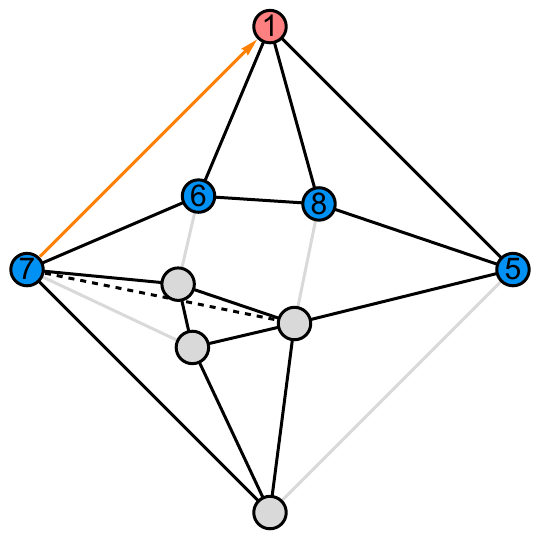}}}\xleftarrow[\text{remove 0}]{\frac{x_{67}^{2}x_{68}^{2}}{x_{78}^2}} \vcenter{\hbox{\includegraphics[scale=0.4]{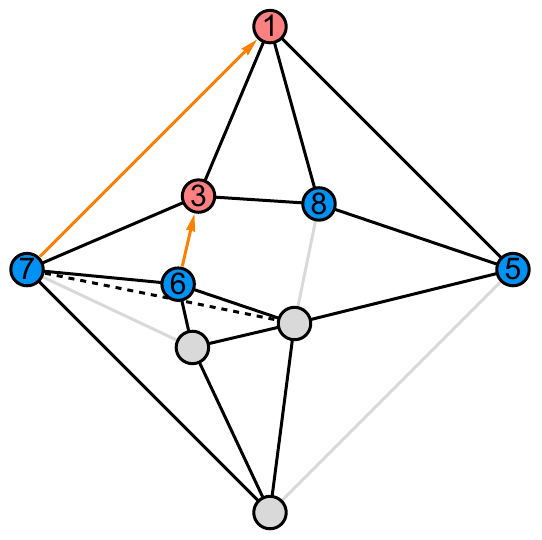}}} \\
        &\!\!\xleftarrow[\text{remove 1}]{{\color{orange}\frac{x_{56}^{2}x_{78}^{2}}{x_{57}^{2}x_{68}^{2}}\times}\frac{x_{86}^{2}x_{85}^{2}}{x_{56}^2}}\!\!\vcenter{\hbox{\includegraphics[scale=0.4]{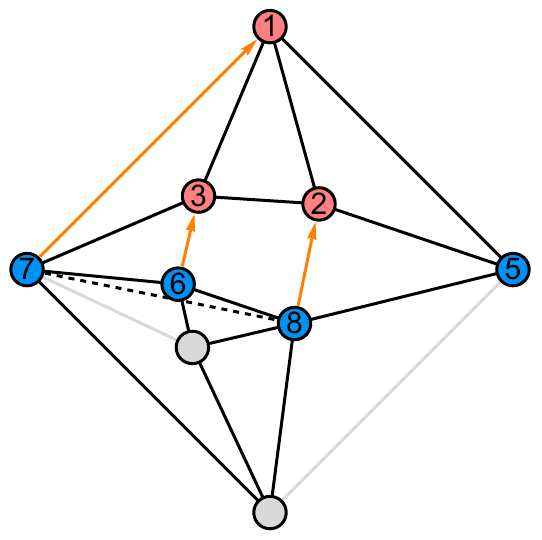}}}\!\!\xleftarrow[\text{remove 0}]{\frac{x_{58}^{2}x_{57}^{2}}{x_{78}^2}}\!\!\vcenter{\hbox{\includegraphics[scale=0.4]{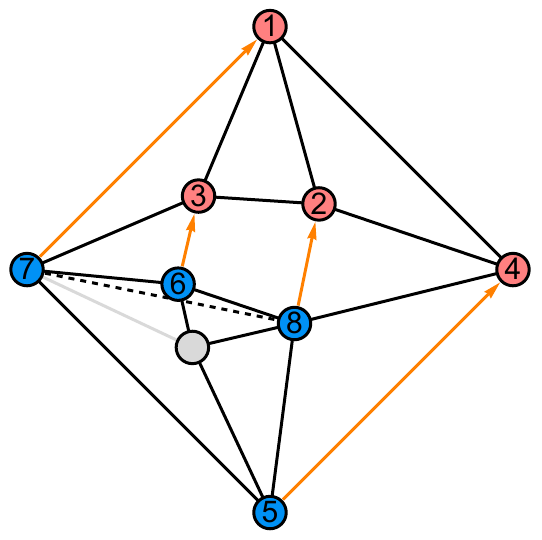}}} \!\!\xleftarrow[\text{remove 0}]{\frac{x_{67}^{2}x_{57}^{2}}{x_{56}^2}} \!\!\vcenter{\hbox{\includegraphics[scale=0.4]{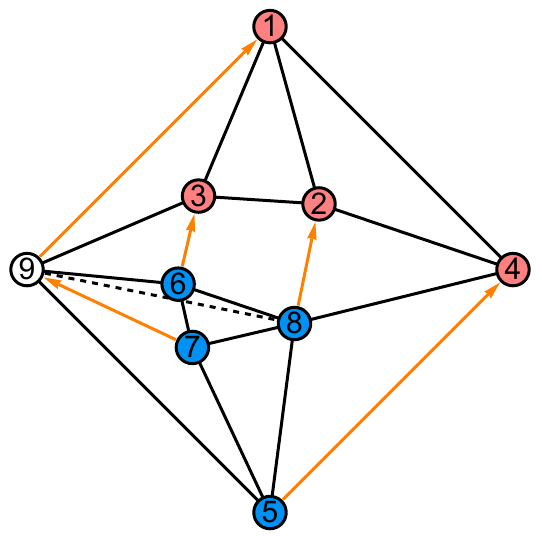}}}
    \end{aligned}
\end{equation}
The light gray line should be pinched in above equations. Here we note that \eqref{eq:I5f5pathrev} is still a boxing sequence with the direction from right-bottom to left-top. It is \textit{not} the inverse boxing (which solves the boxing differential equation) we mentioned in the main text. The key point is that, if from one direction, some $f$-graph need to be re-normalized\footnote{Recall that re-normalization is needed when the external four points are entangled together. We re-normalize the integral to make sure some external points is attached and only attached to one inner points, then the boxing can be carried on. So whether a re-normalization is needed only depend on the connection state of the external four points.} (e.g. the third graph in \eqref{eq:I5f5path}), then in the reverse direction this $f$-graph (the third graph in \eqref{eq:I5f5pathrev}) should also be re-normalized to continue the boxing sequence. The re-normalization factor is only relevant to the four external points. However, since the flow in \eqref{eq:I5f5pathrev} is the reverse of \eqref{eq:I5f5path}, the ``squall lines'' of these two boxing sequences will always coincide with each other with $x_{1},x_{2},x_{3},x_{4}$ replaced by $x_{7},x_{8},x_{6},x_{5}$ as indicated in Fig.~\ref{fig:squallline}.
This ensures that in each step of the two boxing sequences, the word removed is the same. Only the order is reversed. Then \eqref{eq:I5f5pathrev} indicates that the DCI integral with $x_{7},x_{8},x_{6},x_{5}$ as external points is $B_{10100}(\mz)$. In contrast, \eqref{eq:I5f5path} states that the DCI integrals with $x_{1},x_{2},x_{3},x_{4}$ as external points corresponds to $B_{10010}(\mz)$. Then we have shown that if the binary DCI integral $B_{10010}(\mz)$ can be identified in the $f$-graph, so will $B_{10100}(\mz)$. They are dual to each other. 

Above proof for this special case can be directly generalized to arbitrary weight-preserving $f$-graph as long as it corresponds to some binary DCI integral $B_{n_1,n_2,\ldots,n_{r}}(\mz)$ with $n_{i}\ge 2$. We first identify the boxing sequence of $B_{n_1,n_2,\ldots,n_{r}}(\mz)$ and then the boxing sequence of $B_{n_r,n_{r-1},\ldots,n_{1}}(\mz)$ can be identified by reversing the process. They are guaranteed to appear in the same time. Proposition~\ref{prop:reverse} is proved. Finally, we comment on the condition $n_{i}\ge 2$ in $B_{n_1,n_2,\ldots,n_{r}}(\mz)$. Only such integrals can be embedding into a \textit{planar} $f$-graphs by taking four cycles. Our proof relies on this. A single `1' in the words usually indicates non-planarity. See App.~\ref{app:Bxxx1} for more discussions.

\section{From binary DCI integrals to \texorpdfstring{$f$-graphs}{f-graphs}}\label{app:Bxxx1}
In the main text, we have discussed how to construct $f$-graphs for binary DCI integrals whose words end with '0' and in this appendix, we will show that how to construct $f$-graphs for binary DCI integrals whose words end with '1'.

There are two ways to find an $f$-graph that generates binary DCI integrals whose words end with '1'. The first way is to directly supplement the binary DCI integrals to an $f$-graph. Let us take the zigzag DCI integrals $B_{101}$ as an example:
\begin{equation}\label{eq:b101f}
        \begin{aligned}
        &\vcenter{\hbox{\includegraphics[scale=0.35]{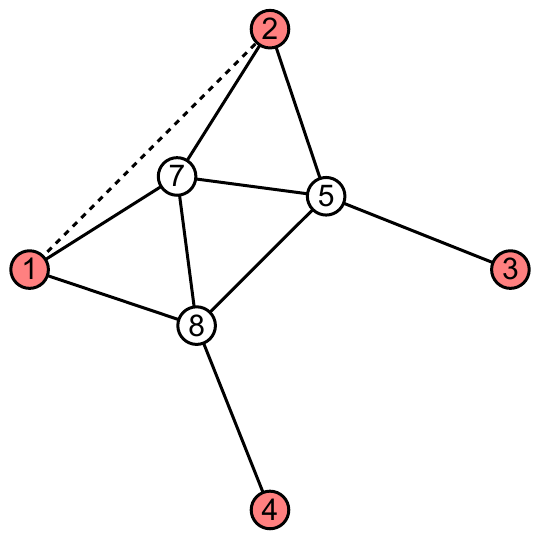}}}\xrightarrow[\frac{1}{x_{12}^2x_{23}^{2}x_{34}^{2}x_{14}^2x_{13}^{2}x_{24}^{2}}]{\text{supplementing with}}\vcenter{\hbox{\includegraphics[scale=0.35]{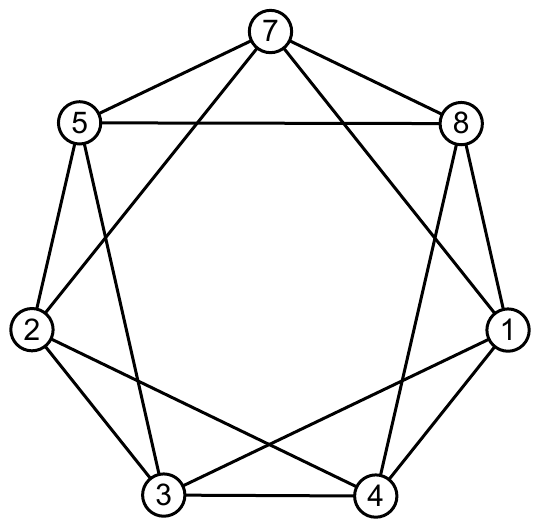}}}.
        \end{aligned}
\end{equation}
The resulting $f$-graph is non-planar. For general binary DCI integrals whose words end with '1', they have two main characters that follow from the same argument about those whose words end with '0'. First, there is (at least) one external leg, for example $x_{4}$ without loss of generality, which is attached to a single inner point without any other attachments. Second, $x_1$ and $x_{3}$ are \textit{not} connected with each other by a dashed line. In such cases, $f$-graphs from directly supplementing the corresponding binary DCI integrals like \eqref{eq:b101f} will be non-planar due to two crossed lines (denominators) $x_{13}^{2}$ and $x_{24}^{2}$. However, we prefer to generate these DCI integrals from planar $f$-graphs, which are now better understood. This can be done by taking another substructure rather than four-cycles of planar $f$-graphs. For example, the same zigzag DCI integrals $B_{101}$ can be obtained by
\begin{equation}\label{eq:gen1}
        \begin{aligned}
        &\vcenter{\hbox{\includegraphics[scale=0.35]{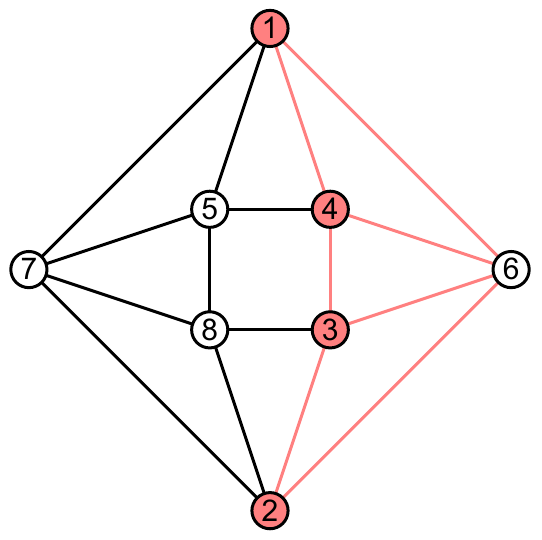}}}\xrightarrow[\xi_{4}]{\text{multiplying}}\vcenter{\hbox{\includegraphics[scale=0.35]{fig/B101.pdf}}}\times \vcenter{\hbox{\includegraphics[scale=0.35]{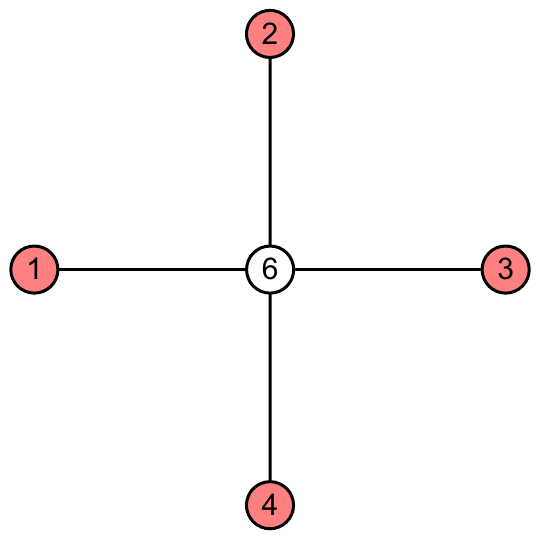}}}.
        \end{aligned}
\end{equation}
The external points are colored pink. Though such conformal integrals will not contribute to the four-point planar amplitudes, they do contribute to the four-point correlators. The substructure with red lines generally exists in planar $f$-graphs. Then we can generate $L$-loop binary DCI integrals whose words end with '1' from $(L+1)$-loop f-graphs. For example, all the odd-loop zigzag DCI integrals can be generated from antiprism $f$-graphs in such a way. An interesting identity we can derive from above operation is,
\begin{equation}\label{eq:zigzagid}
    P(B_{1010}(\mz))=P^{\prime}(B_{101}(\mz)\times B_{1}(\mz)),
\end{equation}
since the ``external points'' will all be integrated out and the periods of $f$-graph are independent of which four points being chosen as ``external points''.

 For binary DCI integrals whose corresponding words end with '0', we have defined a canonical series of $f$-graphs, $\mathcal{F}$. For the $L$-loop canonical binary DCI integrals whose corresponding word end with '1', we can multiply the integrand with $1/(x_{12}^{2}x_{23}^{2}x_{34}^{2}x_{14}^{2}x_{15}^{2}x_{25}^{2}x_{35}^{2}x_{45}^{2})$ where $5$ is an inner point, which brings it to an $(L+1)$-loop $f$-graph. We have checked up to ten loops that all these $f$-graphs belong to $\mathcal{F}$ as well. It can be directly proven at the graph level that any $f$-graph which can generate $B_{w},w=w^{\prime}1$ by multiplying $(x_{12}^{2}x_{23}^{2}x_{34}^{2}x_{14}^{2}x_{15}^{2}x_{25}^{2}x_{35}^{2}x_{45}^{2})$ can also generate $B_{w0}=B_{w^{\prime}10}$ by taking the four-cycle of $1,4,3,5$, that is, multiplying $x_{14}^{2}x_{43}^{2}x_{35}^{2}x_{15}^{2}x_{13}^{2}x_{45}^{2}$. For example,
\begin{equation}
    \begin{aligned}
        &\vcenter{\hbox{\vbox{\hbox to 4cm{\hfill \includegraphics[scale=0.3]{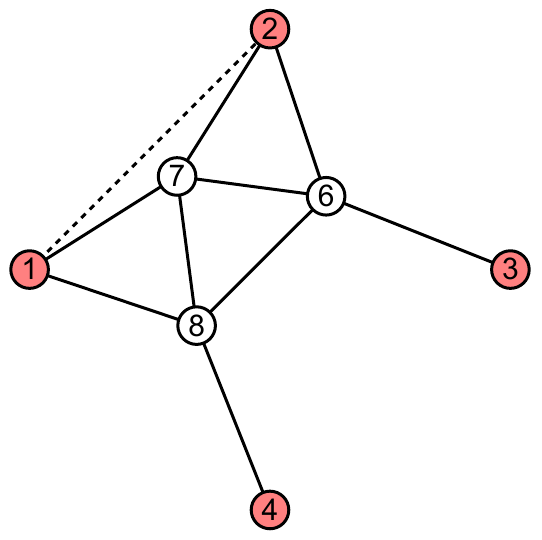} \hfill}\hbox to 4cm{\hfill \scriptsize $B_{101}(\mz)$ \hfill}}}}\!\!\!\!\!\!\!\xrightarrow[1/(x_{12}^{2}x_{23}^{2}x_{34}^{2}x_{14}^{2}x_{15}^{2}x_{25}^{2}x_{35}^{2}x_{45}^{2})]{\text{multiply}}\vcenter{\hbox{\includegraphics[scale=0.35]{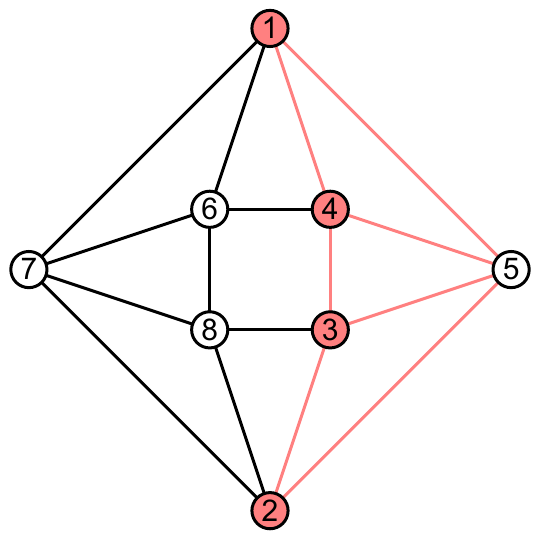}}}=\vcenter{\hbox{\includegraphics[scale=0.35]{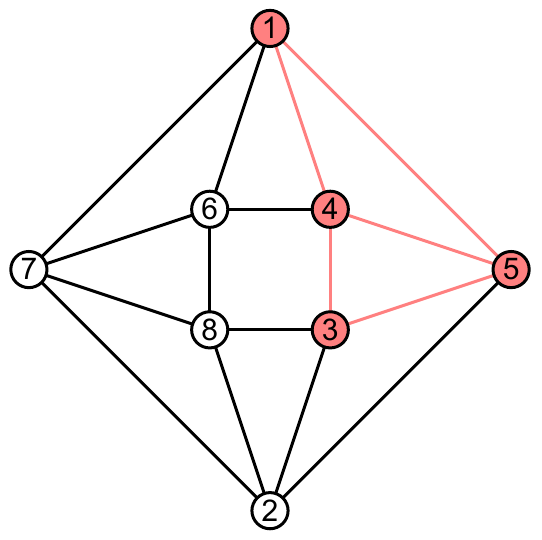}}} \\
        &\xrightarrow[x_{14}^{2}x_{43}^{2}x_{35}^{2}x_{15}^{2}x_{13}^{2}x_{45}^{2}]{\text{multiply}} \vcenter{\hbox{\vbox{\hbox to 4cm{\hfill \includegraphics[scale=0.4]{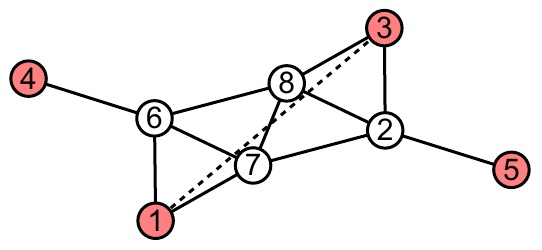} \hfill}\hbox to 4cm{\hfill \scriptsize $B_{1010}(\mz)$ \hfill}}}}\xrightarrow[\text{renaming 5 as 2 finally}]{\frac{x_{35}^{2}x_{15}^{2}}{x_{13}^{2}}\square_{5}}\vcenter{\hbox{\vbox{\hbox to 4cm{\hfill \includegraphics[scale=0.3]{fig/B101a.pdf} \hfill}\hbox to 4cm{\hfill \scriptsize $B_{101}(\mz)$ \hfill}}}}
    \end{aligned}
\end{equation}
Therefore, $\mathcal{F}$ can generate any binary Steinmann DCI integrals we want.

Though we have constructed a canonical series of $f$-graphs for binary Steinmann SVHPLs, we should keep in mind that one binary Steinmann SVHPL can usually correspond to many different $f$-graphs. This is natural, since $f$-graphs contain the information of integrands corresponding to a binary Steinmann SVHPL. Let us still take the zigzag series as examples. Starting from six loops\footnote{For lower loops, the constraints from conformal property are strong enough to give a single $f$-graph for zigzag DCI integrals.}, there is more than one $f$-graph that can generate zigzag DCI integrals. In six loops, except for the antiprism $f$-graph, there is another $f$-graph which can also generate $B_{101010}(\mz)$. It has been given in \eqref{eq:f31l6}.
In eight loops, there are totally 4 $f$-graphs with nonzero coefficients that can generate $B_{10101010}(\mz)$. In general, there will be more and more $f$-graphs corresponding to the same SVHPL. 

    \bibliographystyle{JHEP}
    \bibliography{ref.bib}
\end{document}